\documentclass[journal,draftcls,onecolumn,12pt,twoside]{IEEEtranTCOM}
\normalsize
\usepackage{cite}
\ifCLASSINFOpdf
  \usepackage[pdftex]{graphicx}
\else
  \usepackage[dvips]{graphicx}
\fi
\usepackage[cmex10]{amsmath}
\usepackage{amsfonts}
\usepackage{amsthm,amssymb}
\usepackage{mathrsfs}
\usepackage{algorithm}
\usepackage{algpseudocodex}
\usepackage{array}
\usepackage[caption=false]{subfig}

\usepackage{bm}
\usepackage{tikz}
\usepackage{pgfplots}
\pgfplotsset{compat=newest}
\usepackage{ifthen}
\usetikzlibrary{arrows.meta, shapes.geometric, decorations.pathmorphing, backgrounds, positioning, fit, petri, calc}
\usepackage{pgffor}
\usetikzlibrary{plotmarks}
\usepgfplotslibrary{patchplots}
\usepackage{grffile}
\usepackage{tablefootnote}
\usepackage{nicematrix}
\usepackage[multiple]{footmisc}
\DeclareMathOperator{\erf}{erf}
\DeclareMathOperator{\erfc}{erfc}
\DeclareMathOperator{\supp}{supp}

\newcounter{example}
\newenvironment{example}[1][]{\refstepcounter{example}\par\medskip
   \noindent \textbf{Example~\arabic{example}. #1} \rmfamily}{\medskip}
   
\newtheorem{proposition}{Proposition} 
\newtheorem{definition}{Definition} 
\newcommand{\rchange}[1]{{#1}}

\newcommand*\circled[1]{\tikz[baseline=(char.base)]{
            \node[shape=circle,draw,minimum size=1.6em,inner sep=0pt] (char) {#1};}}

\begin{document}
\title{
  A Random Coding Approach to Performance \\
  Analysis of the Ordered Statistic Decoding \\
  With Local Constraints
}
\author{Jifan~Liang and Xiao~Ma
\thanks{
  This work was supported in part by the National Key R\&D Program of China~(No. 2021YFA1000500). \emph{(Corresponding author: Xiao Ma.)}
}
\thanks{
  The authors are with the School of Computer Science and Engineering and the Guangdong Key Laboratory of Information Security Technology, Sun Yat-sen University, Guangzhou 510006, China (e-mail: liangjf56@mail2.sysu.edu.cn; maxiao@mail.sysu.edu.cn).
}
}

\markboth{IEEE Transactions on Information Theory}%
{Submitted paper}
\maketitle

\begin{abstract}
This paper is concerned with the ordered statistic decoding with local constraints~(LC-OSD) of binary linear block codes, which is a near maximum-likelihood decoding algorithm. 
Compared with the conventional OSD, the LC-OSD significantly reduces both the maximum and the average number of searches. 
The former is achieved by performing 
the serial list Viterbi algorithm~(SLVA) or a two-way flipping pattern tree~(FPT) algorithm with local constraints on the test error patterns, while the latter is achieved by incorporating tailored early termination criteria. 
The main objective of this paper is to explore the relationship between the performance of the LC-OSD and decoding parameters, such as the constraint degree and the maximum list size. 
To this end, we approximate the local parity-check matrix as a totally random matrix and then estimate the performance of the LC-OSD by analyzing with a saddlepoint approach the performance of random codes over the channels associated with the most reliable bits~(MRBs). 
The random coding approach enables us to derive an upper bound on the performance and predict the average rank of the transmitted codeword in the list delivered by the LC-OSD. This allows us to balance the constraint degree and the maximum list size for the average~(or maximum) time complexity reduction. 
Simulation results show that the approximation by random coding approach is numerically effective and powerful. 
Simulation results also show that the RS codes decoded by the LC-OSD can approach the random coding union~(RCU) bounds, verifying the efficiency and universality of the LC-OSD.
\end{abstract}


\IEEEpeerreviewmaketitle

\section{Introduction} 
\label{sec:intro}
As a central problem in coding theory, the maximum-likelihood decoding~(MLD) of a general binary linear code is considered to have no efficient algorithms due to its NP-completeness~\cite{berlekamp1978inherent}.
The ordered statistics decoding~(OSD) investigated by Fossorier and Lin~\cite{Fossorier1995OSD} is a near-optimal list decoding algorithm but requires a large number of searches, leading to high complexity.
The OSD algorithm first sorts the reliability vector and then selects $k$ linearly independent positions with the most reliabilities, referred to as the most reliable basis.
By re-encoding, the OSD lists all codewords with Hamming distance no greater than $t$~(a preset nonnegative integer) in the MRB and delivers the most likely candidate codeword as output.
As a list decoding algorithm, three issues about OSD naturally arise:
1) for what candidate codewords to search, 
2) in which order to search, and
3) when to stop the search.
To solve these issues, many improvements to OSD have been proposed.
For reducing search space, segmentation-discarding OSD~(SD-OSD)~\cite{Yue2019Segmentation} and linear-equation OSD~(LE-OSD)~\cite{yue2022linear} have been proposed;
for early stopping, probability-based OSD~(PB-OSD)~\cite{yue2021probability} can significantly reduce the number of searches. 

Recently, a variant of OSD called OSD with local constraints~(LC-OSD) was proposed further to reduce the decoding complexity~\cite{Wang2022ITW, Liang2022CommL}.
The basic idea of the LC-OSD is to consider an extended subset of reliable positions by including $\delta$ extra positions next to the MRB, which was proposed in a report~\cite{greenberger1979efficient} of NASA as early as 1979 and widely used in the information set decoding~(ISD) algorithms of cryptography~\cite{may2011decoding, becker2012decoding, may2015computing}.
The local constraints can be characterized as a trellis specified by a local parity-check matrix, and a list of candidate codewords can be generated by performing the serial list Viterbi algorithm~(SLVA)~\cite{Wang2022ITW}. 
Aided by tailored early stopping criteria, the LC-OSD can reduce the number of \rchange{searches/re-encodings} to within ten in the high signal-to-noise ratio~(SNR) region~\cite{Liang2022CommL}.
We further proposed an algorithm, the two-way flipping pattern tree, for replacing the SLVA in the LC-OSD algorithm to reduce the overhead of the SLVA in the high SNR~(low average number of searches) region.

The main contributions of the manuscript are listed below.
\begin{enumerate}
  \item 
  Theoretically, we derived an upper bound on the performance gap between the LC-OSD and the MLD performance for a general code.
  The performance gap is estimated by the random coding approach. 
  This implies that the gap between the LC-OSD and the MLD is less relevant to the code structure, suggesting the universality of the LC-OSD. 
  In particular, simulation results show that RS codes can approach the random coding union~(RCU) bound in a wide range of code rates.
  We also proposed a semi-simulation method to speed up the calculation of the proposed bound, which is powered by the saddlepoint method~\cite{Martinez2011Saddlepoint, Font2018Saddlepoint}.
  \item 
  Based on the idea of ``search-on-demand,'' we provided in Sec.~\ref{sec:lva} an implementation of SLVA with an intuitive example.
  For a $\mathscr{C}[N, K]$ binary linear codes, the time complexity of this implementation is $\mathcal{O}(N(2^{N-K} + \ell))$, which is highly reduced from the PLVA~(parallel list Viterbi algorithm), where $\ell$ denotes the average number of searches.
  This implementation is asymptotically lower than that of~\cite{seshadri1994list} and~\cite{Nill1995List} with complexities of $\mathcal{O}(N(2^{N-K} + \ell^2))$ and $\mathcal{O}(N(2^{N-K} + \ell) + \ell^2)$, respectively.
  \item 
  The list-generating algorithm~(LGA) is a procedure generating a list of codewords, which is used in searching for MRB codewords~(Sec.~\ref{sec:mrb-codes}).
  As a substitution, we provide an algorithm, the two-way flipping pattern tree~(tFPT) algorithm in Sec.~\ref{sec:fpt}, to reduce the overhead of producing the first candidate.
  The tFPT algorithm uses a similar technology as the box-and-match~(BMA) algorithm~\cite{Valembois2004Box}, which reduces computation in the high SNR region.
  More details about tFPT algorithm are given in Sec.~\ref{sec:fpt}.
\end{enumerate}


The rest of this paper is organized as follows. 
In Sec.~\ref{sec:lc-osd}, we illustrate the system model, describe the implementation of the LC-OSD, and analyze the complexity of the LC-OSD.
In Sec.~\ref{sec:analysis}, we derive the theoretical results and present some related examples.
In Sec.~\ref{sec:2-list-alg}, we present two LGAs, the SLVA and the tFPT and their performance and/or complexity~(throughput).
\section{LC-OSD}
\label{sec:lc-osd}
\subsection{System Model}

Let $\mathbb{F}_2 = \{0, 1\}$ be the binary field%
\footnote{In this paper, when it is clear from the context, we also convert naturally $0$ and $1$ in $\mathbb{F}_2$ to the real numbers.}
 and $\mathscr{C}[n, k]$ be a binary linear code of length $n$ and dimension $k$.
Let $\mathbf{G}$ be a generator matrix of $\mathscr{C}$ and $\mathbf{H}$ be a parity-check matrix of $\mathscr{C}$.
The generator matrix of size $k\times n$ and the parity-check matrix of size $(n-k)\times n$ are related by 
\begin{equation}
  \mathbf{G}\mathbf{H}^{\mathrm{T}} = \mathbf{O}.
\end{equation}
Let $\bm{u}\in \mathbb{F}_2^k$ be an information vector to be transmitted, which is first encoded into $\bm{c}\in \mathbb{F}_2^n$ by
\begin{equation}
  \bm{c} = \bm{u}\mathbf{G},
\end{equation}
and then modulated by the binary phase shift keying~(BPSK) into a signal vector $\bm{x} \in \mathbb{R}^n$ as
\begin{equation}
  x_i = (-1)^{c_i},\qquad 1 \leq i \leq n.
\end{equation}
Then the signal vector $\bm{x}$ is transmitted over an AWGN channel, resulting in a received vector $\bm{y}\in\mathbb{R}^n$ given by
\begin{equation}
  \bm{y} = \bm{x} + \bm{w},
\end{equation}
where $\bm{w}\sim \mathrm{Normal}(\mathbf{0}, \sigma^2\mathbf{I}_n)$ is a sample vector of white Gaussian noise.
The bit-wise hard-decision vector $\bm{z}\in\mathbb{F}_2^n$ is given by
\begin{equation}
  \label{equ:calc-z}
  z_i \triangleq
  \begin{cases} 
    0, & \mbox{if }y_i\geq0\\
    1, & \mbox{if }y_i<0\\
  \end{cases},
  \qquad 1 \leq i \leq n.
\end{equation}
The log-likelihood ratio~(LLR) vector, denoted by $\bm{r}\in\mathbb{R}^n$, is defined as
\begin{equation}
  \label{equ:calc-r}
  r_i \triangleq \log{\frac{f_{y\mid c}(y_i\mid 0)}{f_{y\mid c}(y_i\mid 1)}} = \frac{2y_i}{\sigma^2},\qquad 1 \leq i \leq n,
\end{equation}
where $f_{y\mid c}(\cdot\mid\cdot)$ is the conditional probability density function.
The reliability of $z_i$ refers to as $|r_i|$.
Also, the test error pattern~(TEP) $\bm{e}\in\mathbb{F}_2^n$ for a test codeword $\bm{v}\in\mathbb{F}_2^n$ is given by
\begin{equation}
  \bm{e} \triangleq \bm{z} - \bm{v}.
\end{equation}
The MLD is to find one of the codewords $\bm{v}^*$ such that%
\footnote{If two or more codewords achieve the maximum, we simply select one at random.}%
\begin{equation}
  \label{equ:ml-codeword}
  \bm{v}^* = \underset{\bm{v}\in\mathscr{C}}{\arg\max}\,f_{\bm{y}\mid\bm{c}}(\bm{y}\mid\bm{v}),
\end{equation}
which is equivalent to
\begin{equation}
  \bm{v}^* = \underset{\bm{v}\in\mathscr{C}}{\arg\min}\,\log\frac{f_{\bm{y}\mid\bm{c}}(\bm{y}\mid\bm{z})}{f_{\bm{y}\mid\bm{c}}(\bm{y}\mid\bm{v})}.
\end{equation}

\begin{definition}{(Soft weight%
  \footnote{This is called weighted Hamming distance in~\cite{yue2021revisit} and ellipsoidal weight in~\cite{valembois2002comparison}.}%
  )}
  \label{def:soft-weight}
  The soft weight of the TEP $\bm{e}\in\mathbb{F}_2^n$ is defined as
  \begin{equation}
    \label{equ:soft-weight}
      \Gamma(\bm{e}) 
      \triangleq \log\frac{f_{\bm{y}\mid\bm{c}}(\bm{y}\mid\bm{z})}{f_{\bm{y}\mid\bm{c}}(\bm{y}\mid\bm{z}-\bm{e})} 
      = \sum_{i=1}^{n} \log{\frac{f_{y\mid c}(y_i\mid z_i)}{f_{y\mid c}(y_i\mid z_i-e_i)}} 
      = \sum_{i=1}^{n} e_i|r_i|.
  \end{equation}
  For convenience, we also write the soft weight of $\bm{e}$ as an inner product. That is, 
  \begin{equation}
    \Gamma(\bm{e}) = \langle\bm{e},|\bm{r}|\rangle
  \end{equation}
  where $|\cdot|$ denotes the element-wise absolute value.
\end{definition}
We see that the MLD is equivalent to the lightest-soft-weight decoding.

\subsection{OSD With Local Constraints}
\label{sec:osd-with-lc}
In this subsection, we focus on an arbitrary binary linear code $\mathscr{C}[n, k]$ specified by a parity-check matrix $\mathbf{H}\in \mathbb{F}_2^{(n-k)\times n}$.
For a preset constraint degree $\delta\,(0\leq \delta \leq n-k)$ and a maximum number of searches $\ell_{\textrm{max}}$, we restate the LC-OSD algorithm in the following. 

\begin{enumerate}
  \item 
  Upon receiving $\bm{y}$, calculate the bit-wise hard-decision vector $\bm{z}$ by~(\ref{equ:calc-z}) and the LLR vector $\bm{r}$ by~(\ref{equ:calc-r}).

  \item 
  Sort the LLR vector $\bm{r}$ into $\bm{r}\mathbf{\Pi}$ such that the first $(n-k-\delta)$ \rchange{columns of the corresponding permuted parity-check matrix, $\mathbf{H\Pi}$,} are linearly independent with least reliabilities.
  To be precise, $\mathbf{\Pi}$ is a permutation matrix satisfying:

  \begin{enumerate}
    \item the first $(n-k-\delta)$ columns of $\mathbf{H\Pi}$ is column full rank, i.e., $\mathrm{rank}\,(\mathbf{H\Pi})_{[1,n-k-\delta]} = n-k-\delta$, and 
    \label{enum:permut-independent}
    \item $\sum_{i=1}^{n-k-\delta} (|\bm{r}|\mathbf{\Pi})_i$ is as small as possible, where $(|\bm{r}|\mathbf{\Pi})_i$ stands for the $i$-th element of the permuted reliability vector $(|\bm{r}|\mathbf{\Pi})$.
    \label{enum:permut-lrb}
  \end{enumerate}
  \rchange{This method is equivalent to that of the box-and-match~(BMA)~\cite{Valembois2004Box} algorithm and} can be accomplished by a greedy algorithm, \rchange{summarized in Algorithm~\ref{algo:pp-sort} for completeness}.

  \begin{algorithm}[H]
    \renewcommand{\algorithmicrequire}{\textbf{Input:}}
    \renewcommand{\algorithmicensure}{\textbf{Output:}}
    \caption{Find a permutation matrix for \rchange{the LC-OSD}}
    \label{algo:pp-sort}
    \begin{algorithmic}[1]
      \Require The parity-check matrix $\mathbf{H}\in \mathbb{F}_2^{(n-k)\times n}$, the reliability vector $|\bm{r}|\in\mathbb{R}^n$, the constraint degree $\delta\,(0\leq \delta \leq n-k)$.
      \Ensure A permutation matrix $\mathbf{\Pi}$ satisfying the requirements (\ref{enum:permut-independent}--\ref{enum:permut-lrb}).
      
      \Function{Get-Permutation}{$\mathbf{H}, |\bm{r}|, \delta$}
      \State $\mathbf{\Pi} \leftarrow$ an ascending sorted permutation of vector $|\bm{r}|$ 
      \Comment{That is, $(|\bm{r}|\mathbf{\Pi})$ is non-decreasing.}
      \State $i'\leftarrow n-k-\delta+1$ 
      \For{$i=1,2,\ldots,n-k-\delta$}
        \While{$\mathrm{rank}\,(\mathbf{H\Pi})_{[1,i]} < i$} 
        \Comment{$(\mathbf{H\Pi})_{[1,i]}$ is formed by the first $i$ columns of $\mathbf{H\Pi}$.}
          \State $\mathrm{swap}(\mathbf{\Pi}_i, \mathbf{\Pi}_{i'})$
          \Comment{Swap the columns $i$ and $i'$ of $\mathbf{\Pi}$.}
          \State $i'\leftarrow i'+1$
        \EndWhile
      \EndFor
      \State \Return $\mathbf{\Pi}$
      \EndFunction
    \end{algorithmic}
  \end{algorithm}

  For convenience, we denote the permuted $\bm{z}$ by $\widetilde{\bm{z}}$, i.e., 
  \begin{equation}
    \label{equ:permute}
    \widetilde{\bm{z}} \triangleq \bm{z}\mathbf{\Pi}, 
  \end{equation}
  and use similar notations for vectors $\bm{r}, \bm{e}, \bm{v}$ etc.

  \item 
   Perform the Gaussian elimination for $\mathbf{H\Pi}$, resulting in $\widetilde{\mathbf{H}}=\mathbf{QH\Pi}$ of form
   \begin{equation}
     \label{equ:gauss-elim}
     \widetilde{\mathbf{H}}=
     \begin{bNiceArray}{cw{c}{2cm}|cw{c}{1cm}}[margin, first-row, last-col]
       \Block{1-2}{_{n-k-\delta\textrm{ columns}}} & & \Block{1-2}{_{k+\delta\textrm{ columns}}} \\
       \Block{3-2}{\mathbf{I}} & &\Block{3-2}{\mathbf{P}_1} & & \Block{3-1}{^{\rotate n-k-\delta\textrm{ rows}}} \\
       & & & \\
       & & & \\
       \hline
       \Block{1-2}{\mathbf{O}}& &\Block{1-2}{\mathbf{P}_2} & & \Block{1-1}{^{\rotate \delta\textrm{ rows}}}\\
       \end{bNiceArray}
   \end{equation}
  where $\mathbf{I}$ denotes the identity matrix of order $n - k - \delta$.
  \item 
  \label{enum:preprocess-max}
  A test vector ${\bm{v}} = {\bm{z}} - {\bm{e}}$ is a codeword of $\mathscr{C}$ if and only if the TEP ${\bm{e}}$ satisfies the parity-check equation
  \begin{equation}
    {{\bm{v}}}{{\mathbf{H}}}^{{\mathrm{T}}} 
    = ({\bm{z}} - {\bm{e}}){{\mathbf{H}}}^{{\mathrm{T}}}
    = \mathbf{0},
  \end{equation}
  or equivalently, in the permuted form,
  \begin{equation}
    \label{equ:c-parity}
    {\widetilde{\bm{v}}}{\widetilde{\mathbf{H}}}^{{\mathrm{T}}} 
    = (\widetilde{\bm{z}} - \widetilde{\bm{e}}){\widetilde{\mathbf{H}}}^{{\mathrm{T}}}
    = \mathbf{0}.
  \end{equation}
  So, searching a (permuted) codeword $\widetilde{\bm{v}}$ is equivalent to searching a TEP $\widetilde{\bm{e}}$ because $\widetilde{\bm{z}}$ is fixed upon receiving $\bm{y}$. 
  If we divide $\widetilde{\bm{e}}$~(and $\widetilde{\bm{z}}$) into two parts of lengths $(n-k-\delta)$ and $(k+\delta)$, namely
  \begin{equation}
    \label{equ:two-parts}
    \widetilde{\bm{e}}=
    \begin{bNiceArray}{cw{c}{2cm}|cw{c}{1cm}}[margin, first-row]
      \Block{1-2}{_{n-k-\delta\textrm{ bits}}} & & \Block{1-2}{_{k+\delta\textrm{ bits}}} \\
      \Block{1-2}{{\widetilde{\bm{e}}}_{\textrm{L}}} & &\Block{1-2}{{\widetilde{\bm{e}}}_{\textrm{R}}} & \\
      \end{bNiceArray},~
    \widetilde{\bm{z}}=
    \begin{bNiceArray}{cw{c}{2cm}|cw{c}{1cm}}[margin, first-row]
      \Block{1-2}{_{n-k-\delta\textrm{ bits}}} & & \Block{1-2}{_{k+\delta\textrm{ bits}}} \\
      \Block{1-2}{{\widetilde{\bm{z}}}_{\textrm{L}}} & &\Block{1-2}{{\widetilde{\bm{z}}}_{\textrm{R}}} & \\
      \end{bNiceArray},
  \end{equation}
  \rchange{where the left part is known as the least reliable bits~(LRBs) and the right part is the MRBs.}
  Then by expanding~(\ref{equ:c-parity}), we can obtain equations with unknowns $\widetilde{\bm{e}}=({\widetilde{\bm{e}}}_{\textrm{L}},{\widetilde{\bm{e}}}_{\textrm{R}})$,
  \begin{align}
    \label{equ:el-parity}
    {\widetilde{\bm{e}}}_{\textrm{L}} + {\widetilde{\bm{e}}}_{\textrm{R}}\mathbf{P}_1^{{\mathrm{T}}}  &= \widetilde{\bm{s}}_1, \\
    \label{equ:er-parity}
    {\widetilde{\bm{e}}}_{\textrm{R}} \mathbf{P}_2^{{\mathrm{T}}} &= \widetilde{\bm{s}}_2,
  \end{align}
  where the right-hand sides~(RHSs) of~(\ref{equ:el-parity}) and~(\ref{equ:er-parity}) are given by
  \begin{align}
    \label{equ:def-s1}
    \widetilde{\bm{s}}_1 &\triangleq {\widetilde{\bm{z}}}_{\textrm{L}} + {\widetilde{\bm{z}}}_{\textrm{R}} \mathbf{P}_1^{{\mathrm{T}}}, \\
    \label{equ:def-s2}
    \widetilde{\bm{s}}_2 &\triangleq {\widetilde{\bm{z}}}_{\textrm{R}} \mathbf{P}_2^{{\mathrm{T}}},
  \end{align}
  which will be calculated and stored before searching codewords to avoid replicated computations.

  As in~(\ref{equ:soft-weight}), the soft weight
  of $\widetilde{\bm{e}}$ is denoted as%
  \footnote{Strictly speaking, it should use $\widetilde{\Gamma}(\widetilde{\bm{e}})$ instead of $\Gamma(\widetilde{\bm{e}})$ for $\widetilde{\bm{r}}$.}
  \begin{equation}
    \Gamma(\widetilde{\bm{e}}) \triangleq \langle\widetilde{\bm{e}},|\widetilde{\bm{r}}|\rangle.
  \end{equation}
  Similarly, 
  \begin{equation}
    \Gamma(\widetilde{\bm{e}}_{\textrm{L}}) \triangleq \langle\widetilde{\bm{e}}_{\textrm{L}}, |\widetilde{\bm{r}}_{\textrm{L}}|\rangle, 
    \Gamma(\widetilde{\bm{e}}_{\textrm{R}}) \triangleq \langle\widetilde{\bm{e}}_{\textrm{R}}, |\widetilde{\bm{r}}_{\textrm{R}}|\rangle.
  \end{equation}
  We have
  \begin{equation}
    \label{equ:decompose-gamma-e}
    \Gamma(\widetilde{\bm{e}})=
    \Gamma(\widetilde{\bm{e}}_{\textrm{L}})+
    \Gamma(\widetilde{\bm{e}}_{\textrm{R}}).
  \end{equation}

  \item
  From~(\ref{equ:el-parity}), we see that ${\widetilde{\bm{e}}}_{\textrm{L}}$~(and ${\widetilde{\bm{e}}}$ by~(\ref{equ:two-parts})) are uniquely determined by ${\widetilde{\bm{e}}}_{\textrm{R}}$. 
  Hence, we only need to search for ${\widetilde{\bm{e}}}_{\textrm{R}}$ instead of $\widetilde{\bm{e}}$~(or $\widetilde{\bm{v}}$).
  Intuitively, we should make a list of ${\widetilde{\bm{e}}}_{\textrm{R}}$ 
  \begin{equation}
      {\widetilde{\bm{e}}}_{\textrm{R}}^{(1)}, 
      {\widetilde{\bm{e}}}_{\textrm{R}}^{(2)},
      \ldots,
      {\widetilde{\bm{e}}}_{\textrm{R}}^{(\ell_{\textrm{max}})}
  \end{equation}
  satisfying~(\ref{equ:er-parity}) in an order such that the partial soft weights are non-decreasing, i.e.,
  \begin{equation}
    \label{equ:er-non-decrease}
    \Gamma({\widetilde{\bm{e}}}_{\textrm{R}}^{(1)}) 
    \leq \Gamma({\widetilde{\bm{e}}}_{\textrm{R}}^{(2)})
    \leq \cdots
    \leq \Gamma({\widetilde{\bm{e}}}_{\textrm{R}}^{(\ell_{\textrm{max}})}).
  \end{equation}
  \rchange{
  This can be achieved by an LGA, such as the SLVA~\cite{seshadri1994list} and the tFPT. See Sec.~\ref{sec:2-list-alg} for details.
  }

  \item 
  For each ${\widetilde{\bm{e}}}_{\textrm{R}}^{(j)}\,(1\leq j\leq \ell_{\textrm{max}})$, solve for ${\widetilde{\bm{e}}}^{(j)}$ by~(\ref{equ:el-parity}) and~(\ref{equ:two-parts}).
  
  \item 
  Finally, output the codeword corresponding to one of the ${\widetilde{\bm{e}}}^{(j)}$ with the lightest soft weight.
\end{enumerate}

For the sake of clarity, the LC-OSD is summarized in Algorithm~\ref{algo:LC-OSD} with an optional early stopping criterion~(see Sec.~\ref{sec:stopping-criteria} for details).
\begin{algorithm}[H]
  \renewcommand{\algorithmicrequire}{\textbf{Input:}}
  \renewcommand{\algorithmicensure}{\textbf{Output:}}
  \caption{OSD with local constraints~(LC-OSD)}
  \label{algo:LC-OSD}
  \begin{algorithmic}[1] 
    \Require $\mathbf{H}$, $\bm{y}$, $\delta$, $\ell_{\textrm{max}}$, an early stopping criterion $\Psi$~(optional).
    \Ensure The optimal searched codeword $\bm{v}$.
    \Function{LC-OSD}{$\mathbf{H}, \bm{y}, \delta, \ell_{\textrm{max}}, \Psi$}
      \State $\bm{z}\leftarrow$ the bit-wise hard decision vector of $\bm{y}$
      \Comment{Calculate by~(\ref{equ:calc-z}).}
      \State $\bm{r}\leftarrow$ the LLR vector of $\bm{y}$
      \Comment{Calculate by~(\ref{equ:calc-r}).}
      \State $\mathbf{\Pi}\leftarrow$ \Call{Get-Permutation}{$\mathbf{H}, |\bm{r}|, \delta$}
      \Comment{Get a permutation matrix by Algorithm~\ref{algo:pp-sort}.}
      \State Permute $\bm{z}, \bm{r}$ into $\widetilde{\bm{z}}, \widetilde{\bm{r}}$ by $\mathbf{\Pi}$
      \Comment{Permute as~(\ref{equ:permute}).}
      \State $\widetilde{\mathbf{H}}\leftarrow$ Gaussian eliminated matrix of $\mathbf{H\Pi}$
      \Comment{Elminate into the form as~(\ref{equ:gauss-elim}).}
      \State $\mathbf{P}_1, \mathbf{P}_2 \leftarrow$ the top-right and bottom-right sub-matrices of $\widetilde{\mathbf{H}}$
      \Comment{Extract as~(\ref{equ:gauss-elim}).}
      \State $\widetilde{\bm{s}}_1\leftarrow {\widetilde{\bm{z}}}_{\textrm{L}} + {\widetilde{\bm{z}}}_{\textrm{R}} \mathbf{P}_1^{{\mathrm{T}}}$
      \Comment{Calculate the RHS of~(\ref{equ:def-s1}).}
      \State $\widetilde{\bm{s}}_2\leftarrow {\widetilde{\bm{z}}}_{\textrm{R}} \mathbf{P}_2^{{\mathrm{T}}}$
      \Comment{Calculate the RHS of~(\ref{equ:def-s2}).}

      \State ${\widetilde{\bm{e}}}_{\textrm{opt}} \leftarrow \widetilde{\bm{z}}$
      \Comment{Note that $\widetilde{\bm{z}}$ is a valid TEP for test.}
      \For{${\ell}= 1, 2, \ldots, \ell_{\textrm{max}}$}
          \State ${\widetilde{\bm{e}}}_{\textrm{R}}^{({\ell})}\leftarrow$\Call{LGA}{$\mathbf{P}_2, |{\widetilde{\bm{r}}}_{\textrm{R}}|, \widetilde{\bm{s}}_2, {\ell}$}
          \Comment{Generate the ${\ell}$-th lightest TEP~(Sec.~\ref{sec:2-list-alg}).}
          \label{line:LC-OSD:find-jth}
          \State $\widetilde{\bm{e}}_{\textrm{L}}^{({\ell})} \leftarrow \widetilde{\bm{s}}_1 - {\widetilde{\bm{e}}}_{\textrm{R}}^{(\ell)}\mathbf{P}_1 ^{{\mathrm{T}}}$
          \Comment{Solve from~(\ref{equ:el-parity}).}
          \label{line:LC-OSD:calculate-eL}
          \State $\widetilde{\bm{e}}^{({\ell})} \leftarrow (\widetilde{\bm{e}}_{\textrm{L}}^{({\ell})},\widetilde{\bm{e}}_{\textrm{R}}^{({\ell})})$
          \If{${\Gamma}({\widetilde{\bm{e}}}_{\textrm{opt}}) > {\Gamma}(\widetilde{\bm{e}}^{({\ell})})$}
          \Comment{Update the optimal TEP with LSW.}
            \State ${\widetilde{\bm{e}}}_{\textrm{opt}} \leftarrow \widetilde{\bm{e}}^{({\ell})}$
          \EndIf
          \If{the early stopping criterion $\Psi$ is satisfied}
          \Comment{See Sec.\ref{sec:stopping-criteria} for details.}
            \State \textbf{break}
          \EndIf
      \EndFor
      \State $\bm{v} \leftarrow \bm{z} - {\widetilde{\bm{e}}}_{\textrm{opt}} \mathbf{\Pi}^{-1}$
      \State \Return $\bm{v}$
    \EndFunction
  \end{algorithmic}
\end{algorithm}

\textbf{Remarks.} If $\delta = 0$, Algorithm~\ref{algo:LC-OSD} reduces to the OSD algorithm~\cite{Fossorier1995OSD} with a search order the same as that of the FPT;
if $\delta = n - k$, it is indeed the MLD.

\subsection{Complexity Analysis}
\label{sec:complexity}
In this subsection, we focus on the time and space complexity of the \rchange{LC-OSD~(Algorithm~\ref{algo:LC-OSD})}.
\subsubsection{Time Complexity}
\label{sec:time-complexity}
We analyze the time complexity of the algorithm by evaluating the number of floating point operations~(FLOPs) or binary operations~(BOPs) required by each step.
The time complexity consists of three dominant parts:
\begin{enumerate}
  \item \emph{Gaussian elimination}. Reducing $\mathbf{H}$ into a block upper triangular matrix as shown in~(\ref{equ:gauss-elim}) requires $\mathcal{O}((n-k)(n-k-\delta))$ elementary row operations, each taking $\mathcal{O}(n)$ BOPs.
  \item \emph{Re-encoding}. With initializations done, searching a candidate ${\widetilde{\bm{e}}}_{\textrm{R}}^{(j)}\,(j >1)$ by the SLVA (line~\ref{line:LC-OSD:find-jth}, Algorithm~\ref{algo:LC-OSD}) requires $\mathcal{O}(n)$ FLOPs
  and re-encoding ${\widetilde{\bm{e}}}_{\textrm{L}}^{(j)}$ by~(\ref{equ:el-parity})~(line~\ref{line:LC-OSD:calculate-eL}, Algorithm~\ref{algo:LC-OSD}) requires $\mathcal{O}((n-k-\delta)(k+\delta))$ BOPs~(recalling that $\mathbf{P}_1$ is of size $(n-k-\delta)\times(k+\delta)$).
  \item \rchange{\emph{The LGA}. 
  As analyzed in Sec.~\ref{sec:2-list-alg}, different LGAs may have different time complexities, denoted by $T^{\textrm{LGA}}$.}
\end{enumerate}
To summarize, the overall average time complexity is
\begin{equation}
  \label{equ:Tavg}
  T_{\textrm{avg}}^{\textrm{tot}}
  =  \underbrace{\mathcal{O}(n(n-k)(n-k-\delta))}_{\textrm{Gaussian elimination}}
  + \underbrace{{\ell}_\textrm{avg}\mathcal{O}((n-k-\delta)(k+\delta))}_\textrm{re-encoding}
  + \underbrace{T_{\textrm{avg}}^{\textrm{LGA}}}_\textrm{the LGA},
\end{equation}
where ${\ell}_\textrm{avg}$ denotes the average number%
\footnote{Usually, ${\ell}_\textrm{avg}$ is much less than ${\ell}_\textrm{max}$ due to the early stopping criteria~(Sec.~\ref{sec:stopping-criteria}).}
of searches per frame, and the overall worst-case time complexity is
\begin{equation}
  \label{equ:Tmax}
  T_{\textrm{max}}^{\textrm{tot}}
  =  {\mathcal{O}(n(n-k)(n-k-\delta))}
  + {{\ell}_\textrm{max}\mathcal{O}((n-k-\delta)(k+\delta))}
  + {T_{\textrm{max}}^{\textrm{LGA}}}.
\end{equation}

\subsubsection{Space Complexity}
We analyze the space complexity of the algorithm by calculating the number of bytes used.
The space complexity consists of two dominant parts:
\begin{enumerate}
  \item \emph{Matrices and vectors}. The algorithm needs to store matrices $\mathbf{H}$, $\mathbf{P}_1$ and $\mathbf{P}_2$ of size $\mathcal{O}((n-k)\times n)$ and several vectors of length $\mathcal{O}(n)$.
  \item \rchange{\emph{The LGA}. Different LGAs may have different space complexities, denoted by $S^{\textrm{LGA}}$.}
\end{enumerate}
In summary, the overall space complexity is given by 
\begin{equation}
      S_{\textrm{max}}^{\textrm{tot}}
      = \underbrace{\mathcal{O}(n(n-k))}_{\textrm{matrices and vectors}}
      + \underbrace{S_{\textrm{max}}^{\textrm{LGA}}}_\textrm{the LGA}.
\end{equation}

\section{Performance Analysis of the LC-OSD}
\label{sec:analysis}
\subsection{Early Stopping Criteria}
\label{sec:stopping-criteria}
The analysis in Sec.~\ref{sec:complexity} suggests that the complexity can be reduced by limiting the maximum number of tests along with properly designed early stopping criteria.
Let $\widetilde{\bm{e}}^{(\ell)}=(\widetilde{\bm{e}}_{\textrm{L}}^{(\ell)},\widetilde{\bm{e}}_{\textrm{R}}^{(\ell)})$ be the $\ell$-th searched TEP in the LC-OSD algorithm~(Algorithm~\ref{algo:LC-OSD}) and
denote by ${\widetilde{\bm{e}}}_{\textrm{opt}}^{(j)}$ the up-to-date optimal TEP after the $j$-th iteration.
That is, 
\begin{equation}
  \label{equ:def-e-opt}
  {\widetilde{\bm{e}}}_{\textrm{opt}}^{(j)} 
   = \underset{\widetilde{\bm{e}}^{(\ell)}:1\leq\ell\leq j}{\arg\min}\,{\Gamma}(\widetilde{\bm{e}}^{(\ell)}).
\end{equation}
\begin{example}
\label{ex:gamma-relation}
\begin{figure}[!t]
	\centering
  \input{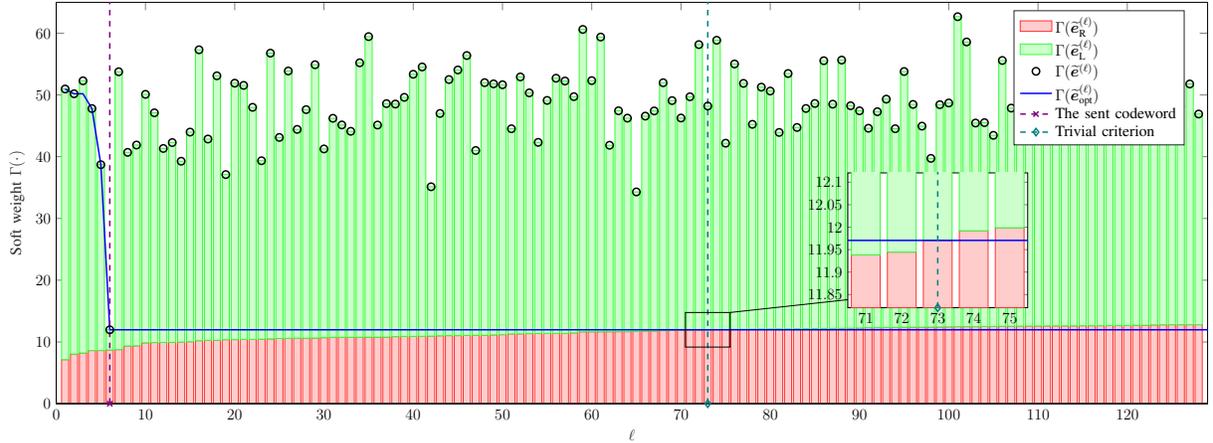}
  \caption{
    Numerical illustration of the relationships among soft weights $\Gamma(\cdot)$.
    The data are obtained by simulating a $\mathscr{C}[128, 64]$ code at $E_{\textrm{b}}/N_0 = 2.5~\textrm{dB}$, where the LC-OSD parameters are $(\delta, \ell_{\textrm{max}}) = (8, 2^{7})$.
    }
	\label{fig:sketch-gamma}
\end{figure}
To illustrate the relationship among
\begin{equation}
  \widetilde{\bm{e}}_{\textrm{R}}^{(\ell)},\,
  \widetilde{\bm{e}}_{\textrm{L}}^{(\ell)},\,
  \widetilde{\bm{e}}^{(\ell)},\,
  {\widetilde{\bm{e}}}_{\textrm{opt}}^{(\ell)},\,
\end{equation}
\rchange{
we have simulated a $\mathscr{C}[128, 64]$ code at $E_{\textrm{b}}/N_0 = 2.5~\textrm{dB}$.
The numerical results are shown in Fig.~\ref{fig:sketch-gamma}, from which we have the following observation.
\begin{enumerate}
  \item For all $\ell$, the heights~(soft weights) of the red bar~(MRBs) and the green bar~(LRBs) add up to that of the small black circle~(whole TEP). This is consistent with~(\ref{equ:decompose-gamma-e}).
  \item The heights of the red bars keep increasing, which is one of the main features~(\ref{equ:er-non-decrease}) of LC-OSD.
  \item The height of the blue solid line~(soft weight of up-to-date optimal TEP) is non-increasing, and all small black circles are located above the blue line as expected by the definition~(\ref{equ:def-e-opt}).
\end{enumerate}
}

\rchange{
Another trivial but not least fact is that whenever the red bar reaches the blue line, i.e.,
}
\begin{equation}
  \label{equ:trivial-1}
  {\Gamma}({\widetilde{\bm{e}}}_{\textrm{opt}}^{(j)}) \leq {\Gamma}({\widetilde{\bm{e}}}_{\textrm{R}}^{(j)})
\end{equation}
for \rchange{the minimum} $j\,(1\leq j\leq \ell_{\textrm{max}})$,
\rchange{the optimal TEP ${\widetilde{\bm{e}}}_{\textrm{opt}}^{(j)}$ is already guaranteed to be ML.}
This leads to the following trivial stopping criterion.

\end{example}
\begin{definition}{(Trivial stopping criterion)}
\label{def:trivial}
If, for some $j$, ${\Gamma}({\widetilde{\bm{e}}}_{\textrm{opt}}^{(j)}) \leq {\Gamma}({\widetilde{\bm{e}}}_{\textrm{R}}^{(j)})$, terminate the LC-OSD algorithm and deliver $\bm{v} = \bm{z} - {\widetilde{\bm{e}}}_{\textrm{opt}}^{(j)} \mathbf{\Pi}^{-1}$ as the output. 
\end{definition}

\begin{proposition}
\label{prop:trivial}
If the LC-OSD is stopped with the trivial stopping criterion being satisfied~(usually requiring a sufficiently large $\ell_{\textrm{max}}$), the output is an ML codeword.
\end{proposition}
\begin{IEEEproof}
For all $\ell > j$, we have
\begin{align}
  {\Gamma}({\widetilde{\bm{e}}}_{\textrm{opt}}^{(j)})
  &\leq {\Gamma}({\widetilde{\bm{e}}}_{\textrm{R}}^{(j)}) && (\textrm{by assumption}) \\
  &\leq {\Gamma}({\widetilde{\bm{e}}}_{\textrm{R}}^{(\ell)}) && ({\Gamma}({\widetilde{\bm{e}}}_{\textrm{R}}^{(j)})\textrm{ is non-decreasing}) \\
  &\leq {\Gamma}({\widetilde{\bm{e}}}^{(\ell)}). && (\textrm{by~(\ref{equ:decompose-gamma-e})})
  \label{equ:proof-2}
\end{align}
On the other hand, definition~(\ref{equ:def-e-opt}) ensures that \rchange{${\Gamma}({\widetilde{\bm{e}}}_{\textrm{opt}}^{(j)}) \leq {\Gamma}({\widetilde{\bm{e}}}^{(\ell)})$} also holds for all $\ell \leq j$.
Therefore ${\widetilde{\bm{e}}}_{\textrm{opt}}^{(j)}$ is the TEP of an ML codeword.
\end{IEEEproof}

This proposition implies that using the trivial stopping criterion would not incur performance loss, \rchange{as confirmed numerically in Fig.~\ref{fig:sketch-gamma}}.
However, as we can see \rchange{from the figure}, the termination raised by the trivial stopping criterion is quite behind the location of the true TEP.
So, if we could know $\Gamma({\widetilde{\bm{e}}_{\textrm{L}}})$ in advance, the LC-OSD can terminate \rchange{significantly earlier}.
Based on this idea, we propose several early stopping criteria.

Let ${\widetilde{\bm{e}}}$ denote the true TEP, i.e., 
\begin{equation}
  {\widetilde{\bm{e}}} \triangleq {\widetilde{\bm{z}}} - {\widetilde{\bm{c}}},
\end{equation}
where ${\widetilde{\bm{c}}} = \bm{c}\mathbf{\Pi}$ is the permuted \rchange{sent} codeword.
\begin{definition}{(Ideal stopping criterion)} 
\label{def:ideal}
Assume that $\Gamma({\widetilde{\bm{e}}_{\textrm{L}}})$ were known. If for some $j$, ${\Gamma}({\widetilde{\bm{e}}}_{\textrm{opt}}^{(j)}) < \Gamma({\widetilde{\bm{e}}_{\textrm{L}}}) + {\Gamma}({\widetilde{\bm{e}}}_{\textrm{R}}^{(j)})$, terminate the LC-OSD algorithm and deliver $\bm{v} = \bm{z} - {\widetilde{\bm{e}}}_{\textrm{opt}}^{(j)} \mathbf{\Pi}^{-1}$ as the output.
\end{definition}

\begin{proposition}
If the LC-OSD were stopped with the ideal stopping criterion being satisfied, the performance of the LC-OSD would be no worse (or even better)%
\footnote{This is not surprising since knowing $\Gamma({\widetilde{\bm{e}}_{\textrm{L}}})$ in advance is an ideal but impractical assumption.}
than that of the MLD.
\end{proposition}
\begin{IEEEproof}
When the LC-OSD is stopped by the ideal stopping criterion, there are two cases. 
\begin{itemize}
  \item The first case is that the true TEP ${\widetilde{\bm{e}}}$ has already been searched, i.e., 
  ${\widetilde{\bm{e}}_{\textrm{R}}} = {\widetilde{\bm{e}}_{\textrm{R}}}^{(\ell)}$ for some $ \ell \leq j$.
  In this case, it is not necessary to continue the remaining search. 
  \item The second case is that the true TEP ${\widetilde{\bm{e}}}$ has not been searched, indicating that $\Gamma({\widetilde{\bm{e}}_{\textrm{R}}}) \geq \Gamma({\widetilde{\bm{e}}_{\textrm{R}}}^{(j)})$.
  By assumption, it follows that 
  \begin{equation}
    {\Gamma}({\widetilde{\bm{e}}}_{\textrm{opt}}^{(j)}) 
    < \Gamma({\widetilde{\bm{e}}_{\textrm{L}}}) + {\Gamma}({\widetilde{\bm{e}}}_{\textrm{R}}^{(j)})
    \leq \Gamma({\widetilde{\bm{e}}_{\textrm{L}}}) + {\Gamma}({\widetilde{\bm{e}}}_{\textrm{R}})
    = \Gamma(\widetilde{\bm{e}}),
  \end{equation}
  which implies that the MLD must be in error and further search is unnecessary either.
\end{itemize}
\end{IEEEproof}

The ideal criterion is useless for decoding because we do not know $\Gamma({\widetilde{\bm{e}}_{\textrm{L}}})$ in advance. 

\rchange{
\begin{definition}{(Approximate ideal stopping criteria)}
\label{def:ai}
If for some $j$, ${\Gamma}({\widetilde{\bm{e}}}_{\textrm{opt}}^{(j)}) < {\tau} + {\Gamma}({\widetilde{\bm{e}}}_{\textrm{R}}^{(j)})$, terminate the LC-OSD algorithm, where the threshold ${\tau}$ is a nonnegative number to be determined.
\end{definition}
}

\rchange{
We can see from the approximate ideal stopping criteria that the greater the threshold ${\tau}$, the earlier the termination and hence the lower the complexity.
However, a large threshold can lead to premature termination and performance loss.
So how large can we set for ${\tau}$ without compromising the performance of LC-OSD?
The following proposition provides a bound on the threshold to promise the MLD performance.
}

\rchange{
\begin{proposition}
  \label{prop:dai}
  If the LC-OSD is stopped with the approximate ideal stopping criterion being satisfied and
  \begin{equation}
    \label{equ:tau-bound}
    {\tau} \leq \Gamma({\widetilde{\bm{e}}_{\textrm{L}}}) + \Delta\Gamma, 
  \end{equation}
  then the performance of the LC-OSD is no worse than that of the MLD.
  Here, the tolerance $\Delta\Gamma$ is positive and defined as 
  \begin{equation}
    \label{equ:def-DeltaGamma}
    \Delta\Gamma 
    \triangleq 
    \begin{cases} 
      {\Gamma}({\widetilde{\bm{e}}}_{\textrm{opt}}^{(j^* - 1)})
    - {\Gamma}({\widetilde{\bm{e}}}_{\textrm{R}}^{(j^* - 1)})
    - {\Gamma}({\widetilde{\bm{e}}}_{\textrm{L}}^{(j^*)}), & \mbox{if } j^* > 1\\
      +\infty, & \mbox{otherwise}\\
    \end{cases},
  \end{equation}
  where $j^*$ is the rank of the ML codeword in the search list of LC-OSD.%
  \footnote{If two or more codewords are ML, we refer to the first one~(i.e., $j^* \triangleq \min \{\arg\min_j {\Gamma}({\widetilde{\bm{e}}}^{(j)})\}$).}
\end{proposition}
}

\rchange{
\begin{IEEEproof}
  Since ${\widetilde{\bm{e}}}^{(j^*)}$ is the ML TEP, all preceding TEPs are heavier.
  Hence, for $j^* >1$, we have
  \begin{align}
    \Delta\Gamma 
    &= {\Gamma}({\widetilde{\bm{e}}}_{\textrm{opt}}^{(j^* - 1)})
    - {\Gamma}({\widetilde{\bm{e}}}_{\textrm{R}}^{(j^* - 1)})
    - {\Gamma}({\widetilde{\bm{e}}}_{\textrm{L}}^{(j^*)}) \\
    \label{equ:prop-dai-proof-a2}
    &\geq {\Gamma}({\widetilde{\bm{e}}}_{\textrm{opt}}^{(j^* - 1)})
    - {\Gamma}({\widetilde{\bm{e}}}_{\textrm{R}}^{(j^*)})
    - {\Gamma}({\widetilde{\bm{e}}}_{\textrm{L}}^{(j^*)}) \\
    \label{equ:prop-dai-proof-a3}
    &={\Gamma}({\widetilde{\bm{e}}}_{\textrm{opt}}^{(j^* - 1)})
    - {\Gamma}({\widetilde{\bm{e}}}^{(j^*)}) \\
    \label{equ:prop-dai-proof-a4}
    &>0.
  \end{align}
  The inequality~(\ref{equ:prop-dai-proof-a2}) follows from the monotonicity of ${\Gamma}({\widetilde{\bm{e}}}_{\textrm{R}}^{(j)})$,
  the equality~(\ref{equ:prop-dai-proof-a3}) follows from~(\ref{equ:decompose-gamma-e}),
  and the inequality~(\ref{equ:prop-dai-proof-a4}) follows from the hypothesis that $j^*$ is the rank of the ML codeword.

  When the sent codeword is not ML, the LC-OSD is naturally no worse than the MLD.
  Therefore, we only need to consider the case when the sent codeword is ML, i.e., 
  \begin{equation}
    \label{equ:sent-ml-hypo}
    {\Gamma}({\widetilde{\bm{e}}}) = {\Gamma}({\widetilde{\bm{e}}}^{(j^*)}).
  \end{equation}
  For $j^* = 1$, the ML codeword is always in the list.
  For $j^* > 1$, we have
  \begin{align}
    {\Gamma}({\widetilde{\bm{e}}}_{\textrm{opt}}^{(j^*-1)}) &- {\tau} - {\Gamma}({\widetilde{\bm{e}}}_{\textrm{R}}^{(j^*-1)}) \\
    \label{equ:prop-dai-proof-b2}
    &\geq {\Gamma}({\widetilde{\bm{e}}}_{\textrm{opt}}^{(j^*-1)})
    - \bigg(\Gamma({\widetilde{\bm{e}}_{\textrm{L}}})
    + {\Gamma}({\widetilde{\bm{e}}}_{\textrm{opt}}^{(j^* - 1)})
    - {\Gamma}({\widetilde{\bm{e}}}_{\textrm{R}}^{(j^* - 1)})
    - {\Gamma}({\widetilde{\bm{e}}}_{\textrm{L}}^{(j^*)})
    \bigg)
    - {\Gamma}({\widetilde{\bm{e}}}_{\textrm{R}}^{(j^*-1)})\\
    & = {\Gamma}({\widetilde{\bm{e}}}_{\textrm{L}}^{(j^*)}) - \Gamma({\widetilde{\bm{e}}_{\textrm{L}}}) \\
    & = \bigg({\Gamma}({\widetilde{\bm{e}}}^{(j^*)}) - {\Gamma}({\widetilde{\bm{e}}}_{\textrm{R}}^{(j^*)})\bigg)
    - \bigg(\Gamma({\widetilde{\bm{e}}}) - \Gamma({\widetilde{\bm{e}}_{\textrm{R}}})\bigg) \\
    \label{equ:prop-dai-proof-b5}
    & = \Gamma({\widetilde{\bm{e}}_{\textrm{R}}}) - {\Gamma}({\widetilde{\bm{e}}}_{\textrm{R}}^{(j^*)}) \\
    \label{equ:prop-dai-proof-b6}
    & \geq 0
  \end{align}
  The equality~(\ref{equ:prop-dai-proof-b5}) follows from~(\ref{equ:sent-ml-hypo}),
  and the inequality~(\ref{equ:prop-dai-proof-b6}) follows from the hypothesis that $j^*$ is the rank of the first ML codeword and the sent is one of the ML codewords.
  The inequality~(\ref{equ:prop-dai-proof-b6}) indicates that the approximate ideal stopping criterion can not be satisfied at $(j^* - 1)$ and before, ensuring the ML codeword is in the search list of the LC-OSD.
\end{IEEEproof}
}

\rchange{
Proposition~\ref{prop:dai} suggests that the threshold of the ideal stopping criterion can be increased without degrading the performance.
}
However, the ideal criterion~(${\tau} = \Gamma({\widetilde{\bm{e}}_{\textrm{L}}})$) is not helpful in practice, we replace ${\tau}$ by its \rchange{expectation (or approximation)}, as presented in the following.

\begin{definition}{(Dynamic approximate ideal~(DAI) stopping criterion)} 
\label{def:dai}
\rchange{
The approximate ideal stopping criterion~(Definition~\ref{def:ai}) with $\tau = {\tau}_{\textrm{DAI}}$ is called the DAI stopping criterion, where the DAI threshold is defined by
}
\begin{equation}
  \label{equ:def-tau-dai}
  {\tau}_{\textrm{DAI}}
  \triangleq \mathbb{E}\bigl[\Gamma({\widetilde{\bm{e}}_{\textrm{L}}})\bigm|\bm{r}\bigr].
\end{equation}
\end{definition}

Now, we show how to calculate the DAI threshold ${\tau}_{\textrm{DAI}}$.
Upon receiving $\bm{y}$, the LLR vector $\bm{r}$ can be calculated out by~(\ref{equ:calc-r}).
Then, by definition, we have
\begin{equation}
  \Gamma({\widetilde{\bm{e}}_{\textrm{L}}}) 
  = \langle\widetilde{\bm{e}}_{\textrm{L}},|\widetilde{\bm{r}}_{\textrm{L}}|\rangle
  = \sum_{i=1}^{n-k-\delta}\widetilde{e}_i|\widetilde{r}_i|,
\end{equation}
and hence
\begin{equation}
  {\tau}_{\textrm{DAI}}
  = \mathbb{E}\bigl[\Gamma({\widetilde{\bm{e}}_{\textrm{L}}})\bigm|\bm{r}\bigr] 
  = \mathbb{E}\biggl[\,\sum_{i=1}^{n-k-\delta}\widetilde{e}_i|\widetilde{r}_i| \bigg| \bm{r}\biggr] 
  = \sum_{i=1}^{n-k-\delta} \mathbb{E}\bigl[\widetilde{e}_i|\widetilde{r}_i| \bigm| \bm{r}\bigr] 
  = \sum_{i=1}^{n-k-\delta} \mathbb{E}\bigl[\widetilde{e}_i \bigm| \bm{r}\bigr]\cdot|\widetilde{r}_i|
  \label{equ:sum-expect}
\end{equation}
The $i$-th bit of the true TEP $\widetilde{e}_i$ is a random variable with
\begin{align}
  \label{equ:prob-e0}
  \mathbb{P}[\widetilde{e}_i=0\mid \bm{r}] &= \frac{1}{1+\exp(-|\widetilde{r}_i|)},\\ 
  \label{equ:prob-e1}
  \mathbb{P}[\widetilde{e}_i=1\mid \bm{r}] &= \frac{1}{1+\exp(|\widetilde{r}_i|)}.
\end{align}
Therefore,
\begin{equation}
  \label{equ:calc-tau-dai}
  {\tau}_{\textrm{DAI}} 
  = \sum_{i=1}^{n-k-\delta} \frac{1}{1+\exp(|\widetilde{r_i}|)}\cdot |\widetilde{r_i}|.
\end{equation}

Given $\bm{r}$, the threshold ${\tau}_{\textrm{DAI}}$ is computable.
The performance of the LC-OSD with the DAI criterion is closely related to whether or not 
${\tau}_{\textrm{DAI}} \leq \Gamma({\widetilde{\bm{e}}_{\textrm{L}}}) + \Delta\Gamma$, 
which in turn depends on the realization of $\bm{r}$. 
Hence, we have the following proposition from Proposition~\ref{prop:dai}.

\rchange{
\begin{proposition}
  The LC-OSD with the DAI stopping criterion performs as well as the MLD with a probability of at least $\mathbb{P}[{\tau}_{\textrm{DAI}} \leq \Gamma({\widetilde{\bm{e}}_{\textrm{L}}}) + \Delta\Gamma]$,
  where the tolerance $\Delta\Gamma$ is defined in Proposition~\ref{prop:dai}.
\end{proposition}
}

\rchange{
\begin{IEEEproof}
  This follows immediately from Proposition~\ref{prop:dai} by taking into the randomness of the noisy channel (or the LLR vector $\bm{r}$).
\end{IEEEproof}
}

\rchange{
To reduce the overhead of online calculation, one may use the following stopping criterion instead, which pre-calculates the SNR-dependent thresholds for use.
}

\begin{definition}{(Static approximate ideal~(SAI) stopping criterion)} 
\label{def:sai}
\rchange{
The approximate ideal stopping criterion~(Definition~\ref{def:ai}) with $\tau = {\tau}_{\textrm{SAI}}$ is called the SAI stopping criterion, where the SAI threshold
}
\begin{equation}
  \label{equ:def-tau-sai}
  {\tau}_{\textrm{SAI}} 
  \triangleq (n-k-\delta) \mathbb{E}\bigl[e|r| \bigm| |r| \leq \alpha\bigr]
\end{equation}
is the approximation of $\mathbb{E}[\Gamma({\widetilde{\bm{e}}_{\textrm{L}}})]$
and $\alpha$ is the $(n-k-\delta)$-th $n$-quantile for the random variable $|r|$.
\end{definition}

Formally, the $(n-k-\delta)$-th $n$-quantile for the random variable $|r|$ can be defined as
\begin{equation}
  \label{equ:alpha}
  \alpha \triangleq F_{|r|}^{-1}\left(\frac{n-k-\delta}{n}\right),
\end{equation}
where $F_{|r|}^{-1}(\cdot)$ is the inverse function of $F_{|r|}(\cdot)$, the cumulative distribution function~(CDF) of $|r|$, i.e., 
\begin{align}
  f_{|r|}(r) &
  = \frac{\sigma }{2 \sqrt{2 \pi }}\exp\left(-\frac{(\sigma ^2 r-2)^2}{8 \sigma^2}\right)
  + \frac{\sigma }{2 \sqrt{2 \pi }}\exp\left(-\frac{(\sigma ^2 r+2)^2}{8 \sigma^2}\right),
  & r &\geq 0, \\
  F_{|r|}(r) &
  = \int_0^{r} f_{|r|}(t) \,\mathrm{d}t
  = \frac{1}{2} \erf\left(\frac{\sigma ^2 r-2}{2 \sqrt{2} \sigma }\right)
  + \frac{1}{2} \erf\left(\frac{\sigma ^2 r+2}{2 \sqrt{2} \sigma }\right),
  & r &\geq 0,
\end{align}
and $\erf(\cdot)$ is the error function defined as 
\begin{equation}
  \erf(x)
  \triangleq \frac{2}{\sqrt{\pi}} \int_0^x \exp(-t^2) \,\mathrm{d}t.
\end{equation}
Therefore, ${\tau}_{\textrm{SAI}}$~(\ref{equ:def-tau-sai}) can be calculated by 
\begin{align}
  {\tau}_{\textrm{SAI}} 
  &= (n-k-\delta) \mathbb{E}\bigl[e|r| \bigm| |r| \leq \alpha\bigr] \\
  &= \frac{n-k-\delta}{\mathbb{P}[|r| \leq \alpha]} \int_0^{\alpha} \frac{r}{1+\exp(r)} f_{|r|}(r) \,\mathrm{d}r \\
  &= \frac{n-k-\delta}{F_{|r|}(\alpha)} \int_0^{\alpha} \frac{r}{1+\exp(r)} f_{|r|}(r) \,\mathrm{d}r\\
  &= n \int_0^{\alpha} \frac{r}{1+\exp(r)} f_{|r|}(r) \,\mathrm{d}r
  \label{equ:calc-tau-sai}
\end{align}
\rchange{
  From the definition~(\ref{equ:def-tau-sai}), the SAI threshold does not depend on the LLR vector $\bm{r}$.
  Therefore, to reduce the computational costs, we can calculate offline and store the SAI thresholds for different SNRs. See Table~\ref{tab:acc-E-Gamma-eL} for an example.
}

\begin{example}
\begin{figure}[!t]
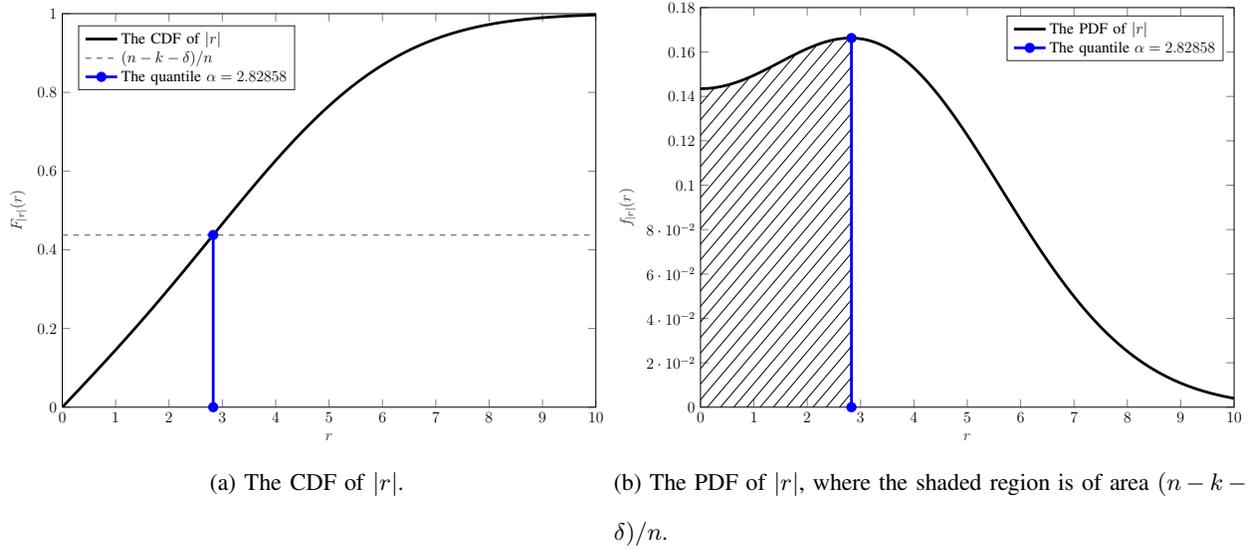

  \centering
  \subfloat[The CDF of $|r|$.\label{fig:sketch-quantile-cdf}]{\input{Figures/sketch_quantile_cdf.tex}} 
  \hfill
  \subfloat[The PDF of $|r|$, where the shaded region is of area $(n-k-\delta)/n$.\label{fig:sketch-quantile-pdf}]{\input{Figures/sketch_quantile_pdf.tex}} 
  \caption{Sketches of $(n-k-\delta)$-th $n$-quantile for $|r|$, where $(n, k, \delta) = (128, 64, 8)$ and $E_{\textrm{b}}/N_0=2.0~\textrm{dB}$.}
  \label{fig:sketch-quantile}
\end{figure}
For a better understanding of \rchange{the quantile} $\alpha$, we have provided in Fig.~\ref{fig:sketch-quantile} an example with the CDF and probability density function~(PDF) of $|r|$.
\end{example}

\begin{example}
To illustrate the accuracy of the approximation, we compare ${\tau}_{\textrm{SAI}}$ with the statistic value of $\mathbb{E}[\Gamma({\widetilde{\bm{e}}_{\textrm{L}}})]$ by $10^7$ simulations in Table~\ref{tab:acc-E-Gamma-eL}.
\rchange{
  We can see from the table that the relative error of ${\tau}_{\textrm{SAI}}$ is small, justifying the negligible performance loss (caused by the SAI stopping).
}

\begin{table}[H]
  \centering
  \caption{The relative error$^*$ of ${\tau}_{\textrm{SAI}}$ with different combination of $(n, k, \delta)$.}
\resizebox{\linewidth}{!}{
\begin{tabular}{c|c c|c c|c c|c c|c c|c c}
  \hline
  $(n, k, \delta)$ & \multicolumn{2}{c|}{$(128, 32, 8)$} & \multicolumn{2}{c|}{$(128, 32, 12)$} & \multicolumn{2}{c|}{$(128, 64, 8)$} & \multicolumn{2}{c|}{$(128, 64, 12)$} & \multicolumn{2}{c|}{$(128, 92, 8)$} & \multicolumn{2}{c}{$(128, 92, 12)$} \\
  \hline
  $E_\textrm{b}/N_0$ & ${\tau}_{\textrm{SAI}}$ & Error & ${\tau}_{\textrm{SAI}}$ & Error & ${\tau}_{\textrm{SAI}}$ & Error & ${\tau}_{\textrm{SAI}}$ & Error & ${\tau}_{\textrm{SAI}}$ & Error & ${\tau}_{\textrm{SAI}}$ & Error \\
  \hline
  $0.0~\textrm{dB}$ & $18.34$ & $+0.18\%$ & $17.30$ & $+0.16\%$ & $12.07$ & $+0.44\%$ & $11.04$ & $+0.40\%$ & $4.528$ & $-0.03\%$ & $3.431$ & $-0.68\%$ \\
  $1.0~\textrm{dB}$ & $18.91$ & $+0.27\%$ & $17.98$ & $+0.27\%$ & $12.34$ & $+0.60\%$ & $11.45$ & $+0.62\%$ & $5.081$ & $+0.80\%$ & $3.983$ & $+0.35\%$ \\
  $2.0~\textrm{dB}$ & $19.01$ & $+0.33\%$ & $18.23$ & $+0.34\%$ & $11.96$ & $+0.66\%$ & $11.29$ & $+0.72\%$ & $5.348$ & $+1.51\%$ & $4.396$ & $+1.45\%$ \\
  $3.0~\textrm{dB}$ & $18.48$ & $+0.33\%$ & $17.88$ & $+0.37\%$ & $10.78$ & $+0.56\%$ & $10.36$ & $+0.66\%$ & $5.031$ & $+1.70\%$ & $4.382$ & $+2.08\%$ \\
  $4.0~\textrm{dB}$ & $17.19$ & $+0.30\%$ & $16.79$ & $+0.34\%$ & $8.880$ & $+0.36\%$ & $8.668$ & $+0.43\%$ & $4.033$ & $+1.26\%$ & $3.718$ & $+1.84\%$ \\
\end{tabular}
}
  \label{tab:acc-E-Gamma-eL}
\end{table}

\end{example}

\begin{example}
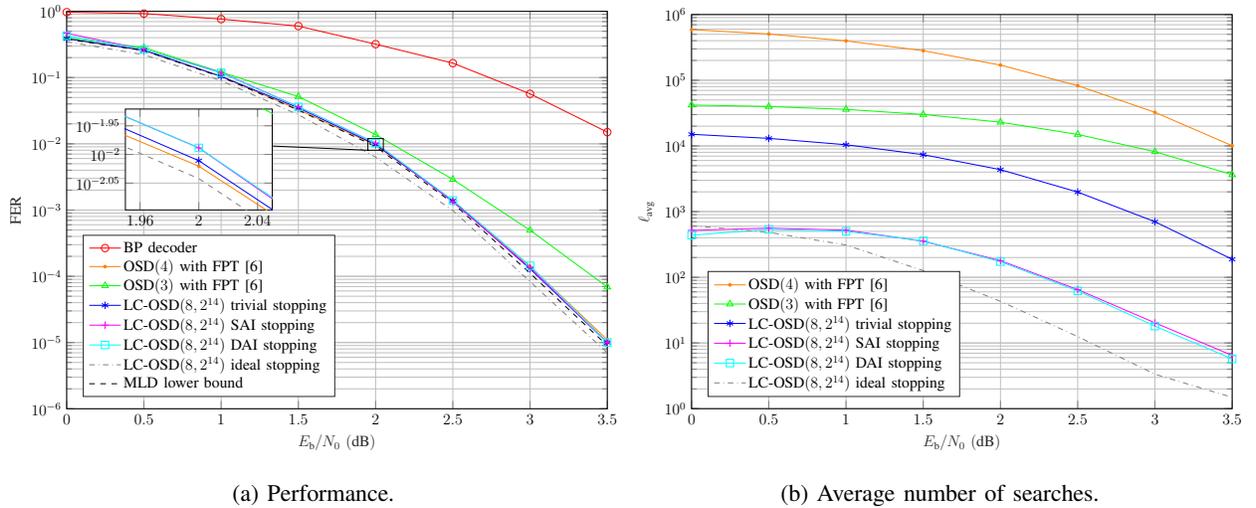
\begin{figure}[!t]
  \centering
  \subfloat[Performance.\label{fig:ldpc-fer}]{
%
%
\definecolor{mycolor1}{rgb}{1.00000,0.00000,1.00000}%
\definecolor{mycolor2}{rgb}{0.00000,1.00000,1.00000}%
\begin{tikzpicture}[%
thick,scale=0.7, every node/.style={scale=0.7}
]

\begin{axis}[%
width=10.267cm,
height=7.554cm,
at={(0cm,0cm)},
scale only axis,
xmin=0,
xmax=3.5,
xlabel style={font=\color{white!15!black}},
xlabel={$E_{\textrm{b}}/N_0~(\textrm{dB})$},
ymode=log,
ymin=1e-06,
ymax=1,
yminorticks=true,
ylabel style={font=\color{white!15!black}},
ylabel={FER},
axis background/.style={fill=white},
xmajorgrids,
ymajorgrids,
yminorgrids,
legend style={at={(0.03,0.03)}, anchor=south west, legend cell align=left, align=left, draw=white!15!black}
]
\addplot [color=red, mark=o, mark options={solid, red}]
  table[row sep=crcr]{%
0	0.9708737864\\
0.5	0.9259259259\\
1	0.7633587786\\
1.5	0.5988023952\\
2	0.3194888179\\
2.5	0.1652892562\\
3	0.0568504832\\
3.5	0.0149970006\\
};
\addlegendentry{BP decoder}

\addplot [color=orange, mark size=0.7pt, mark=*, mark options={solid, orange}]
  table[row sep=crcr]{%
0	0.3875968992\\
0.5	0.2570694087\\
1	0.1041666667\\
1.5	0.0328731098\\
2	0.0095383441\\
2.5	0.0014024262\\
3	0.0001408734\\
3.5	1.1e-05\\
};
\addlegendentry{OSD$(4)$ with FPT~\cite{Wang2022ITW}}

\addplot [color=green, mark=triangle, mark options={solid, green}]
  table[row sep=crcr]{%
0	0.3968253968\\
0.5	0.283286119\\
1	0.1204819277\\
1.5	0.0516795866\\
2	0.0137741047\\
2.5	0.0029026734\\
3	0.0004958129\\
3.5	6.9e-05\\
};
\addlegendentry{OSD$(3)$ with FPT~\cite{Wang2022ITW}}

\addplot [color=blue, mark=asterisk, mark options={solid, blue}]
  table[row sep=crcr]{%
0	0.3861003861\\
0.5	0.2583979328\\
1	0.1041666667\\
1.5	0.0348918353\\
2	0.0097513408\\
2.5	0.001373004\\
3	0.000129768\\
3.5	1e-05\\
};
\addlegendentry{LC-OSD$(8, 2^{14})$ trivial stopping}

\addplot [color=mycolor1, mark=+, mark options={solid, mycolor1}]
  table[row sep=crcr]{%
0	0.4672897196\\
0.5	0.2652519894\\
1	0.119474313\\
1.5	0.0361271676\\
2	0.0102669405\\
2.5	0.0013299818\\
3	0.0001348718\\
3.5	1e-05\\
};
\addlegendentry{LC-OSD$(8, 2^{14})$ SAI stopping}

\addplot [color=mycolor2, mark=square, mark options={solid, mycolor2}]
  table[row sep=crcr]{%
0	0.4291845494\\
0.5	0.2652519894\\
1	0.1179245283\\
1.5	0.0361532899\\
2	0.0102669405\\
2.5	0.0013995018\\
3	0.0001441579\\
3.5	1e-05\\
};
\addlegendentry{LC-OSD$(8, 2^{14})$ DAI stopping}

\addplot [color=gray, dashdotted]
  table[row sep=crcr]{%
0	0.3472222222\\
0.5	0.2212389381\\
1	0.0896860987\\
1.5	0.0274951883\\
2	0.0062916824\\
2.5	0.0009842229\\
3	8.3e-05\\
3.5	7e-06\\
};
\addlegendentry{LC-OSD$(8, 2^{14})$ ideal stopping}

\addplot [color=black, dashed]
  table[row sep=crcr]{%
0	0.3802281369\\
0.5	0.2557544757\\
1	0.1023541453\\
1.5	0.0322997416\\
2	0.0090744102\\
2.5	0.0012750223\\
3	0.000110989\\
3.5	9e-06\\
};
\addlegendentry{MLD lower bound}

\addplot [color=black, line width=0.1pt, forget plot]
  table[row sep=crcr]{%
1.95	0.008\\
2.05	0.008\\
2.05	0.012\\
1.95	0.012\\
1.95	0.008\\
1	0.01\\
};
\end{axis}

\begin{axis}[%
width=2.789cm,
height=1.918cm,
at={(1.118cm,3.778cm)},
scale only axis,
xmin=1.95,
xmax=2.05,
xtick={1.96,    2, 2.04},
ymode=log,
ymin=0.008,
ymax=0.012,
ytick={0.0063096, 0.0070795, 0.0079433, 0.0089125,      0.01,   0.01122,  0.012589},
yminorticks=true,
axis background/.style={fill=white},
xmajorgrids,
ymajorgrids,
yminorgrids
]
\addplot [color=red, mark=o, mark options={solid, red}, forget plot]
  table[row sep=crcr]{%
0	0.9708737864\\
0.5	0.9259259259\\
1	0.7633587786\\
1.5	0.5988023952\\
2	0.3194888179\\
2.5	0.1652892562\\
3	0.0568504832\\
3.5	0.0149970006\\
};
\addplot [color=orange, mark size=0.7pt, mark=*, mark options={solid, orange}, forget plot]
  table[row sep=crcr]{%
0	0.3875968992\\
0.5	0.2570694087\\
1	0.1041666667\\
1.5	0.0328731098\\
2	0.0095383441\\
2.5	0.0014024262\\
3	0.0001408734\\
3.5	1.1e-05\\
};
\addplot [color=green, mark=triangle, mark options={solid, green}, forget plot]
  table[row sep=crcr]{%
0	0.3968253968\\
0.5	0.283286119\\
1	0.1204819277\\
1.5	0.0516795866\\
2	0.0137741047\\
2.5	0.0029026734\\
3	0.0004958129\\
3.5	6.9e-05\\
};
\addplot [color=blue, mark=asterisk, mark options={solid, blue}, forget plot]
  table[row sep=crcr]{%
0	0.3861003861\\
0.5	0.2583979328\\
1	0.1041666667\\
1.5	0.0348918353\\
2	0.0097513408\\
2.5	0.001373004\\
3	0.000129768\\
3.5	1e-05\\
};
\addplot [color=mycolor1, mark=+, mark options={solid, mycolor1}, forget plot]
  table[row sep=crcr]{%
0	0.4672897196\\
0.5	0.2652519894\\
1	0.119474313\\
1.5	0.0361271676\\
2	0.0102669405\\
2.5	0.0013299818\\
3	0.0001348718\\
3.5	1e-05\\
};
\addplot [color=mycolor2, mark=square, mark options={solid, mycolor2}, forget plot]
  table[row sep=crcr]{%
0	0.4291845494\\
0.5	0.2652519894\\
1	0.1179245283\\
1.5	0.0361532899\\
2	0.0102669405\\
2.5	0.0013995018\\
3	0.0001441579\\
3.5	1e-05\\
};
\addplot [color=black, dashdotted, forget plot]
  table[row sep=crcr]{%
0	0.3472222222\\
0.5	0.2212389381\\
1	0.0896860987\\
1.5	0.0274951883\\
2	0.0062916824\\
2.5	0.0009842229\\
3	8.3e-05\\
3.5	7e-06\\
};
\addplot [color=gray, dashed, forget plot]
  table[row sep=crcr]{%
0	0.3802281369\\
0.5	0.2557544757\\
1	0.1023541453\\
1.5	0.0322997416\\
2	0.0090744102\\
2.5	0.0012750223\\
3	0.000110989\\
3.5	9e-06\\
};
\end{axis}
\end{tikzpicture}%
} 
  \hfill
  \subfloat[Average number of searches.\label{fig:ldpc-timing}]{
%
%
\definecolor{mycolor1}{rgb}{1.00000,0.00000,1.00000}%
\definecolor{mycolor2}{rgb}{0.00000,1.00000,1.00000}%
\begin{tikzpicture}[%
thick,scale=0.7, every node/.style={scale=0.7}
]

\begin{axis}[%
width=10.267cm,
height=7.491cm,
at={(0cm,0cm)},
scale only axis,
xmin=0,
xmax=3.5,
xlabel style={font=\color{white!15!black}},
xlabel={$E_{\textrm{b}}/N_0~(\textrm{dB})$},
ymode=log,
ymin=1,
ymax=1000000,
yminorticks=true,
ylabel style={font=\color{white!15!black}},
ylabel={$\ell_{\textrm{avg}}$},
axis background/.style={fill=white},
xmajorgrids,
ymajorgrids,
yminorgrids,
legend style={at={(0.03,0.03)}, anchor=south west, legend cell align=left, align=left, draw=white!15!black}
]
\addplot [color=orange, mark size=0.7pt, mark=*, mark options={solid, orange}]
  table[row sep=crcr]{%
0	589639.77\\
0.5	505845.66\\
1	397125.82\\
1.5	281790.25\\
2	170034.57\\
2.5	82646.58\\
3	32481.13\\
3.5	10075.49\\
};
\addlegendentry{OSD$(4)$ with FPT~\cite{Wang2022ITW}}

\addplot [color=green, mark=triangle, mark options={solid, green}]
  table[row sep=crcr]{%
0	41999.71\\
0.5	39752.04\\
1	35992.63\\
1.5	30131.75\\
2	22967.44\\
2.5	14972.38\\
3	8173.36\\
3.5	3645.26\\
};
\addlegendentry{OSD$(3)$ with FPT~\cite{Wang2022ITW}}

\addplot [color=blue, mark=asterisk, mark options={solid, blue}]
  table[row sep=crcr]{%
0	15026.65\\
0.5	13030.66\\
1	10423.34\\
1.5	7356.36\\
2	4355.79\\
2.5	1984.26\\
3	701.71\\
3.5	188.14\\
};
\addlegendentry{LC-OSD$(8, 2^{14})$ trivial stopping}

\addplot [color=mycolor1, mark=+, mark options={solid, mycolor1}]
  table[row sep=crcr]{%
0	517.63\\
0.5	560.01\\
1	522.4\\
1.5	353.88\\
2	179.33\\
2.5	65.02\\
3	20.29\\
3.5	6.47\\
};
\addlegendentry{LC-OSD$(8, 2^{14})$ SAI stopping}

\addplot [color=mycolor2, mark=square, mark options={solid, mycolor2}]
  table[row sep=crcr]{%
0	437.48\\
0.5	535.03\\
1	501.6\\
1.5	359.3\\
2	172.84\\
2.5	62.06\\
3	18.07\\
3.5	5.73\\
};
\addlegendentry{LC-OSD$(8, 2^{14})$ DAI stopping}

\addplot [color=gray, dashdotted]
  table[row sep=crcr]{%
0	612.32\\
0.5	483.4\\
1	310.37\\
1.5	126.4\\
2	43.12\\
2.5	12.52\\
3	3.35\\
3.5	1.49\\
};
\addlegendentry{LC-OSD$(8, 2^{14})$ ideal stopping}

\end{axis}
\end{tikzpicture}%
} 
  \caption{
    Simulation results of the LDPC code $\mathscr{C}_{\textrm{LDPC}}[128,64]$ constructed in~\cite{Baldi2016OSD}. 
  }
  \label{fig:ldpc}
\end{figure}
Consider \rchange{the} rate-$1/2$ LDPC code $\mathscr{C}_{\textrm{LDPC}}[128,64]$ constructed in~\cite{Baldi2016OSD}. 
The constraint degree $\delta=8$ and the maximum number of searches $\ell_{\textrm{max}}=2^{14}$.
Several decoders are compared in terms of the frame error rate~(FER) and the average search number, as shown in Fig.~\ref{fig:ldpc}, where the OSD with FTP~\cite{Wang2022ITW} using trivial stopping criterion is a low-complexity implementation for OSD that has the same performance as OSD.
The MLD lower bound in Fig.~\ref{fig:ldpc-fer} is a simulated bound as presented in~\cite{TANG2022Chase} and calculated by utilizing the LC-OSD~(see Algorithm~\ref{algo:MLD-error-counter}).
\end{example}

\begin{algorithm}[H]
  \renewcommand{\algorithmicrequire}{\textbf{Input:}}
  \renewcommand{\algorithmicensure}{\textbf{Output:}}
  \caption{Lower bound on error counter for MLD}
  \label{algo:MLD-error-counter}
  \begin{algorithmic}[1]
    \Require $\mathbf{H}$, $\delta$, $\ell_{\textrm{max}}$, transmission-receiving pair $(\bm{c}, \bm{y})$
    \Ensure MLD error is detected or not.
    \Function{MLD-Error-Counter}{$\mathbf{H}, \delta, \ell_{\textrm{max}}, \bm{c}, \bm{y}$}
      \State $\widehat{\bm{c}} \leftarrow$ \Call{LC-OSD}{$\mathbf{H}, \bm{y}, \delta, \ell_{\textrm{max}}$}
      \Comment{Search for the optimal codeword by the LC-OSD.}
      \State \Return $f_{\bm{y}\mid\bm{c}}(\bm{y}\mid\widehat{\bm{c}}) > f_{\bm{y}\mid\bm{c}}(\bm{y}\mid\bm{c})$
      \Comment{An MLD error is detected if $\widehat{\bm{c}}$ is more likely than $\bm{c}$.}
    \EndFunction
  \end{algorithmic}
\end{algorithm}
\begin{figure}[!t]
  \centering
  \subfloat[Performance. The box $\boxed{k}$ with the dimension $k$ inside is marked for each result cluster.\label{fig:rs31-fer}]{\input{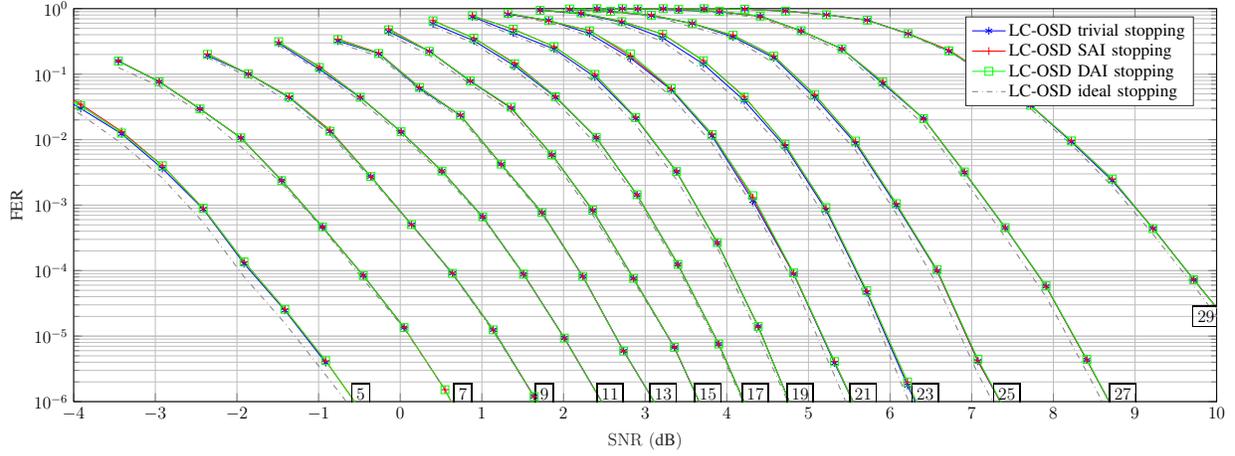}
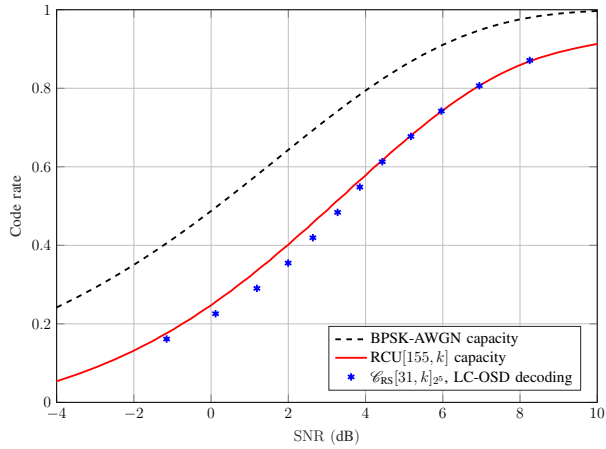
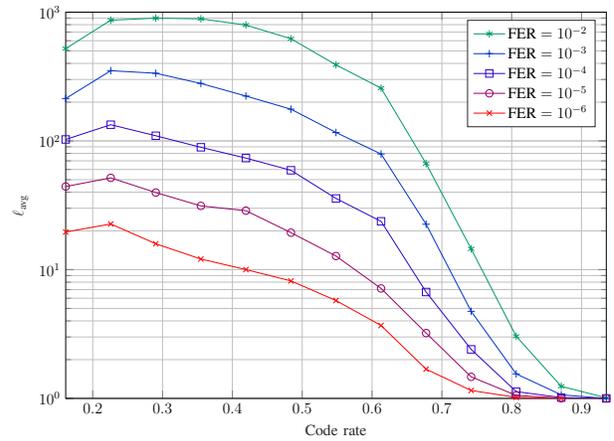}
  \\
  \subfloat[The finite-length capacity and the required SNR by the RS codes with the LC-OSD. Here, the LC-OSD is with the DAI criterion, and the target FER is $10^{-5}$.\label{fig:rs31-cap}]{
%
%
\begin{tikzpicture}[%
thick,scale=0.7, every node/.style={scale=0.7}
]

\begin{axis}[%
width=10.2666667cm,
height=7.4623567cm,
at={(0cm,0cm)},
scale only axis,
xmin=-4,
xmax=10,
xtick={-4, -2,  0,  2,  4,  6,  8, 10},
xlabel style={font=\color{white!15!black}},
xlabel={$\mathrm{SNR}~(\textrm{dB})$},
ymin=0,
ymax=1,
ylabel style={font=\color{white!15!black}},
ylabel={Code rate},
axis background/.style={fill=white},
xmajorgrids,
ymajorgrids,
legend style={at={(0.97,0.03)}, anchor=south east, legend cell align=left, align=left, draw=white!15!black}
]
\addplot [color=black, dashed, line width=1.0pt]
  table[row sep=crcr]{%
-4	0.241777485557792\\
-3.5	0.266070494931655\\
-3	0.292232993679784\\
-2.5	0.320287780762773\\
-2	0.350229551205787\\
-1.5	0.382019082300729\\
-1	0.415577216766474\\
-0.5	0.450778895575637\\
0	0.48744761446564\\
0.5	0.525350842575779\\
1	0.564197142093943\\
1.5	0.603635926075678\\
2	0.643260903844185\\
2.5	0.682618166556375\\
3	0.721219456566649\\
3.5	0.758560481955829\\
4	0.79414343886877\\
4.5	0.827502409063906\\
5	0.858229545568293\\
5.5	0.885997940981153\\
6	0.910574910612285\\
6.5	0.931822327540001\\
7	0.949691155209556\\
7.5	0.964226743252211\\
8	0.975594157613141\\
8.5	0.984100457740059\\
9	0.990165662159793\\
9.5	0.994246100195212\\
10	0.996807284398243\\
};
\addlegendentry{BPSK-AWGN capacity}

\addplot [color=red, line width=1.0pt]
  table[row sep=crcr]{%
-6.15959313428983	0.00645161290322581\\
-5.66887398971014	0.0129032258064516\\
-5.32166713820325	0.0193548387096774\\
-5.0241818148193	0.0258064516129032\\
-4.76650935383122	0.032258064516129\\
-4.51671880165526	0.0387096774193548\\
-4.29372731069889	0.0451612903225806\\
-4.0742842147414	0.0516129032258065\\
-3.86901057236057	0.0580645161290323\\
-3.67380696530691	0.0645161290322581\\
-3.47809800529004	0.0709677419354839\\
-3.30357703398309	0.0774193548387097\\
-3.12825278088662	0.0838709677419355\\
-2.95498902955733	0.0903225806451613\\
-2.80057533501658	0.0967741935483871\\
-2.64139277477313	0.103225806451613\\
-2.48254818986454	0.109677419354839\\
-2.34478638366048	0.116129032258065\\
-2.2032095837567	0.12258064516129\\
-2.0561379194903	0.129032258064516\\
-1.92374802321565	0.135483870967742\\
-1.80016061787427	0.141935483870968\\
-1.66955947964111	0.148387096774194\\
-1.5361608533258	0.154838709677419\\
-1.41856460314049	0.161290322580645\\
-1.30746280492931	0.167741935483871\\
-1.18959711159375	0.174193548387097\\
-1.06636970837632	0.180645161290323\\
-0.953969751249117	0.187096774193548\\
-0.853271802373347	0.193548387096774\\
-0.748561394615601	0.2\\
-0.638348696019627	0.206451612903226\\
-0.52429000087366	0.212903225806452\\
-0.427117681029066	0.219354838709677\\
-0.335793259544048	0.225806451612903\\
-0.237700401440305	0.232258064516129\\
-0.136743356683565	0.238709677419355\\
-0.0306634702398234	0.245161290322581\\
0.0608479582757349	0.251612903225806\\
0.144842500489312	0.258064516129032\\
0.236683261662535	0.264516129032258\\
0.327996629107366	0.270967741935484\\
0.423164386281246	0.27741935483871\\
0.520370265905709	0.283870967741935\\
0.598079888665298	0.290322580645161\\
0.680415495424007	0.296774193548387\\
0.766268556661849	0.303225806451613\\
0.85204035817139	0.309677419354839\\
0.944114218671943	0.316129032258064\\
1.03144705552541	0.32258064516129\\
1.1052947061948	0.329032258064516\\
1.18365498390395	0.335483870967742\\
1.26345509495993	0.341935483870968\\
1.34611083695442	0.348387096774194\\
1.43230240144761	0.354838709677419\\
1.51878366632547	0.361290322580645\\
1.58827296100504	0.367741935483871\\
1.66230617651858	0.374193548387097\\
1.73735346569203	0.380645161290323\\
1.81594919522497	0.387096774193548\\
1.89769563328826	0.393548387096774\\
1.98454394827028	0.4\\
2.0537816316269	0.406451612903226\\
2.12311596938243	0.412903225806452\\
2.1942595693972	0.419354838709677\\
2.26807076576846	0.425806451612903\\
2.34765874420635	0.432258064516129\\
2.42853793922329	0.438709677419355\\
2.50929962483496	0.445161290322581\\
2.5729894611419	0.451612903225806\\
2.64086434013179	0.458064516129032\\
2.7109584355498	0.464516129032258\\
2.78565968328382	0.470967741935484\\
2.86369571976114	0.47741935483871\\
2.94353598686082	0.483870967741935\\
3.01924597827853	0.490322580645161\\
3.08167186606975	0.496774193548387\\
3.14839148422755	0.503225806451613\\
3.21889336405512	0.509677419354839\\
3.2956350219765	0.516129032258065\\
3.3717081476129	0.52258064516129\\
3.45162114492493	0.529032258064516\\
3.52398323681385	0.535483870967742\\
3.58619165828263	0.541935483870968\\
3.65349233031549	0.548387096774194\\
3.72546325130426	0.554838709677419\\
3.8025527599945	0.561290322580645\\
3.87916058220635	0.567741935483871\\
3.957997162749	0.574193548387097\\
4.02871177425635	0.580645161290323\\
4.0934549648789	0.587096774193548\\
4.16274398543482	0.593548387096774\\
4.2366762467501	0.6\\
4.31521129077463	0.606451612903226\\
4.39308303256476	0.612903225806452\\
4.47094799652511	0.619354838709677\\
4.54161001773273	0.625806451612903\\
4.61026670273504	0.632258064516129\\
4.68341341703892	0.638709677419355\\
4.76074272976843	0.645161290322581\\
4.8416722063007	0.651612903225806\\
4.921266320468	0.658064516129032\\
5.00170548853779	0.664516129032258\\
5.07170895055105	0.670967741935484\\
5.146616204572	0.67741935483871\\
5.22479453447735	0.683870967741935\\
5.30788658224297	0.690322580645161\\
5.39219982475347	0.696774193548387\\
5.47588493238368	0.703225806451613\\
5.55409532076242	0.709677419354839\\
5.63241147731325	0.716129032258065\\
5.71547331914398	0.72258064516129\\
5.80234195923406	0.729032258064516\\
5.89151848886305	0.735483870967742\\
5.98337860164315	0.741935483870968\\
6.06728230078359	0.748387096774194\\
6.15405400469053	0.754838709677419\\
6.24545184506815	0.761290322580645\\
6.33995032599032	0.767741935483871\\
6.43821354789858	0.774193548387097\\
6.53705074659393	0.780645161290323\\
6.63309181995551	0.787096774193548\\
6.73356418205273	0.793548387096774\\
6.8392154248169	0.8\\
6.94645961505008	0.806451612903226\\
7.0592820355656	0.812903225806452\\
7.17400774025713	0.819354838709677\\
7.29191071107937	0.825806451612903\\
7.41427552039241	0.832258064516129\\
7.54294546933286	0.838709677419355\\
7.68273788660005	0.845161290322581\\
7.82529041144603	0.851612903225806\\
7.97059146930372	0.858064516129032\\
8.14056331991659	0.864516129032258\\
8.31761420753795	0.870967741935484\\
8.49720611319603	0.87741935483871\\
8.72139059641145	0.883870967741936\\
8.94776491468055	0.890322580645161\\
9.21704321548221	0.896774193548387\\
9.49887834710557	0.903225806451613\\
9.82870544999792	0.909677419354839\\
10.1722072602889	0.916129032258065\\
10.5308549299066	0.92258064516129\\
10.8958856436518	0.929032258064516\\
11.2544281045062	0.935483870967742\\
11.6047631607015	0.941935483870968\\
11.942592048873	0.948387096774194\\
12.2617328537912	0.954838709677419\\
12.5706125319992	0.961290322580645\\
12.8595921245145	0.967741935483871\\
13.1370941945362	0.974193548387097\\
13.40374811129	0.980645161290323\\
13.6537723642652	0.987096774193548\\
13.8944038926497	0.993548387096774\\
};
\addlegendentry{${\textrm{RCU}}[155, k]$ capacity}

\addplot [color=blue, line width=1.0pt, only marks, mark=asterisk, mark options={solid, blue}]
  table[row sep=crcr]{%
-1.15193538374402	0.161290322580645\\
0.11642831058663	0.225806451612903\\
1.18603076657383	0.290322580645161\\
1.99527537250348	0.354838709677419\\
2.63615194213866	0.419354838709677\\
3.27605641080867	0.483870967741935\\
3.85167049984583	0.548387096774194\\
4.43346177179555	0.612903225806452\\
5.17550173636032	0.67741935483871\\
5.96284839593041	0.741935483870968\\
6.94690129722602	0.806451612903226\\
8.25194430392264	0.870967741935484\\
10.2738174714077	0.935483870967742\\
};
\addlegendentry{$\mathscr{C}_{\textrm{RS}}[31, k]_{2^5}$, LC-OSD decoding}

\end{axis}
\end{tikzpicture}%
} 
  \hfill
  \subfloat[Average number of searches of LC-OSD with the DAI criterion at the different FERs~(corresponding to different SNRs for different code rates).\label{fig:rs31-lavg}]{
%
%
\definecolor{mycolor1}{rgb}{0.00000,0.44700,0.74100}%
\begin{tikzpicture}[%
thick,scale=0.7, every node/.style={scale=0.7}
]

\begin{axis}[%
width=10.2666667cm,
height=7.3407553cm,
at={(0cm,0cm)},
scale only axis,
unbounded coords=jump,
xmin=0.161290322580645,
xmax=0.935483870967742,
xlabel style={font=\color{white!15!black}},
xlabel={Code rate},
ymode=log,
ymin=1,
ymax=1000,
yminorticks=true,
ylabel style={font=\color{white!15!black}},
ylabel={$\ell_{\textrm{avg}}$},
axis background/.style={fill=white},
xmajorgrids,
ymajorgrids,
yminorgrids,
legend style={legend cell align=left, align=left, draw=white!15!black}
]
\addplot [color=mycolor1, forget plot]
  table[row sep=crcr]{%
0	0\\
};
\addplot [color=green!60!blue, mark=asterisk, mark options={solid, green!60!blue}]
  table[row sep=crcr]{%
0.161290322580645	520.936203873636\\
0.225806451612903	866.202209673151\\
0.290322580645161	898.617862464606\\
0.354838709677419	887.94438847526\\
0.419354838709677	794.590828671342\\
0.483870967741935	621.730210652661\\
0.548387096774194	389.877173454055\\
0.612903225806452	257.052695073239\\
0.67741935483871	66.6323200599986\\
0.741935483870968	14.5047971787246\\
0.806451612903226	3.03772429197003\\
0.870967741935484	1.2412532531269\\
0.935483870967742	1.0103548058503\\
};
\addlegendentry{$\textrm{FER}=10^{-2}$}

\addplot [color=green!20!blue, mark=+, mark options={solid, green!20!blue}]
  table[row sep=crcr]{%
0.161290322580645	213.399098642411\\
0.225806451612903	351.095224486802\\
0.290322580645161	335.34068514233\\
0.354838709677419	279.480583401312\\
0.419354838709677	223.182140517946\\
0.483870967741935	176.364658766833\\
0.548387096774194	116.258022683163\\
0.612903225806452	79.1423426556943\\
0.67741935483871	22.5352726347095\\
0.741935483870968	4.74850284998479\\
0.806451612903226	1.5505182969543\\
0.870967741935484	1.06633553089257\\
0.935483870967742	1\\
};
\addlegendentry{$\textrm{FER}=10^{-3}$}

\addplot [color=red!20!blue, mark=square, mark options={solid, red!20!blue}]
  table[row sep=crcr]{%
0.161290322580645	102.629406811547\\
0.225806451612903	133.639341487405\\
0.290322580645161	109.667113736026\\
0.354838709677419	89.2118851097349\\
0.419354838709677	73.5607638661849\\
0.483870967741935	59.1413947455915\\
0.548387096774194	35.7364347395352\\
0.612903225806452	23.7328503357436\\
0.67741935483871	6.70532570953995\\
0.741935483870968	2.40418023651647\\
0.806451612903226	1.12892807930306\\
0.870967741935484	1.01522962237353\\
0.935483870967742	1\\
};
\addlegendentry{$\textrm{FER}=10^{-4}$}

\addplot [color=red!60!blue, mark=o, mark options={solid, red!60!blue}]
  table[row sep=crcr]{%
0.161290322580645	44.2460219097682\\
0.225806451612903	51.6545719214004\\
0.290322580645161	39.7250677457863\\
0.354838709677419	31.2900666096078\\
0.419354838709677	28.7269590735813\\
0.483870967741935	19.3868184294212\\
0.548387096774194	12.7561567972357\\
0.612903225806452	7.14689060783074\\
0.67741935483871	3.21853616464839\\
0.741935483870968	1.47129105974942\\
0.806451612903226	1.05583636155829\\
0.870967741935484	1.00316752711929\\
0.935483870967742	1\\
};
\addlegendentry{$\textrm{FER}=10^{-5}$}

\addplot [color=red, mark=x, mark options={solid, red}]
  table[row sep=crcr]{%
0.161290322580645	19.6018622080019\\
0.225806451612903	22.6374716750428\\
0.290322580645161	15.8910186887258\\
0.354838709677419	12.0920475953084\\
0.419354838709677	10.0145610083079\\
0.483870967741935	8.17908457098232\\
0.548387096774194	5.75704854525394\\
0.612903225806452	3.68450449979247\\
0.67741935483871	1.68813015017888\\
0.741935483870968	1.15211406746748\\
0.806451612903226	1.01880637239206\\
0.870967741935484	1\\
0.935483870967742	nan\\
};
\addlegendentry{$\textrm{FER}=10^{-6}$}

\end{axis}
\end{tikzpicture}%
} 
  \caption{
    Simulation results of the RS codes $\mathscr{C}_{\textrm{RS}}[31, k]_{2^5}$, which are mapped into $\mathbb{F}_2^{155}$ for LC-OSD$(8, 2^{14})$.
  }
  \label{fig:rs31}
\end{figure}

\rchange{
\begin{example}
  \label{ex:rs31}
  To demonstrate the effect of various criteria on the LC-OSD, we present the simulation results for Reed-Solomon~(RS) codes with different code rates in Fig.~\ref{fig:rs31}.
  The simulated RS codes are defined over the field $\mathbb{F}_{2^5}$, denoted by $\mathscr{C}_{\textrm{RS}}[31, k]_{2^5}$ with $k$ as the dimension. 
  They are mapped into $\mathbb{F}_2^{155}$ and decoded by the LC-OSD. 
  The performances are shown in Fig.~\ref{fig:rs31-fer}, from which we see that the approximate ideal stopping~(including SAI and DAI) criteria perform as well as the trivial stopping criterion. 
  In Fig.~\ref{fig:rs31-cap}, we compare the RS codes with the RCU bounds, from which we can see that the achieving code rates of the RS codes are close to the RCU bounds.
  This suggests that the RS codes are good codes in the finite length region.
  Further, we present the average list size $\ell_{\textrm{avg}}$ for different code rates and target FER.
  As shown in Fig.~\ref{fig:rs31-cap}, when the RS codes run on their working regions~($\textrm{FER} \approx 10^{-5}$), the average numbers of searches are reduced to tens or even a few.
  Equipped with the tailored stopping criteria, LC-OSD is indeed a sub-ML decoding algorithm for short block codes.
  In Fig.~\ref{fig:rs31-lavg}, we show the average number of searches of LC-OSD at different FERs, which correspond to different SNRs for different code rates. For example, to achieve $\textrm{FER}=10^{-4}$, the required SNRs range from about $-1.75~\textrm{dB}$ for the lowest code rate to about $9.5~\textrm{dB}$ for the highest code rate.
  The results show that, in the low FER region, the average number of searches can be very low, suggesting potential applications in future communications where the target reliability is $99.99999\%$.
\end{example}}

For the reason that the LC-OSD with the DAI stopping criterion has much lower time complexity without incurring \rchange{noticeable} performance loss%
\footnote{This was also verified by simulations in~\cite{Liang2022CommL} and holds for a wide class of short codes.},
we use the LC-OSD with the DAI stopping criterion \rchange{for simulations} in the rest of this paper. 

\subsection{Influences of Decoding Parameters}
\label{sec:infulence}
Recalling that $\delta = 0$ implies the original OSD algorithm and $\delta = n-k$ implies the MLD, 
we can safely infer that, as the increase of $\delta$, the maximum number of searches can be reduced.
Therefore, in this subsection, we discuss the relationship between FER and the decoding parameters $(\delta,\ell_{\textrm{max}})$.

\subsubsection{The Channel of MRBs}
The channel of each MRB is nearly%
\footnote{This is because the linear dependency of columns of $\mathbf{H}$ and hence $|\widetilde{\bm{r}}|$ is not always non-decreasing.}
an ``order statistic'' AWGN channel
\begin{equation}
  \widetilde{y}_i=\widetilde{x}_i+\widetilde{w}_i,\qquad n-k-\delta < i \leq n.
\end{equation}
Formally, $|\widetilde{y}_i|\,(n-k-\delta< i \leq n)$ is the $i$-th order statistic of $|y_i|$ \rchange{\cite{Fossorier1995OSD, yue2021revisit}, whose} probability density function given by
\begin{equation}
  f_{|\widetilde{y}_i|}(y) = 
  \frac{n!}{(i-1)!(n-i)!} f_{|y|}(y) (F_{|y|}(y))^{i-1} (1-F_{|y|}(y))^{n-i},
\end{equation}
where $f_{|y|}(\cdot)$ and $F_{|y|}(\cdot)$ are the probability density function and cumulative distribution function of $|y|$, respectively, i.e., 
\begin{align}
  f_{|y|}(y) &
  = \frac{1}{\sqrt{2\pi}\sigma} \exp \bigg(-\frac{(y-1)^2}{2\sigma^2}\bigg)
  + \frac{1}{\sqrt{2\pi}\sigma} \exp \bigg(-\frac{(y+1)^2}{2\sigma^2}\bigg), 
  & y \geq 0,\\
  F_{|y|}(y) &
  = \int_0^{y} f_{|y|}(t) \,\mathrm{d}t
  = \frac{1}{2} \erf\left(\frac{y-1}{\sqrt{2}\sigma}\right) +
  \frac{1}{2} \erf\left(\frac{y+1}{\sqrt{2}\sigma}\right),
  & y \geq 0.
\end{align}

\subsubsection{The MRB Code}
\label{sec:mrb-codes}
After preprocessing, the MRB part of the code $\mathscr{C}[n, k]$ is determined.
\begin{definition}{(MRB code)}
  \label{def:mrb-code}
  The MRB code $\mathscr{C}_{\textrm{MRB}}[k + \delta, k]$ is a code from puncturing $\mathscr{C}[n, k]$, where the LRBs~(see Sec.~\ref{sec:lc-osd}) are punctured.
\end{definition}
\begin{proposition}
  \label{prop:mrb-code}
  Up to a column permutation, the local-constraint matrix $\mathbf{P}_2$~(see Sec.~\ref{sec:lc-osd}) is a parity-check matrix of the MRB code.
\end{proposition}
\begin{IEEEproof}
  The proof follows immediately from~(\ref{equ:c-parity}).
\end{IEEEproof}
From the Definition~\ref{def:mrb-code} and Proposition~\ref{prop:mrb-code}, we can see that the MRB code is a semi-random code due to the randomness of the LLR vector $\bm{r}$.

In the following, instead of $\widetilde{\bm{e}}_{\textrm{R}}$, $\widetilde{\bm{y}}_{\textrm{R}}$, \rchange{$\widetilde{\bm{r}}_{\textrm{R}}$} and $\widetilde{\bm{z}}_{\textrm{R}}$, we use \rchange{for simplicity} the notations ${\bm{e}}$, ${\bm{y}}$, \rchange{${\bm{r}}$} and ${\bm{z}}$ to denote the MRB part of the TEP, the received vector, \rchange{the LLR vector}, and the bit-wise hard decision vector, respectively. 

\rchange{
Denote the number of MRB codewords lighter than the sent MRB codeword by $L$, i.e.,
\begin{equation}
  L \triangleq \bigl|\bigl\{
    {\bm{v}} \in \mathscr{C}_{\textrm{MRB}}: \Gamma({\bm{f}}) < \Gamma({\bm{e}}), \bm{f} = {\bm{z}} - {\bm{v}}
  \bigr\}\bigr|.
\end{equation}
Then the FER of the MRB code} under the optimal list decoder%
\footnote{
Here, the optimal list decoder generates a list containing the $\ell_{\textrm{max}}$ lightest TEPs. 
}
with the list size of $\ell_{\textrm{max}}$ is \rchange{given by}
\begin{equation}
  \label{equ:def-epsilon-mrb}
  \varepsilon_{\textrm{MRB}} \triangleq \mathbb{P}[L \geq \ell_{\textrm{max}}].
\end{equation}

\subsubsection{\rchange{The Random Code and Performance Approximation}}
\label{sec:random-codes}
\rchange{
\begin{definition}{(Random code)}
  Let $\widehat{\mathscr{C}}$ be a multi-subset of $\mathbb{F}_2^N$ with $2^K$ elements.
  If all elements of $\widehat{\mathscr{C}}$ are independently and uniformly drawn from $\mathbb{F}_2^N$, then we call $\widehat{\mathscr{C}}$ as a random code, denoted as $\widehat{\mathscr{C}}[N, K]$.
\end{definition}
Similarly, the list FER performance of the random code, denoted $\widehat{\mathscr{C}}_{\textrm{MRB}}[k+\delta, k]$, is defined as
\begin{equation}
  \label{equ:def-epsilon-mrb-hat}
  \widehat{\varepsilon}_{\textrm{MRB}}
  \triangleq \mathbb{P}[\widehat{L} \geq \ell_{\textrm{max}}],
\end{equation}
where $\widehat{L}$ denotes the number of codeword lighter the sent codeword in $\widehat{\mathscr{C}}_{\textrm{MRB}}$, i.e.,
\begin{equation}
  \widehat{L} \triangleq \bigl|\bigl\{
    {\bm{v}} \in \widehat{\mathscr{C}}_{\textrm{MRB}}: \Gamma({\bm{f}}) < \Gamma({\bm{e}}), \bm{f} = {\bm{z}} - {\bm{v}}
  \bigr\}\bigr|.
\end{equation}
}

Here, we introduce a set
\begin{equation}
  {\mathcal{D}} \triangleq \bigl\{{\bm{f}}\in\mathbb{F}_2^{k+\delta}: \Gamma({\bm{f}}) < \Gamma({\bm{e}}) \bigr\}
\end{equation}
of all sequences \rchange{lighter than the correct TEP ${\bm{e}}$ and denote its cardinality as 
\begin{equation}
  D \triangleq |\mathcal{D}| 
  = \sum_{{\bm{f}}\in{\mathbb{F}_2^{k+\delta}}} 
  \mathbb{I}\bigl[\Gamma({\bm{f}}) < \Gamma({\bm{e}})\bigr],
\end{equation}
where $\mathbb{I}[\cdot]$ is the indicator function.
Due to the symmetry, we can safely assume that the all-zero codeword $\bm{0}$~(of ${\mathscr{C}}_{\textrm{MRB}}$ or $\widehat{\mathscr{C}}_{\textrm{MRB}}$) is sent.

The cardinality $D$ of the set ${\mathcal{D}}$ depends only on the LLR vector $\bm{r}$, denoted $D(\bm{r})$, which does not rely on any specific code.
One way to calculate $D(\bm{r})$ is to directly count the number by listing all vectors in the set $\mathcal{D}$ with algorithms such as the FPT~(Sec.~\ref{sec:fpt}).
This is suitable for short codes~($n\leq 64$) but is time-consuming for longer codes.
In the latter case, we use the saddlepoint method to approximate the cardinality $D(\bm{r})$.

To do so, we first introduce a probability model over $\mathbb{F}_2^{k+\delta}$.
Formally, let $\bm{f}$ be a random sequence distributed uniformly over $\mathbb{F}_2^{k+\delta}$.
That is, the probability of any $\bm{f}^*\in \mathbb{F}_2^{k+\delta}$ is
\begin{equation}
  \mathbb{P}[\bm{f} = \bm{f}^*] = 2^{-(k + \delta)}. 
\end{equation}
With this probability model, we can rewrite the cardinality as
\begin{align}
  D(\bm{r})
  &= 2^{k+\delta} \cdot \mathbb{P}[\Gamma({\bm{f}}) < \Gamma({\bm{e}}) | \bm{r}] \\
  &= 2^{k+\delta} \cdot \mathbb{P}\left[\left. \sum_{i=1}^{k+\delta} f_i |r_i| < \sum_{i=1}^{k+\delta} e_i |r_i| \right|\bm{r}\right] \\
  &= 2^{k+\delta} \cdot \mathbb{P}\left[\left. \sum_{i=1}^{k+\delta} (f_i - e_i) |r_i| < 0 \right|\bm{r}\right] \\
  &= 2^{k+\delta} \cdot \mathbb{P}\left[\left. \sum_{i=1}^{k+\delta} W_i < 0 \right|\bm{r}\right] \\
  &= 2^{k+\delta} \cdot \mathbb{P}\left[\left. W < 0 \right|\bm{r}\right],
\end{align}
where $W_i = (f_i - e_i) |r_i|$ and $W = \sum_{i=1}^{k+\delta} W_i$.
Recalling that $e_i = \mathbb{I}[r_i < 0]$, we can show that $W_i~(1\leq i \leq k+\delta)$ are independent random variables with probabilities
\begin{align}
  \mathbb{P}[W_i = 0 | \bm{r}] &= \mathbb{P}[f_i = e_i | \bm{r}] = 1/2, \\
  \mathbb{P}[W_i = r_i | \bm{r}] &= \mathbb{P}[f_i \neq e_i | \bm{r}] = 1/2.
\end{align}
Therefore, the calculation of the cardinality is equivalent to evaluating the tail probability, which can be solved by saddlepoint approximation~\cite{Martinez2011Saddlepoint, Font2018Saddlepoint}.
To be precise, let $\kappa(s)$ be the cumulant generating function of the summation $W$, then 
\begin{align}
  \label{equ:def-kappa}
  \kappa(s) 
  &\triangleq \log \mathbb{E}[e^{s W}] \\
  &= \sum_{i=1}^{k+\delta} \log \mathbb{E}[e^{s W_i}] \\
  &= \sum_{i=1}^{k+\delta} \log \Big(\frac{1}{2}+\frac{1}{2}e^{s r_i}\Big).
\end{align}
By using saddlepoint method~\cite{Font2018Saddlepoint}, (given $\bm{r}$,) the PDF of $W$ can be approximated by
\begin{equation}
  f_W(w) \approx \exp \left({\kappa(\hat{s}) - \hat{s}w}\right)\cdot
  \frac{1}{\sqrt{2\pi \kappa''(\hat{s})}}
  \exp\left({-\frac{(w-\kappa'(\hat{s}))^2}{2 \kappa''(\hat{s})}}\right),
\end{equation}
where $\kappa'$~(or $\kappa''$) is the first~(or second) derivative of $\kappa$ and $\hat{s}$ is the solution%
\footnote{It can be proven that $\kappa'(s)$ is an increasing function and has exactly one root if $(\max_i r_i)(\min_i r_i) < 0$.}
 of $\kappa'(s)=0$.
Therefore, the cardinality can be approximated by
\begin{align}
  D(\bm{r})
  &= 2^{k+\delta} \cdot \mathbb{P}\left[\left. W < 0 \right|\bm{r}\right] \\
  &= 2^{k+\delta} \cdot \int_{-\infty}^{0} f_W(w) \mathrm{d}w \\
  \label{equ:approx-Dr}
  &\approx 2^{k+\delta} \cdot \frac{1}{2} 
  \exp\left({\kappa(\hat{s})-\hat{s}\kappa'(\hat{s})+\frac{1}{2}\hat{s}^2\kappa''(\hat{s})}\right)
  \cdot \erfc\bigg(\frac{\kappa'(\hat{s}) - \hat{s}\kappa''(\hat{s})}{\sqrt{2\kappa''(\hat{s})}}\bigg),
\end{align}
where $\kappa(s)$ is defined in (\ref{equ:def-kappa}) and $\erfc(\cdot)$ is the complementary error function defined as
\begin{equation}
  \erfc(x) \triangleq 1 - \erf(x).
\end{equation}
}

\rchange{
\begin{example}
\label{ex:two-techs-Dr}
\begin{figure}[!t]
	\centering
%
%
\definecolor{mycolor1}{rgb}{0.00000,0.44700,0.74100}%
\begin{tikzpicture}[%
thick,scale=0.8, every node/.style={scale=0.8}
]

\begin{axis}[%
width=10.4cm,
height=7.561316499999999cm,
at={(0cm,0cm)},
scale only axis,
xmode=log,
xmin=1,
xmax=1000000,
xminorticks=true,
xlabel style={font=\color{white!15!black}},
xlabel={$d$},
ymode=log,
ymin=0.0001,
ymax=1,
yminorticks=true,
ylabel style={font=\color{white!15!black}},
ylabel={The CCDF $F_D^c(d)$},
axis background/.style={fill=white},
xmajorgrids,
xminorgrids,
ymajorgrids,
yminorgrids,
legend style={at={(0.03,0.03)}, anchor=south west, legend cell align=left, align=left, draw=white!15!black},
legend columns=2
]
\addplot [color=mycolor1, forget plot]
  table[row sep=crcr]{%
0	0\\
};
\addplot [color=green, dashed, line width=1.0pt]
  table[row sep=crcr]{%
1	0.967116\\
1.46779926762207	0.967116\\
2.15443469003188	0.963243\\
3.16227766016838	0.959593\\
4.64158883361278	0.955984\\
6.81292069057961	0.949259\\
10	0.939937\\
14.6779926762207	0.926588\\
21.5443469003188	0.911639\\
31.6227766016838	0.896026\\
46.4158883361278	0.88214\\
68.1292069057961	0.870578\\
99.9999999999999	0.857996\\
146.779926762207	0.841908\\
215.443469003188	0.822694\\
316.227766016838	0.800452\\
464.158883361278	0.776494\\
681.292069057961	0.751404\\
999.999999999999	0.7274\\
1467.79926762207	0.703858\\
2154.43469003188	0.679342\\
3162.27766016838	0.652917\\
4641.58883361277	0.624988\\
6812.92069057961	0.59529\\
9999.99999999999	0.565454\\
14677.9926762207	0.536002\\
21544.3469003188	0.507023\\
31622.7766016838	0.477965\\
46415.8883361277	0.448777\\
68129.206905796	0.41956\\
99999.9999999999	0.390509\\
146779.926762207	0.362056\\
215443.469003188	0.334679\\
316227.766016837	0.308214\\
464158.883361277	0.282673\\
681292.06905796	0.258168\\
999999.999999998	0.23484\\
};
\addlegendentry{Counting}

\addplot [color=green!75!blue, line width=1.0pt, only marks, mark=o, mark options={solid, green!75!blue}]
  table[row sep=crcr]{%
1	0.972513\\
2.15443469003188	0.97251\\
4.64158883361278	0.961481\\
10	0.922941\\
21.5443469003188	0.898226\\
46.4158883361278	0.884962\\
100	0.86604\\
215.443469003188	0.81626\\
464.158883361278	0.76656\\
1000	0.728189\\
2154.43469003188	0.680614\\
4641.58883361278	0.619762\\
10000	0.561593\\
21544.3469003189	0.506354\\
46415.8883361278	0.446463\\
100000	0.387758\\
215443.469003189	0.333003\\
464158.883361278	0.281029\\
1000000	0.233142\\
};
\addlegendentry{Saddlepoint~(\ref{equ:approx-Dr}), $E_{\textrm{b}}/N_0 = 0~\textrm{dB}$}

\addplot [color=teal, dashed, line width=1.0pt]
  table[row sep=crcr]{%
1	0.86835\\
1.46779926762207	0.86835\\
2.15443469003188	0.857446\\
3.16227766016838	0.847021\\
4.64158883361278	0.837146\\
6.81292069057961	0.818669\\
10	0.793702\\
14.6779926762207	0.758473\\
21.5443469003188	0.720069\\
31.6227766016838	0.682268\\
46.4158883361278	0.65081\\
68.1292069057961	0.628685\\
99.9999999999999	0.60569\\
146.779926762207	0.578163\\
215.443469003188	0.546292\\
316.227766016838	0.510705\\
464.158883361278	0.473655\\
681.292069057961	0.437247\\
999.999999999999	0.404516\\
1467.79926762207	0.375693\\
2154.43469003188	0.34795\\
3162.27766016838	0.320159\\
4641.58883361277	0.291718\\
6812.92069057961	0.263668\\
9999.99999999999	0.237125\\
14677.9926762207	0.21264\\
21544.3469003188	0.190582\\
31622.7766016838	0.170344\\
46415.8883361277	0.151182\\
68129.206905796	0.133158\\
99999.9999999999	0.116419\\
146779.926762207	0.101564\\
215443.469003188	0.088032\\
316227.766016837	0.07605\\
464158.883361277	0.065343\\
681292.06905796	0.055803\\
999999.999999998	0.047291\\
};
\addlegendentry{Counting}

\addplot [color=green!25!blue, line width=1.0pt, only marks, mark=square, mark options={solid, green!25!blue}]
  table[row sep=crcr]{%
1	0.883086\\
2.15443469003188	0.883085\\
4.64158883361278	0.856265\\
10	0.748299\\
21.5443469003188	0.68542\\
46.4158883361278	0.655083\\
100	0.622368\\
215.443469003188	0.537193\\
464.158883361278	0.457543\\
1000	0.404917\\
2154.43469003188	0.350492\\
4641.58883361278	0.286956\\
10000	0.233375\\
21544.3469003189	0.190219\\
46415.8883361278	0.149944\\
100000	0.114845\\
215443.469003189	0.087274\\
464158.883361278	0.064758\\
1000000	0.0467109999999999\\
};
\addlegendentry{Saddlepoint~(\ref{equ:approx-Dr}), $E_{\textrm{b}}/N_0 = 1~\textrm{dB}$}

\addplot [color=blue, dashed, line width=1.0pt]
  table[row sep=crcr]{%
1	0.651847\\
1.46779926762207	0.651847\\
2.15443469003188	0.634339\\
3.16227766016838	0.617312\\
4.64158883361278	0.601505\\
6.81292069057961	0.572255\\
10	0.533505\\
14.6779926762207	0.480153\\
21.5443469003188	0.424589\\
31.6227766016838	0.372084\\
46.4158883361278	0.331995\\
68.1292069057961	0.308893\\
99.9999999999999	0.28758\\
146.779926762207	0.262653\\
215.443469003188	0.235247\\
316.227766016838	0.206608\\
464.158883361278	0.178044\\
681.292069057961	0.151719\\
999.999999999999	0.129948\\
1467.79926762207	0.113097\\
2154.43469003188	0.098853\\
3162.27766016838	0.0853930000000001\\
4641.58883361277	0.072591\\
6812.92069057961	0.060651\\
9999.99999999999	0.050078\\
14677.9926762207	0.041248\\
21544.3469003188	0.034184\\
31622.7766016838	0.028263\\
46415.8883361277	0.0232250000000001\\
68129.206905796	0.018886\\
99999.9999999999	0.015144\\
146779.926762207	0.012019\\
215443.469003188	0.00962600000000002\\
316227.766016837	0.007637\\
464158.883361277	0.00605299999999998\\
681292.06905796	0.00472499999999998\\
999999.999999998	0.00365499999999996\\
};
\addlegendentry{Counting}

\addplot [color=red!25!blue, line width=1.0pt, only marks, mark=+, mark options={solid, red!25!blue}]
  table[row sep=crcr]{%
1	0.674811\\
2.15443469003188	0.674808\\
4.64158883361278	0.639909\\
10	0.465236\\
21.5443469003188	0.3729\\
46.4158883361278	0.334367\\
100	0.305582\\
215.443469003188	0.229531\\
464.158883361278	0.164827\\
1000	0.129236\\
2154.43469003188	0.100907\\
4641.58883361278	0.070764\\
10000	0.048337\\
21544.3469003189	0.0340819999999999\\
46415.8883361278	0.023002\\
100000	0.014775\\
215443.469003189	0.00947200000000004\\
464158.883361278	0.00598399999999999\\
1000000	0.00356800000000002\\
};
\addlegendentry{Saddlepoint~(\ref{equ:approx-Dr}), $E_{\textrm{b}}/N_0 = 2~\textrm{dB}$}

\addplot [color=violet, dashed, line width=1.0pt]
  table[row sep=crcr]{%
1	0.368117\\
1.46779926762207	0.368117\\
2.15443469003188	0.351836\\
3.16227766016838	0.336554\\
4.64158883361278	0.322406\\
6.81292069057961	0.296515\\
10	0.262487\\
14.6779926762207	0.216839\\
21.5443469003188	0.1708\\
31.6227766016838	0.13031\\
46.4158883361278	0.101405\\
68.1292069057961	0.088769\\
99.9999999999999	0.079362\\
146.779926762207	0.069051\\
215.443469003188	0.058119\\
316.227766016838	0.047175\\
464.158883361278	0.036712\\
681.292069057961	0.02742\\
999.999999999999	0.020826\\
1467.79926762207	0.01636\\
2154.43469003188	0.013239\\
3162.27766016838	0.010583\\
4641.58883361277	0.00821700000000003\\
6812.92069057961	0.00617299999999998\\
9999.99999999999	0.00451000000000001\\
14677.9926762207	0.00330699999999995\\
21544.3469003188	0.00241999999999998\\
31622.7766016838	0.00177799999999995\\
46415.8883361277	0.001359\\
68129.206905796	0.000990000000000046\\
99999.9999999999	0.000749\\
146779.926762207	0.000512999999999986\\
215443.469003188	0.000363000000000002\\
316227.766016837	0.000264000000000042\\
464158.883361277	0.000183999999999962\\
681292.06905796	0.000128000000000017\\
999999.999999998	8.69999999999482e-05\\
};
\addlegendentry{Counting}

\addplot [color=purple, line width=1.0pt, only marks, mark=asterisk, mark options={solid, purple}]
  table[row sep=crcr]{%
1	0.389033\\
2.15443469003188	0.389033\\
4.64158883361278	0.366369\\
10	0.203789\\
21.5443469003188	0.127698\\
46.4158883361278	0.101045\\
100	0.0885899999999999\\
215.443469003188	0.056882\\
464.158883361278	0.031541\\
1000	0.020262\\
2154.43469003188	0.013835\\
4641.58883361278	0.00796600000000003\\
10000	0.00422\\
21544.3469003189	0.00241000000000002\\
46415.8883361278	0.00135200000000002\\
100000	0.000716999999999968\\
215443.469003189	0.000354999999999994\\
464158.883361278	0.000183999999999962\\
1000000	8.60000000000305e-05\\
};
\addlegendentry{Saddlepoint~(\ref{equ:approx-Dr}), $E_{\textrm{b}}/N_0 = 3~\textrm{dB}$}

\end{axis}
\end{tikzpicture}%
  \caption{
    The CCDF of $D(\bm{r})$, where the randomness comes from the LLR vector $\bm{r}$. Here $(n, k, \delta) = (128, 64, 8)$.
    }
	\label{fig:Ddist}
\end{figure}
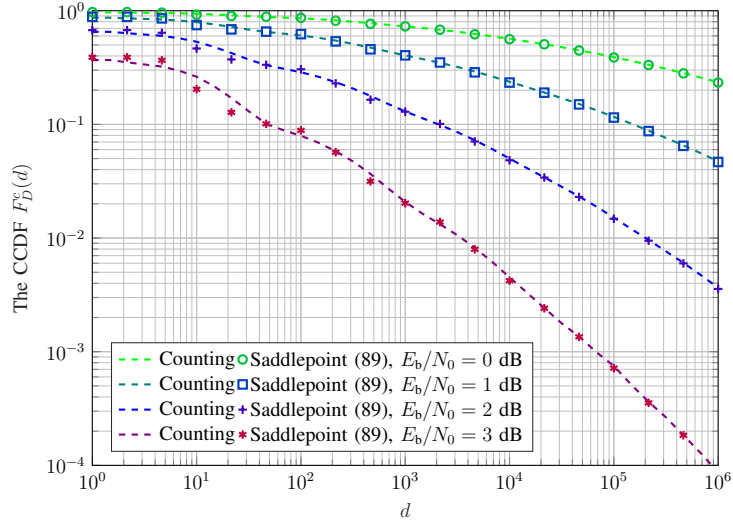
\rchange{
In this example, we employ two techniques to get the complementary CDF~(CCDF) of $D(\bm{r})$, formulated by
\begin{equation}
  F_D^c(d) \triangleq \mathbb{P}[D(\bm{r}) > d],
\end{equation}
in which $\bm{r}$ is a random vector generated by simulations.
The first method utilizes the FPT method~(Sec.~\ref{sec:fpt}) to count the exact number of sequences in $\mathcal{D}$.
The second method employs the saddlepoint approach presented in~(\ref{equ:approx-Dr}).
The simulation results are shown in Fig.~\ref{fig:Ddist}, from which we see that the results from the saddlepoint method match well with those from the counting.
We also see that, as $E_{\textrm{b}}/N_0$ increases, $D(\bm{r})$ decreases in statistical sense, as expected.
For example, the rank of the true TEP is less than $10^2$ for about $70\%$ realizations of $\bm{r}$ at $E_{\textrm{b}}/N_0 = 2~\textrm{dB}$ while about $90\%$ at $E_{\textrm{b}}/N_0 = 3~\textrm{dB}$.
}
\end{example}
}


\rchange{Once the cardinality of ${\mathcal{D}}$ is obtained, $\widehat{L}$ is randomly distributed as the binomial distribution.} 
\begin{equation}
  \widehat{L} \sim 
  \mathrm{Binomial}\bigl(D(\bm{r}), 2^{-\delta}\bigr)
\end{equation}
with the probability mass function~(PMF) given by
\begin{equation}
  \mathbb{P}[\widehat{L} = \ell \mid {D(\bm{r})}] = 
  \binom{D(\bm{r})}{\ell}\bigl(2^{-\delta}\bigr)^{\ell}
  \bigl(1 - 2^{-\delta}\bigr)^{D(\bm{r}) - \ell}
\end{equation}
Hence, we can derive the FER of the random code $\widehat{\mathscr{C}}_{\textrm{MRB}}$ as follows.
\begin{equation}
  \label{equ:mrb-hat-fer}
  \widehat{\varepsilon}_{\textrm{MRB}}
  = \mathbb{P}[\widehat{L} \geq \ell_{\textrm{max}}] 
  = \mathbb{E}_{\bm{r}}[\mathbb{P}[\widehat{L} \geq \ell_{\textrm{max}}\mid D(\bm{r})]].
\end{equation}

\rchange{
\begin{example}
\label{ex:list-fer}
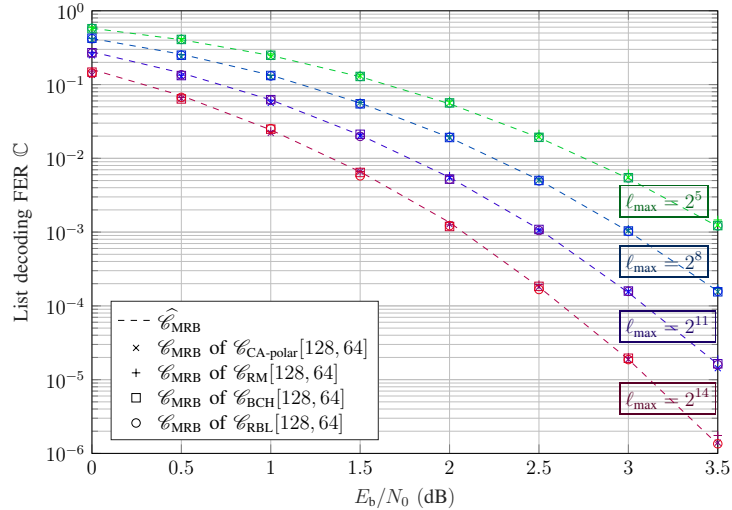
\begin{figure}[!t]
  \centering
%
%
\definecolor{mycolor1}{rgb}{0.00000,0.44700,0.74100}%
\definecolor{mycolor2}{rgb}{0.00000,0.40000,0.10000}%
\definecolor{mycolor3}{rgb}{0.00000,0.15000,0.35000}%
\definecolor{mycolor4}{rgb}{0.10000,0.00000,0.40000}%
\definecolor{mycolor5}{rgb}{0.35000,0.00000,0.15000}%
\begin{tikzpicture}[%
thick,scale=0.8, every node/.style={scale=0.8}
]

\begin{axis}[%
width=10.4cm,
height=7.361142999999999cm,
at={(0cm,0cm)},
scale only axis,
xmin=0,
xmax=3.5,
xlabel style={font=\color{white!15!black}},
xlabel={$E_{\textrm{b}}/N_0~(\textrm{dB})$},
ymode=log,
ymin=1e-06,
ymax=1,
yminorticks=true,
ylabel style={font=\color{white!15!black}},
ylabel={List decoding FER $\mathbb{C}$},
axis background/.style={fill=white},
xmajorgrids,
ymajorgrids,
yminorgrids,
legend style={at={(0.03,0.03)}, anchor=south west, legend cell align=left, align=left, draw=white!15!black},
legend columns=1
]
\addplot [color=mycolor1, forget plot]
  table[row sep=crcr]{%
0	0\\
};
\addplot [color=black, dashed]
  table[row sep=crcr]{%
1	1e-10\\
};
\addlegendentry{$\widehat{\mathscr{C}}_{\textrm{MRB}}$}

\addplot [color=black, only marks, mark=x, mark options={solid, black}]
  table[row sep=crcr]{%
1	1e-10\\
};
\addlegendentry{$\mathscr{C}_{\textrm{MRB}}$ of $\mathscr{C}_{\textrm{CA-polar}}[128,64]$}

\addplot [color=black, only marks, mark=+, mark options={solid, black}]
  table[row sep=crcr]{%
1	1e-10\\
};
\addlegendentry{$\mathscr{C}_{\textrm{MRB}}$ of $\mathscr{C}_{\textrm{RM}}[128,64]$}

\addplot [color=black, only marks, mark=square, mark options={solid, black}]
  table[row sep=crcr]{%
1	1e-10\\
};
\addlegendentry{$\mathscr{C}_{\textrm{MRB}}$ of $\mathscr{C}_{\textrm{BCH}}[128,64]$}

\addplot [color=black, only marks, mark=o, mark options={solid, black}]
  table[row sep=crcr]{%
1	1e-10\\
};
\addlegendentry{$\mathscr{C}_{\textrm{MRB}}$ of $\mathscr{C}_{\textrm{RBL}}[128,64]$}

\addplot [color=green!80!blue, dashed, forget plot]
  table[row sep=crcr]{%
0	0.580263571\\
0.5	0.4057516274\\
1	0.2479584702\\
1.5	0.1277659379\\
2	0.0549372359\\
2.5	0.0191578548\\
3	0.0053413365\\
3.5	0.00119441\\
};
\node[right, align=left, inner sep=0, font=\color{mycolor2}]
at (axis cs:2.95,0.0026303) {$\boxed{{\ell}_{\textrm{max}} = 2^{5}}$};
\addplot [color=green!30!blue, dashed, forget plot]
  table[row sep=crcr]{%
0	0.4206469224\\
0.5	0.2575999719\\
1	0.1342711011\\
1.5	0.0563218696\\
2	0.0192003245\\
2.5	0.0051033321\\
3	0.0010229405\\
3.5	0.0001583916\\
};
\node[right, align=left, inner sep=0, font=\color{mycolor3}]
at (axis cs:2.95,0.0003981) {$\boxed{{\ell}_{\textrm{max}} = 2^{8}}$};
\addplot [color=red!20!blue, dashed, forget plot]
  table[row sep=crcr]{%
0	0.2775135072\\
0.5	0.1435930404\\
1	0.0621176921\\
1.5	0.0206662377\\
2	0.0055255168\\
2.5	0.0010989444\\
3	0.0001512519\\
3.5	1.62198e-05\\
};
\node[right, align=left, inner sep=0, font=\color{mycolor4}]
at (axis cs:2.95,0.0000525) {$\boxed{{\ell}_{\textrm{max}} = 2^{11}}$};
\addplot [color=red!70!blue, dashed, forget plot]
  table[row sep=crcr]{%
0	0.1604375944\\
0.5	0.070766897\\
1	0.0243408102\\
1.5	0.0065841431\\
2	0.0013527799\\
2.5	0.0001806324\\
3	1.88228e-05\\
3.5	1.3271e-06\\
};
\node[right, align=left, inner sep=0, font=\color{mycolor5}]
at (axis cs:2.95,0.0000055) {$\boxed{{\ell}_{\textrm{max}} = 2^{14}}$};
\addplot [color=green, only marks, mark=x, mark options={solid, green}, forget plot]
  table[row sep=crcr]{%
0	0.5764546685\\
0.5	0.3990958993\\
1	0.2372900088\\
1.5	0.1318083872\\
2	0.0562343118\\
2.5	0.0195284229\\
3	0.0055061226\\
3.5	0.0012439\\
};
\addplot [color=green!90!blue, only marks, mark=+, mark options={solid, green!90!blue}, forget plot]
  table[row sep=crcr]{%
0	0.5908745247\\
0.5	0.4205264912\\
1	0.2607230036\\
1.5	0.1356533855\\
2	0.0600696514\\
2.5	0.0213803302\\
3	0.0062128034\\
3.5	0.00146195\\
};
\addplot [color=green!70!blue, only marks, mark=square, mark options={solid, green!70!blue}, forget plot]
  table[row sep=crcr]{%
0	0.5721649485\\
0.5	0.4091344618\\
1	0.249844624\\
1.5	0.1288081068\\
2	0.0564219035\\
2.5	0.0192651666\\
3	0.0054654098\\
3.5	0.00122565\\
};
\addplot [color=green!60!blue, only marks, mark=o, mark options={solid, green!60!blue}, forget plot]
  table[row sep=crcr]{%
0	0.579512894\\
0.5	0.4089231801\\
1	0.2483560951\\
1.5	0.1288183687\\
2	0.0563630916\\
2.5	0.0193447964\\
3	0.0054376029\\
3.5	0.00121465\\
};
\addplot [color=teal, only marks, mark=x, mark options={solid, teal}, forget plot]
  table[row sep=crcr]{%
0	0.401217862\\
0.5	0.2370035518\\
1	0.1272104332\\
1.5	0.0571523289\\
2	0.019094821\\
2.5	0.0050671233\\
3	0.001047414\\
3.5	0.0001551\\
};
\addplot [color=green!40!blue, only marks, mark=+, mark options={solid, green!40!blue}, forget plot]
  table[row sep=crcr]{%
0	0.4228136882\\
0.5	0.2555814728\\
1	0.132628153\\
1.5	0.0571608665\\
2	0.0202553886\\
2.5	0.0052662966\\
3	0.0010940634\\
3.5	0.0001721\\
};
\addplot [color=green!20!blue, only marks, mark=square, mark options={solid, green!20!blue}, forget plot]
  table[row sep=crcr]{%
0	0.4248895434\\
0.5	0.2503992335\\
1	0.1327532629\\
1.5	0.0548174361\\
2	0.0191590066\\
2.5	0.0050110217\\
3	0.0010360842\\
3.5	0.0001552\\
};
\addplot [color=green!10!blue, only marks, mark=o, mark options={solid, green!10!blue}, forget plot]
  table[row sep=crcr]{%
0	0.4183381089\\
0.5	0.2499161355\\
1	0.1310065756\\
1.5	0.0550424183\\
2	0.0193176745\\
2.5	0.004953962\\
3	0.0010208476\\
3.5	0.000158\\
};
\addplot [color=blue, only marks, mark=x, mark options={solid, blue}, forget plot]
  table[row sep=crcr]{%
0	0.248985115\\
0.5	0.129157249\\
1	0.0564765694\\
1.5	0.0201889607\\
2	0.0054684699\\
2.5	0.0010529676\\
3	0.0001556534\\
3.5	1.43e-05\\
};
\addplot [color=red!10!blue, only marks, mark=+, mark options={solid, red!10!blue}, forget plot]
  table[row sep=crcr]{%
0	0.2737642586\\
0.5	0.1339553482\\
1	0.0614901778\\
1.5	0.0213316525\\
2	0.005722712\\
2.5	0.0010743982\\
3	0.0001596431\\
3.5	1.875e-05\\
};
\addplot [color=red!30!blue, only marks, mark=square, mark options={solid, red!30!blue}, forget plot]
  table[row sep=crcr]{%
0	0.2702503682\\
0.5	0.1325455126\\
1	0.0623990056\\
1.5	0.0211677658\\
2	0.0052278932\\
2.5	0.0010860739\\
3	0.0001588799\\
3.5	1.64e-05\\
};
\addplot [color=red!40!blue, only marks, mark=o, mark options={solid, red!40!blue}, forget plot]
  table[row sep=crcr]{%
0	0.2679083095\\
0.5	0.1358604495\\
1	0.0615832069\\
1.5	0.0198338014\\
2	0.0052162002\\
2.5	0.0010504867\\
3	0.000157147\\
3.5	1.59e-05\\
};
\addplot [color=violet, only marks, mark=x, mark options={solid, violet}, forget plot]
  table[row sep=crcr]{%
0	0.1353179973\\
0.5	0.0645786245\\
1	0.0221043324\\
1.5	0.0066302006\\
2	0.0012806721\\
2.5	0.0001813898\\
3	1.88215e-05\\
3.5	1.4e-06\\
};
\addplot [color=red!60!blue, only marks, mark=+, mark options={solid, red!60!blue}, forget plot]
  table[row sep=crcr]{%
0	0.1520912548\\
0.5	0.0666444518\\
1	0.0232477043\\
1.5	0.0066350396\\
2	0.0012549807\\
2.5	0.0001939347\\
3	2.01189e-05\\
3.5	1.75e-06\\
};
\addplot [color=red!80!blue, only marks, mark=square, mark options={solid, red!80!blue}, forget plot]
  table[row sep=crcr]{%
0	0.147275405\\
0.5	0.0638773555\\
1	0.0248601616\\
1.5	0.0064339714\\
2	0.0011922219\\
2.5	0.0001839245\\
3	1.95064e-05\\
3.5	9.5e-07\\
};
\addplot [color=red!90!blue, only marks, mark=o, mark options={solid, red!90!blue}, forget plot]
  table[row sep=crcr]{%
0	0.1432664756\\
0.5	0.06709158\\
1	0.0252908447\\
1.5	0.0057908909\\
2	0.0012230247\\
2.5	0.0001667439\\
3	1.86968e-05\\
3.5	1.35e-06\\
};
\end{axis}
\end{tikzpicture}%
  \caption{
    List decoding FERs $\varepsilon_{\textrm{MRB}}$ of CA-polar, RM, eBCH, and RBL codes~(markers) with varying list size $\ell_{\textrm{max}}$, comparing with the estimation by random code~(dashed lines), where $(n, k, \delta) = (128, 64, 8)$.
    }
  \label{fig:list-fer}
\end{figure}
In this example, we present in Fig.~\ref{fig:list-fer} the list decoding FERs%
\footnote{In this paper, the list decoding FER is defined as the probability of the sent codeword not presented in the list, denoted $\varepsilon_{\textrm{MRB}}$.}
of four types of codes: CA-polar, RM, eBCH, and RBL codes.
The results show that the list decoding FERs for different codes are well estimated by the proposed approximation~(\ref{equ:mrb-hat-fer}).
\end{example}
}

\rchange{
\textbf{Remarks.} 
It is worth pointing out that the above analysis, starting from the definition~(\ref{equ:def-epsilon-mrb-hat}) of $\widehat{\varepsilon}_{\textrm{MRB}}$, does not depend on any specific code. 
Since we do not have a closed-form expression for the distribution of $D(\bm{r})$, the approximation~(\ref{equ:mrb-hat-fer}) still relies on simulating $\bm{r}$.
However, the proposed analysis approach is very different from the conventional simulations, at least in the sense that first, the simulation does not require decoding, and second, the simulation results do not rely on any specific codes but are applicable to analyzing all specific codes with the same code length and code rate, as confirmed by Example~\ref{ex:list-fer}.
}

Also notice that, $\varepsilon_{\textrm{MRB}}$ and $\widehat{\varepsilon}_{\textrm{MRB}}$ are functions of $\ell_{\textrm{max}}$, denoted as $\varepsilon_{\textrm{MRB}}(\ell_{\textrm{max}})$ and $\widehat{\varepsilon}_{\textrm{MRB}}(\ell_{\textrm{max}})$ respectively.
Then we can write the PMF of $L$~(and $\widehat{L}$) by
\begin{align}
  \label{equ:L-pmf}
  \mathbb{P}[L=\ell] 
  &= \mathbb{P}[L \geq \ell] - \mathbb{P}[L \geq \ell + 1]
  = \varepsilon_{\textrm{MRB}}(\ell) - \varepsilon_{\textrm{MRB}}(\ell + 1), \\
  \label{equ:L-pmf-approx}
  \mathbb{P}[\widehat{L}=\ell] 
  &= \mathbb{P}[\widehat{L} \geq \ell] - \mathbb{P}[\widehat{L} \geq \ell + 1]
  = \widehat{\varepsilon}_{\textrm{MRB}}(\ell) - \widehat{\varepsilon}_{\textrm{MRB}}(\ell + 1).
\end{align}

\begin{example}
\label{ex:l-hist}
\begin{figure}[!t]
  \centering
  \subfloat[The eBCH code {$\mathscr{C}_{\textrm{eBCH}}[128,64]$}.\label{fig:l-hist-snr2-ebch64}]{
%
%
\definecolor{mycolor1}{rgb}{0.00000,0.44700,0.74100}%
\definecolor{mycolor2}{rgb}{0.85000,0.32500,0.09800}%
\begin{tikzpicture}[%
thick,scale=0.7, every node/.style={scale=0.7}
]

\begin{axis}[%
width=10.4cm,
height=7.501cm,
at={(0cm,0cm)},
scale only axis,
xmin=-817.5,
xmax=17167.5,
xlabel style={font=\color{white!15!black}},
xlabel={$L$~(or $\widehat{L}$)},
ymode=log,
ymin=1e-05,
ymax=1,
yminorticks=true,
ylabel style={font=\color{white!15!black}},
ylabel={Probability},
axis background/.style={fill=white},
legend style={legend cell align=left, align=left, draw=white!15!black},
log origin=infty
]
\addplot [color=mycolor1, forget plot]
  table[row sep=crcr]{%
0	0\\
};
\addplot[ybar interval, fill=mycolor2, fill opacity=0.6, draw=black, area legend] table[row sep=crcr] {%
x	y\\
0	0.98442667794495\\
327	0.00566317184081921\\
654	0.00246832588790444\\
981	0.00143521755951737\\
1308	0.000938788882240469\\
1635	0.000698684814449018\\
1962	0.000533075086289277\\
2289	0.000421533830442873\\
2616	0.000358654199678473\\
2943	0.000301181416145407\\
3270	0.000257726384693576\\
3597	0.000222481750889326\\
3924	0.00019244371071525\\
4251	0.000165809981760902\\
4578	0.000154395526494753\\
4905	0.000130365094355492\\
5232	0.000124157232719516\\
5559	0.000114545059863812\\
5886	9.75235037651682e-05\\
6213	9.25171637361555e-05\\
6540	8.85120917129453e-05\\
6867	7.42940860305491e-05\\
7194	6.92877460015363e-05\\
7521	6.76857171922523e-05\\
7848	6.48821667760051e-05\\
8175	5.52699939203007e-05\\
8502	5.36679651110166e-05\\
8829	5.48694867179796e-05\\
9156	4.78606106773618e-05\\
9483	4.48568066599542e-05\\
9810	4.66590890703987e-05\\
10137	4.1652749041386e-05\\
10464	3.62459018100522e-05\\
10791	3.74474234170153e-05\\
11118	3.58453946077312e-05\\
11445	3.60456482088917e-05\\
11772	3.36426049949656e-05\\
12099	3.50443802030892e-05\\
12426	3.30418441914841e-05\\
12753	2.64334753531873e-05\\
13080	3.0238293775237e-05\\
13407	2.60329681508663e-05\\
13734	2.52319537462242e-05\\
14061	2.28289105322981e-05\\
14388	2.20278961276561e-05\\
14715	1.54195272893592e-05\\
15042	2.02256137172115e-05\\
15369	1.56197808905198e-05\\
15696	2.0025360116051e-05\\
16023	2.18276425264956e-05\\
16350	2.18276425264956e-05\\
};
\addlegendentry{Statistic value}

\addplot [color=blue, line width=1.0pt, mark size=0.8pt, mark=*, mark options={solid, blue}]
  table[row sep=crcr]{%
163.5	0.984611907770321\\
490.5	0.005541089090482\\
817.5	0.00244611029226351\\
1144.5	0.00143543452182599\\
1471.5	0.000952967638004049\\
1798.5	0.00068580178009479\\
2125.5	0.000529335829719369\\
2452.5	0.000420791993420226\\
2779.5	0.000342554811885788\\
3106.5	0.000285363823760619\\
3433.5	0.000242131959889222\\
3760.5	0.000208557482804292\\
4087.5	0.000186079638564749\\
4414.5	0.000168081271267166\\
4741.5	0.000149150475061666\\
5068.5	0.000133413379086767\\
5395.5	0.000120173360279095\\
5722.5	0.000108916315587194\\
6049.5	9.92556809569965e-05\\
6376.5	9.08959517250617e-05\\
6703.5	8.36078886268809e-05\\
7030.5	7.72113019562347e-05\\
7357.5	7.15628669738693e-05\\
7684.5	6.65473506635417e-05\\
8011.5	6.20711957621513e-05\\
8338.5	6.22481121800465e-05\\
8665.5	5.84823780938799e-05\\
8992.5	5.48672962215299e-05\\
9319.5	5.15932145997504e-05\\
9646.5	4.8617623073725e-05\\
9973.5	4.5904516998079e-05\\
10300.5	4.34232286825274e-05\\
10627.5	4.11474979223329e-05\\
10954.5	3.90547270788382e-05\\
11281.5	3.71253798226152e-05\\
11608.5	3.53424925984183e-05\\
11935.5	3.36912751937059e-05\\
12262.5	3.21587822344315e-05\\
12589.5	3.07336415120295e-05\\
12916.5	2.94058281304238e-05\\
13243.5	2.81664758126517e-05\\
13570.5	2.70077185113862e-05\\
13897.5	2.59225568630112e-05\\
14224.5	2.4904745110834e-05\\
14551.5	2.39486949735987e-05\\
14878.5	2.30493936057814e-05\\
15205.5	2.22023333272833e-05\\
15532.5	2.14034512234108e-05\\
15859.5	2.06490770550957e-05\\
16186.5	1.9935888192211e-05\\
};
\addlegendentry{Predicted by~(\ref{equ:L-pmf-approx})}

\end{axis}
\end{tikzpicture}%
} 
  \hfill
  \subfloat[The RBL code {$\mathscr{C}_{\textrm{RBL}}[128, 64]$}.\label{fig:l-hist-snr2-random64}]{
%
%
\definecolor{mycolor1}{rgb}{0.00000,0.44700,0.74100}%
\definecolor{mycolor2}{rgb}{0.85000,0.32500,0.09800}%
\begin{tikzpicture}[%
thick,scale=0.7, every node/.style={scale=0.7}
]

\begin{axis}[%
width=10.4cm,
height=7.501cm,
at={(0cm,0cm)},
scale only axis,
xmin=-817.5,
xmax=17167.5,
xlabel style={font=\color{white!15!black}},
xlabel={$L$~(or $\widehat{L}$)},
ymode=log,
ymin=1e-05,
ymax=1,
yminorticks=true,
ylabel style={font=\color{white!15!black}},
ylabel={Probability},
axis background/.style={fill=white},
legend style={legend cell align=left, align=left, draw=white!15!black},
log origin=infty
]
\addplot [color=mycolor1, forget plot]
  table[row sep=crcr]{%
0	0\\
};
\addplot[ybar interval, fill=mycolor2, fill opacity=0.6, draw=black, area legend] table[row sep=crcr] {%
x	y\\
0	0.984437691186999\\
327	0.0056946314940998\\
654	0.00247062416453472\\
981	0.00138961344554644\\
1308	0.000965860447884489\\
1635	0.000695707898807944\\
1962	0.000522481838799633\\
2289	0.000429160053870301\\
2616	0.000362473027300627\\
2943	0.000309604033263409\\
3270	0.000245119881445286\\
3597	0.000201262647755321\\
3924	0.000185442001812365\\
4251	0.000161210379545307\\
4578	0.000152398880539104\\
4905	0.000127766735589945\\
5232	0.00011374844171644\\
5559	0.000111145044282789\\
5886	0.000101732761253436\\
6213	9.3922568952483e-05\\
6540	8.73139446978307e-05\\
6867	7.87027070326777e-05\\
7194	7.48977415527263e-05\\
7521	6.56857198644231e-05\\
7848	6.38833677949724e-05\\
8175	6.10797090202715e-05\\
8502	5.48716074477193e-05\\
8829	5.24684713551184e-05\\
9156	4.72616764878166e-05\\
9483	4.50588017362658e-05\\
9810	4.44580177131156e-05\\
10137	4.42577563720655e-05\\
10464	3.94514841868639e-05\\
10791	4.00522682100141e-05\\
11118	3.20418145680113e-05\\
11445	3.72486094353131e-05\\
11772	3.32433826143117e-05\\
12099	3.1441030544861e-05\\
12426	3.44449506606121e-05\\
12753	2.68350197007094e-05\\
13080	2.843711042911e-05\\
13407	2.66347583596594e-05\\
13734	2.2829792879708e-05\\
14061	2.32303155618082e-05\\
14388	2.26295315386579e-05\\
14715	2.30300542207581e-05\\
15042	2.44318836081086e-05\\
15369	2.32303155618082e-05\\
15696	1.9825872763957e-05\\
16023	1.94253500818568e-05\\
16350	1.94253500818568e-05\\
};
\addlegendentry{Statistic value}

\addplot [color=blue, line width=1.0pt, mark size=0.8pt, mark=*, mark options={solid, blue}]
  table[row sep=crcr]{%
163.5	0.984611907770321\\
490.5	0.005541089090482\\
817.5	0.00244611029226351\\
1144.5	0.00143543452182599\\
1471.5	0.000952967638004049\\
1798.5	0.00068580178009479\\
2125.5	0.000529335829719369\\
2452.5	0.000420791993420226\\
2779.5	0.000342554811885788\\
3106.5	0.000285363823760619\\
3433.5	0.000242131959889222\\
3760.5	0.000208557482804292\\
4087.5	0.000186079638564749\\
4414.5	0.000168081271267166\\
4741.5	0.000149150475061666\\
5068.5	0.000133413379086767\\
5395.5	0.000120173360279095\\
5722.5	0.000108916315587194\\
6049.5	9.92556809569965e-05\\
6376.5	9.08959517250617e-05\\
6703.5	8.36078886268809e-05\\
7030.5	7.72113019562347e-05\\
7357.5	7.15628669738693e-05\\
7684.5	6.65473506635417e-05\\
8011.5	6.20711957621513e-05\\
8338.5	6.22481121800465e-05\\
8665.5	5.84823780938799e-05\\
8992.5	5.48672962215299e-05\\
9319.5	5.15932145997504e-05\\
9646.5	4.8617623073725e-05\\
9973.5	4.5904516998079e-05\\
10300.5	4.34232286825274e-05\\
10627.5	4.11474979223329e-05\\
10954.5	3.90547270788382e-05\\
11281.5	3.71253798226152e-05\\
11608.5	3.53424925984183e-05\\
11935.5	3.36912751937059e-05\\
12262.5	3.21587822344315e-05\\
12589.5	3.07336415120295e-05\\
12916.5	2.94058281304238e-05\\
13243.5	2.81664758126517e-05\\
13570.5	2.70077185113862e-05\\
13897.5	2.59225568630112e-05\\
14224.5	2.4904745110834e-05\\
14551.5	2.39486949735987e-05\\
14878.5	2.30493936057814e-05\\
15205.5	2.22023333272833e-05\\
15532.5	2.14034512234108e-05\\
15859.5	2.06490770550957e-05\\
16186.5	1.9935888192211e-05\\
};
\addlegendentry{Predicted by~(\ref{equ:L-pmf-approx})}

\end{axis}
\end{tikzpicture}%
} 
  \caption{The statistical histogram of $L$, the number of codewords better than the sent codeword, in the case of $\delta=8$ and $E_{\textrm{b}}/N_0 = 2.0~\textrm{dB}$.}
  \label{fig:l-hist-snr2}
\end{figure}
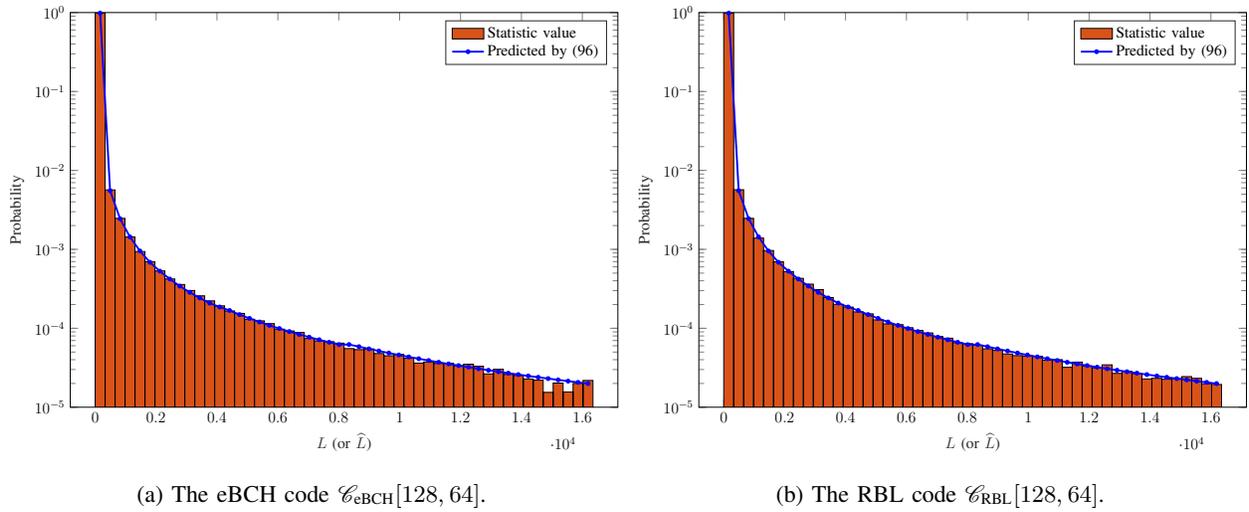
To illustrate the accuracy of the approximation, we use
$\mathbb{P}[L=\ell] \approx \mathbb{P}[\widehat{L}=\ell]$
to estimate the PMF of $L$ and compare it with the statistic value by simulating two different codes, an extended Bose-Chaudhuri-Hocquenghem~(eBCH) code $\mathscr{C}_{\textrm{eBCH}}[128,64]$~(Fig.~\ref{fig:l-hist-snr2-ebch64}) and a random binary linear~(RBL) code $\mathscr{C}_{\textrm{RBL}}[128, 64]$ whose parity-check matrix is randomly constructed~(Fig.~\ref{fig:l-hist-snr2-random64}).
As seen from Fig.~\ref{fig:l-hist-snr2}, the statistic values are quite close to the predicted values from~(\ref{equ:def-epsilon-mrb-hat}) and~(\ref{equ:L-pmf-approx}) and are less relevant to the structure of the code. 
Hence, we have $\mathbb{E}[{L}] \approx \mathbb{E}[\widehat{L}]$ and use $\widehat{\varepsilon}_{\textrm{MRB}}$ to estimate $\varepsilon_{\textrm{MRB}}$ in the following analysis.
\end{example}

Conditional on the transmitted codeword being in the list, we have
\begin{equation}
  \label{equ:cond-L}
  \mathbb{E}[{L}\mid {L} < \ell_{\textrm{max}}]
  \approx \mathbb{E}[\widehat{L}\mid \widehat{L} < \ell_{\textrm{max}}],
\end{equation}
where the RHS can be calculated as
\begin{equation}
  \label{equ:approx-cond-L}
  \mathbb{E}[\widehat{L}\mid \widehat{L} < \ell_{\textrm{max}}]
  = \frac{1}{1 - \widehat{\varepsilon}_{\textrm{MRB}}(\ell_{\textrm{max}})} \Biggl(
    \sum_{\ell = 1}^{\ell_{\textrm{max}} - 1} \widehat{\varepsilon}_{\textrm{MRB}}(\ell)- 
    (\ell_{\textrm{max}} - 1) \cdot \widehat{\varepsilon}_{\textrm{MRB}}(\ell_{\textrm{max}})
    \Biggr).
\end{equation}
\begin{example}
\label{ex:cc-rank64}
We have simulated the left-hand side~(LHS) of~(\ref{equ:cond-L}) of the eBCH code {$\mathscr{C}_{\textrm{eBCH}}[128,64]$} as well as the RBL code {$\mathscr{C}_{\textrm{RBL}}[128, 64]$}. The results are shown in Fig.~\ref{fig:cc-rank64}. Also shown in Fig.~\ref{fig:cc-rank64} are the numerical values calculated by~(\ref{equ:approx-cond-L}). We can see that the simulated values are close to each other and also to the calculated values, validating the assumption on the randomness of ${\mathscr{C}}_{\textrm{MRB}}$.
\begin{figure}[!t]
  \centering
  \subfloat[LC-OSD with {$\ell_{\textrm{max}} = 2^{10}$} and different $\delta$.\label{fig:cc-rank64-lmax10}]{
%
%
\definecolor{mycolor1}{rgb}{0.00000,0.44700,0.74100}%
\definecolor{mycolor2}{rgb}{0.00000,0.77778,0.22222}%
\definecolor{mycolor3}{rgb}{0.00000,0.33333,0.66667}%
\definecolor{mycolor4}{rgb}{0.00000,0.11111,0.88889}%
\definecolor{mycolor5}{rgb}{0.11111,0.00000,0.88889}%
\begin{tikzpicture}[%
thick,scale=0.67, every node/.style={scale=0.67}
]

\begin{axis}[%
width=10.761791299999999cm,
height=8.148999999999999cm,
at={(0cm,0cm)},
scale only axis,
xmin=0,
xmax=3.5,
xlabel style={font=\color{white!15!black}},
xlabel={$E_{\textrm{b}}/N_0~(\textrm{dB})$},
ymode=log,
ymin=0.1,
ymax=200,
yminorticks=true,
ylabel style={font=\color{white!15!black}},
ylabel={$\mathbb{E}[{L}\mid {L} < \ell_{\textrm{max}}]$},
axis background/.style={fill=white},
xmajorgrids,
ymajorgrids,
yminorgrids,
legend style={at={(0.03,0.03)}, anchor=south west, legend cell align=left, align=left, draw=white!15!black}
]
\addplot [color=mycolor1, forget plot]
  table[row sep=crcr]{%
0	0\\
};
\addplot [color=green, dashed, line width=1.0pt]
  table[row sep=crcr]{%
0	118.045131915359\\
0.5	81.9495609411521\\
1	54.8194888706725\\
1.5	29.820351934147\\
2	14.0211874731446\\
2.5	5.39787838366574\\
3	1.7231812715246\\
3.5	0.462386335541267\\
};
\addlegendentry{LC-OSD$(6,2^{10})$ Predicted by~(\ref{equ:approx-cond-L})}

\addplot [color=mycolor2, line width=1.0pt, only marks, mark=+, mark options={solid, mycolor2}]
  table[row sep=crcr]{%
0	122.241566001127\\
0.5	86.7557126897883\\
1	55.6831320729448\\
1.5	30.5807360414036\\
2	14.2785055060835\\
2.5	5.66958580356516\\
3	1.87386296544616\\
3.5	0.549753458087875\\
};
\addlegendentry{LC-OSD$(6,2^{10})$ {$\mathscr{C}_{\textrm{eBCH}}[128,64]$}}

\addplot [color=teal!80!mycolor2, line width=1.0pt, only marks, mark=square, mark options={solid, teal!80!mycolor2}]
  table[row sep=crcr]{%
0	121.645907083111\\
0.5	87.3758560340892\\
1	55.713008284182\\
1.5	30.6328376897793\\
2	14.2017043439492\\
2.5	5.49407892095861\\
3	1.91246259619907\\
3.5	0.568796879687969\\
};
\addlegendentry{LC-OSD$(6,2^{10})$ {$\mathscr{C}_{\textrm{RBL}}[128,64]$}}

\addplot [color=mycolor3, dashed, line width=1.0pt]
  table[row sep=crcr]{%
0	103.415188546203\\
0.5	72.6637787989934\\
1	45.2411008493195\\
1.5	23.282380206892\\
2	9.74708154257229\\
2.5	3.31093625885403\\
3	0.901978899946878\\
3.5	0.2029776429878\\
};
\addlegendentry{LC-OSD$(8,2^{10})$ Predicted by~(\ref{equ:approx-cond-L})}

\addplot [color=mycolor4, line width=1.0pt, only marks, mark=+, mark options={solid, mycolor4}]
  table[row sep=crcr]{%
0	109.95439566906\\
0.5	75.1311834499544\\
1	45.6931910956295\\
1.5	23.5911011273005\\
2	9.84448275166838\\
2.5	3.37773837785858\\
3	0.978072983354674\\
3.5	0.258018061264289\\
};
\addlegendentry{LC-OSD$(8,2^{10})$ {$\mathscr{C}_{\textrm{eBCH}}[128,64]$}}

\addplot [color=mycolor5, line width=1.0pt, only marks, mark=square, mark options={solid, mycolor5}]
  table[row sep=crcr]{%
0	110.2511411491\\
0.5	74.5954101871456\\
1	45.2381247755542\\
1.5	23.2477399004216\\
2	9.77702259824337\\
2.5	3.33270872580063\\
3	1.08578659384409\\
3.5	0.271885437708754\\
};
\addlegendentry{LC-OSD$(8,2^{10})$ {$\mathscr{C}_{\textrm{RBL}}[128,64]$}}

\addplot [color=red!25!mycolor5, dashed, line width=1.0pt]
  table[row sep=crcr]{%
0	100.082293876712\\
0.5	65.4600355473076\\
1	36.8958262403228\\
1.5	17.8256495851042\\
2	6.73210580114989\\
2.5	2.01658335779926\\
3	0.471920883613688\\
3.5	0.0888548431347215\\
};
\addlegendentry{LC-OSD$(10,2^{10})$ Predicted by~(\ref{equ:approx-cond-L})}

\addplot [color=red!50!mycolor5, line width=1.0pt, only marks, mark=+, mark options={solid, red!50!mycolor5}]
  table[row sep=crcr]{%
0	99.6004722517185\\
0.5	65.8200247073357\\
1	37.4725481395995\\
1.5	18.1086476766521\\
2	6.81263005518864\\
2.5	2.05122764170679\\
3	0.463962071655764\\
3.5	0.118223546706401\\
};
\addlegendentry{LC-OSD$(10,2^{10})$ {$\mathscr{C}_{\textrm{eBCH}}[128,64]$}}

\addplot [color=red!75!mycolor5, line width=1.0pt, only marks, mark=square, mark options={solid, red!75!mycolor5}]
  table[row sep=crcr]{%
0	99.0569068122943\\
0.5	64.9073744923783\\
1	37.979183568897\\
1.5	17.9264294037082\\
2	6.81207574492905\\
2.5	2.13622848905307\\
3	0.563914587188078\\
3.5	0.11731\\
};
\addlegendentry{LC-OSD$(10,2^{10})$ {$\mathscr{C}_{\textrm{RBL}}[128,64]$}}

\addplot [color=black, line width=0.1pt, forget plot]
  table[row sep=crcr]{%
2.15	17\\
1.85	17\\
1.85	6\\
2.15	6\\
2.15	17\\
3	80\\
};
\end{axis}

\begin{axis}[%
width=2.8723043cm,
height=3.5903804cm,
at={(7.7122535cm,4.3791006cm)},
scale only axis,
xmin=1.85,
xmax=2.15,
ymode=log,
ymin=6,
ymax=17,
yminorticks=true,
axis background/.style={fill=white},
xmajorgrids,
ymajorgrids,
yminorgrids
]
\addplot [color=mycolor1, forget plot]
  table[row sep=crcr]{%
0	0\\
};
\addplot [color=green, dashed, line width=1.0pt, forget plot]
  table[row sep=crcr]{%
0	118.045131915359\\
0.5	81.9495609411521\\
1	54.8194888706725\\
1.5	29.820351934147\\
2	14.0211874731446\\
2.5	5.39787838366574\\
3	1.7231812715246\\
3.5	0.462386335541267\\
};
\addplot [color=mycolor2, line width=1.0pt, only marks, mark=+, mark options={solid, mycolor2}, forget plot]
  table[row sep=crcr]{%
0	122.241566001127\\
0.5	86.7557126897883\\
1	55.6831320729448\\
1.5	30.5807360414036\\
2	14.2785055060835\\
2.5	5.66958580356516\\
3	1.87386296544616\\
3.5	0.549753458087875\\
};
\addplot [color=green!50!mycolor4, line width=1.0pt, only marks, mark=square, mark options={solid, green!50!mycolor4}, forget plot]
  table[row sep=crcr]{%
0	121.645907083111\\
0.5	87.3758560340892\\
1	55.713008284182\\
1.5	30.6328376897793\\
2	14.2017043439492\\
2.5	5.49407892095861\\
3	1.91246259619907\\
3.5	0.568796879687969\\
};
\addplot [color=mycolor3, dashed, line width=1.0pt, forget plot]
  table[row sep=crcr]{%
0	103.415188546203\\
0.5	72.6637787989934\\
1	45.2411008493195\\
1.5	23.282380206892\\
2	9.74708154257229\\
2.5	3.31093625885403\\
3	0.901978899946878\\
3.5	0.2029776429878\\
};
\addplot [color=mycolor4, line width=1.0pt, only marks, mark=+, mark options={solid, mycolor4}, forget plot]
  table[row sep=crcr]{%
0	109.95439566906\\
0.5	75.1311834499544\\
1	45.6931910956295\\
1.5	23.5911011273005\\
2	9.84448275166838\\
2.5	3.37773837785858\\
3	0.978072983354674\\
3.5	0.258018061264289\\
};
\addplot [color=mycolor5, line width=1.0pt, only marks, mark=square, mark options={solid, mycolor5}, forget plot]
  table[row sep=crcr]{%
0	110.2511411491\\
0.5	74.5954101871456\\
1	45.2381247755542\\
1.5	23.2477399004216\\
2	9.77702259824337\\
2.5	3.33270872580063\\
3	1.08578659384409\\
3.5	0.271885437708754\\
};
\addplot [color=red!25!mycolor5, dashed, line width=1.0pt, forget plot]
  table[row sep=crcr]{%
0	100.082293876712\\
0.5	65.4600355473076\\
1	36.8958262403228\\
1.5	17.8256495851042\\
2	6.73210580114989\\
2.5	2.01658335779926\\
3	0.471920883613688\\
3.5	0.0888548431347215\\
};
\addplot [color=red!50!mycolor5, line width=1.0pt, only marks, mark=+, mark options={solid, red!50!mycolor5}, forget plot]
  table[row sep=crcr]{%
0	99.6004722517185\\
0.5	65.8200247073357\\
1	37.4725481395995\\
1.5	18.1086476766521\\
2	6.81263005518864\\
2.5	2.05122764170679\\
3	0.463962071655764\\
3.5	0.118223546706401\\
};
\addplot [color=red!75!mycolor5, line width=1.0pt, only marks, mark=square, mark options={solid, red!75!mycolor5}, forget plot]
  table[row sep=crcr]{%
0	99.0569068122943\\
0.5	64.9073744923783\\
1	37.979183568897\\
1.5	17.9264294037082\\
2	6.81207574492905\\
2.5	2.13622848905307\\
3	0.563914587188078\\
3.5	0.11731\\
};
\end{axis}
\end{tikzpicture}%
} 
  \hfill
  \subfloat[LC-OSD with {$\ell_{\textrm{max}} = 2^{14}$} and different $\delta$.\label{fig:cc-rank64-lmax14}]{
%
%
\definecolor{mycolor1}{rgb}{0.00000,0.44700,0.74100}%
\definecolor{mycolor2}{rgb}{0.00000,0.77778,0.22222}%
\definecolor{mycolor3}{rgb}{0.00000,0.33333,0.66667}%
\definecolor{mycolor4}{rgb}{0.00000,0.11111,0.88889}%
\definecolor{mycolor5}{rgb}{0.11111,0.00000,0.88889}%
\begin{tikzpicture}[%
thick,scale=0.67, every node/.style={scale=0.67}
]

\begin{axis}[%
width=10.761791299999999cm,
height=8.148999999999999cm,
at={(0cm,0cm)},
scale only axis,
xmin=0,
xmax=3.5,
xlabel style={font=\color{white!15!black}},
xlabel={$E_{\textrm{b}}/N_0~(\textrm{dB})$},
ymode=log,
ymin=0.1,
ymax=2000,
yminorticks=true,
ylabel style={font=\color{white!15!black}},
ylabel={$\mathbb{E}[{L}\mid {L} < \ell_{\textrm{max}}]$},
axis background/.style={fill=white},
xmajorgrids,
ymajorgrids,
yminorgrids,
legend style={at={(0.03,0.03)}, anchor=south west, legend cell align=left, align=left, draw=white!15!black}
]
\addplot [color=mycolor1, forget plot]
  table[row sep=crcr]{%
0	0\\
};
\addplot [color=green, dashed, line width=1.0pt]
  table[row sep=crcr]{%
0	1213.41391165016\\
0.5	745.895492180879\\
1	384.665475834962\\
1.5	163.640710864601\\
2	57.3247979790046\\
2.5	16.0044744261272\\
3	3.60489679022765\\
3.5	0.718103504557454\\
};
\addlegendentry{LC-OSD$(6,2^{14})$ Predicted by~(\ref{equ:approx-cond-L})}

\addplot [color=mycolor2, line width=1.0pt, only marks, mark=+, mark options={solid, mycolor2}]
  table[row sep=crcr]{%
0	1209.99897065131\\
0.5	746.572284414903\\
1	388.732858663415\\
1.5	168.344800088827\\
2	56.2524321453977\\
2.5	15.2754291544956\\
3	3.92984579074744\\
3.5	1.10266102661027\\
};
\addlegendentry{LC-OSD$(6,2^{14})$ {$\mathscr{C}_{\textrm{eBCH}}[128,64]$}}

\addplot [color=teal!80!mycolor2, line width=1.0pt, only marks, mark=square, mark options={solid, teal!80!mycolor2}]
  table[row sep=crcr]{%
0	1214.46760528893\\
0.5	756.325721403363\\
1	390.835112043226\\
1.5	164.598677034942\\
2	56.5923679884528\\
2.5	16.7624768669034\\
3	4.24558245582456\\
3.5	0.77308\\
};
\addlegendentry{LC-OSD$(6,2^{14})$ {$\mathscr{C}_{\textrm{RBL}}[128,64]$}}

\addplot [color=mycolor3, dashed, line width=1.0pt]
  table[row sep=crcr]{%
0	1075.58165611681\\
0.5	602.180901539278\\
1	301.732134639994\\
1.5	117.192594928301\\
2	37.3965322342512\\
2.5	9.23366352690446\\
3	1.75866184259488\\
3.5	0.299906565834333\\
};
\addlegendentry{LC-OSD$(8,2^{14})$ Predicted by~(\ref{equ:approx-cond-L})}

\addplot [color=mycolor4, line width=1.0pt, only marks, mark=+, mark options={solid, mycolor4}]
  table[row sep=crcr]{%
0	1066.76392453189\\
0.5	618.461484706768\\
1	308.177312448137\\
1.5	121.048208534622\\
2	35.924751959793\\
2.5	8.4283842445338\\
3	1.92229689187568\\
3.5	0.53736\\
};
\addlegendentry{LC-OSD$(8,2^{14})$ {$\mathscr{C}_{\textrm{eBCH}}[128,64]$}}

\addplot [color=mycolor5, line width=1.0pt, only marks, mark=square, mark options={solid, mycolor5}]
  table[row sep=crcr]{%
0	1072.81725562838\\
0.5	628.838709677419\\
1	300.170221926093\\
1.5	121.678850780676\\
2	37.5521930702984\\
2.5	9.35222396703308\\
3	1.97004\\
3.5	0.30258\\
};
\addlegendentry{LC-OSD$(8,2^{14})$ {$\mathscr{C}_{\textrm{RBL}}[128,64]$}}

\addplot [color=red!25!mycolor5, dashed, line width=1.0pt]
  table[row sep=crcr]{%
0	989.141266382724\\
0.5	524.435298612367\\
1	242.854745878073\\
1.5	84.8382035810087\\
2	24.5192693210436\\
2.5	5.26543262928545\\
3	0.888617855857663\\
3.5	0.127382243608095\\
};
\addlegendentry{LC-OSD$(10,2^{14})$ Predicted by~(\ref{equ:approx-cond-L})}

\addplot [color=red!50!mycolor5, line width=1.0pt, only marks, mark=+, mark options={solid, red!50!mycolor5}]
  table[row sep=crcr]{%
0	946.787642061894\\
0.5	532.322536868778\\
1	237.929872458335\\
1.5	89.5104477312207\\
2	22.5844150645035\\
2.5	4.78631222058868\\
3	1.18212182121821\\
3.5	0.19296\\
};
\addlegendentry{LC-OSD$(10,2^{14})$ {$\mathscr{C}_{\textrm{eBCH}}[128,64]$}}

\addplot [color=red!75!mycolor5, line width=1.0pt, only marks, mark=square, mark options={solid, red!75!mycolor5}]
  table[row sep=crcr]{%
0	951.253143743776\\
0.5	533.576773154817\\
1	236.206191136299\\
1.5	87.3608700020092\\
2	26.3642623377403\\
2.5	5.70937384364593\\
3	1.01975\\
3.5	0.11731\\
};
\addlegendentry{LC-OSD$(10,2^{14})$ {$\mathscr{C}_{\textrm{RBL}}[128,64]$}}

\addplot [color=black, line width=0.1pt, forget plot]
  table[row sep=crcr]{%
2.15	70\\
1.85	70\\
1.85	20\\
2.15	20\\
2.15	70\\
3	120\\
};
\end{axis}

\begin{axis}[%
width=2.8723043cm,
height=3.5903804cm,
at={(7.7122535cm,4.3791006cm)},
scale only axis,
xmin=1.85,
xmax=2.15,
ymode=log,
ymin=20,
ymax=70,
yminorticks=true,
axis background/.style={fill=white},
xmajorgrids,
ymajorgrids,
yminorgrids
]
\addplot [color=mycolor1, forget plot]
  table[row sep=crcr]{%
0	0\\
};
\addplot [color=green, dashed, line width=1.0pt, forget plot]
  table[row sep=crcr]{%
0	1213.41391165016\\
0.5	745.895492180879\\
1	384.665475834962\\
1.5	163.640710864601\\
2	57.3247979790046\\
2.5	16.0044744261272\\
3	3.60489679022765\\
3.5	0.718103504557454\\
};
\addplot [color=mycolor2, line width=1.0pt, only marks, mark=+, mark options={solid, mycolor2}, forget plot]
  table[row sep=crcr]{%
0	1209.99897065131\\
0.5	746.572284414903\\
1	388.732858663415\\
1.5	168.344800088827\\
2	56.2524321453977\\
2.5	15.2754291544956\\
3	3.92984579074744\\
3.5	1.10266102661027\\
};
\addplot [color=green!50!mycolor4, line width=1.0pt, only marks, mark=square, mark options={solid, green!50!mycolor4}, forget plot]
  table[row sep=crcr]{%
0	1214.46760528893\\
0.5	756.325721403363\\
1	390.835112043226\\
1.5	164.598677034942\\
2	56.5923679884528\\
2.5	16.7624768669034\\
3	4.24558245582456\\
3.5	0.77308\\
};
\addplot [color=mycolor3, dashed, line width=1.0pt, forget plot]
  table[row sep=crcr]{%
0	1075.58165611681\\
0.5	602.180901539278\\
1	301.732134639994\\
1.5	117.192594928301\\
2	37.3965322342512\\
2.5	9.23366352690446\\
3	1.75866184259488\\
3.5	0.299906565834333\\
};
\addplot [color=mycolor4, line width=1.0pt, only marks, mark=+, mark options={solid, mycolor4}, forget plot]
  table[row sep=crcr]{%
0	1066.76392453189\\
0.5	618.461484706768\\
1	308.177312448137\\
1.5	121.048208534622\\
2	35.924751959793\\
2.5	8.4283842445338\\
3	1.92229689187568\\
3.5	0.53736\\
};
\addplot [color=mycolor5, line width=1.0pt, only marks, mark=square, mark options={solid, mycolor5}, forget plot]
  table[row sep=crcr]{%
0	1072.81725562838\\
0.5	628.838709677419\\
1	300.170221926093\\
1.5	121.678850780676\\
2	37.5521930702984\\
2.5	9.35222396703308\\
3	1.97004\\
3.5	0.30258\\
};
\addplot [color=red!25!mycolor5, dashed, line width=1.0pt, forget plot]
  table[row sep=crcr]{%
0	989.141266382724\\
0.5	524.435298612367\\
1	242.854745878073\\
1.5	84.8382035810087\\
2	24.5192693210436\\
2.5	5.26543262928545\\
3	0.888617855857663\\
3.5	0.127382243608095\\
};
\addplot [color=red!50!mycolor5, line width=1.0pt, only marks, mark=+, mark options={solid, red!50!mycolor5}, forget plot]
  table[row sep=crcr]{%
0	946.787642061894\\
0.5	532.322536868778\\
1	237.929872458335\\
1.5	89.5104477312207\\
2	22.5844150645035\\
2.5	4.78631222058868\\
3	1.18212182121821\\
3.5	0.19296\\
};
\addplot [color=red!75!mycolor5, line width=1.0pt, only marks, mark=square, mark options={solid, red!75!mycolor5}, forget plot]
  table[row sep=crcr]{%
0	951.253143743776\\
0.5	533.576773154817\\
1	236.206191136299\\
1.5	87.3608700020092\\
2	26.3642623377403\\
2.5	5.70937384364593\\
3	1.01975\\
3.5	0.11731\\
};
\end{axis}
\end{tikzpicture}%
} 
  \caption{Conditional expectation of $L$ of the eBCH code {$\mathscr{C}_{\textrm{eBCH}}[128,64]$} and the RBL code {$\mathscr{C}_{\textrm{RBL}}[128, 64]$} along with their predictions.}
  \label{fig:cc-rank64}
\end{figure}
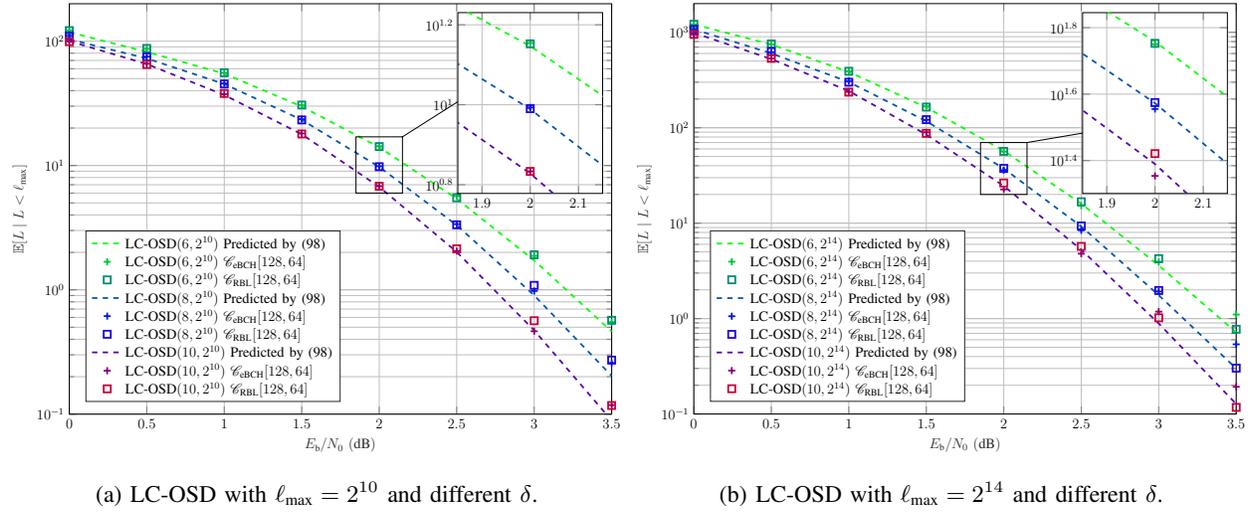
\end{example}

\subsubsection{Performance Bound}
\label{sec:performance-bound}
Denote by $\varepsilon$ and ${\varepsilon}_{\textrm{ML}}$ the FER of the LC-OSD and MLD, respectively.
Then we have
\begin{align}
  \varepsilon 
  & = \mathbb{P}[\{\textrm{${\bm{e}}$ is not in the list}\}] + \mathbb{P}[\{\textrm{${\bm{e}}$ is in the list but is not the most likely one}\}]\\
  & \leq \mathbb{P}[\{\textrm{${\bm{e}}$ is not in the list}\}] + \mathbb{P}[\{\textrm{${\bm{e}}$ is not the most likely one}\}]\\
  &= \varepsilon_{\textrm{MRB}} + {\varepsilon}_{\textrm{ML}}
\end{align}
Hence, the performance gap between the LC-OSD and the MLD can be approximately upper bounded by $\widehat{\varepsilon}_{\textrm{MRB}}$ as
\begin{equation}
  \label{equ:approx-bound}
  \varepsilon - {\varepsilon}_{\textrm{ML}} \leq \varepsilon_{\textrm{MRB}} \approx \widehat{\varepsilon}_{\textrm{MRB}}
\end{equation} 
which does not depend on the structure of the code.

\begin{example}
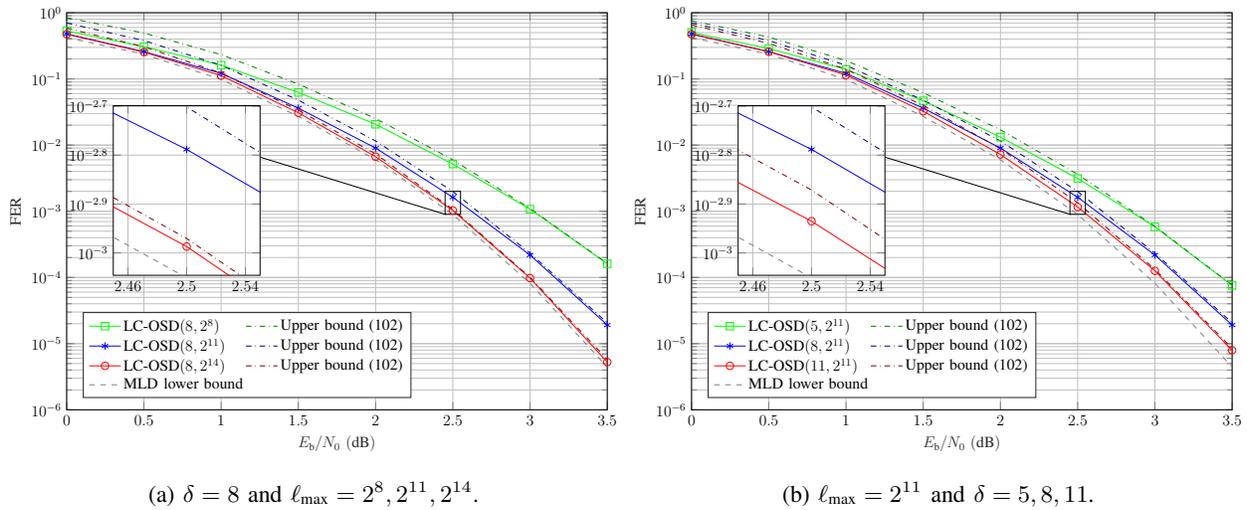
\begin{figure}[!t]
  \centering
  \subfloat[$\delta=8$ and $\ell_{\textrm{max}} = 2^8, 2^{11}, 2^{14}$.\label{fig:bch-lmax-fer}]{
%
%
\begin{tikzpicture}[%
thick,scale=0.7, every node/.style={scale=0.7}
]

\begin{axis}[%
width=10.2666667cm,
height=7.547142cm,
at={(0cm,0cm)},
scale only axis,
xmin=0,
xmax=3.5,
xlabel style={font=\color{white!15!black}},
xlabel={$E_{\textrm{b}}/N_0~(\textrm{dB})$},
ymode=log,
ymin=1e-06,
ymax=1,
yminorticks=true,
ylabel style={font=\color{white!15!black}},
ylabel={FER},
axis background/.style={fill=white},
xmajorgrids,
ymajorgrids,
yminorgrids,
legend style={at={(0.03,0.03)}, anchor=south west, legend cell align=left, align=left, draw=white!15!black},
legend columns=2
]
\addplot [color=green, mark=square, mark options={solid, green}]
  table[row sep=crcr]{%
0	0.5308056872\\
0.5	0.3062261753\\
1	0.1615255188\\
1.5	0.0624903802\\
2	0.0206099476\\
2.5	0.00517729\\
3	0.0010765152\\
3.5	0.0001610522\\
};
\addlegendentry{LC-OSD$(8, 2^{8})$}

\addplot [color=black!50!green, dashdotted]
  table[row sep=crcr]{%
0	0.8425878507\\
0.5	0.4937274808\\
1	0.232696298\\
1.5	0.0834441913\\
2	0.0252137344\\
2.5	0.005992905\\
3	0.001104644\\
3.5	0.0001628241\\
};
\addlegendentry{Upper bound~(\ref{equ:approx-bound})}

\addplot [color=blue, mark=asterisk, mark options={solid, blue}]
  table[row sep=crcr]{%
0	0.4786729858\\
0.5	0.2579415502\\
1	0.1211441391\\
1.5	0.0361705402\\
2	0.0090189398\\
2.5	0.0016278842\\
3	0.0002180408\\
3.5	1.91177e-05\\
};
\addlegendentry{LC-OSD$(8, 2^{11})$}

\addplot [color=black!50!blue, dashdotted]
  table[row sep=crcr]{%
0	0.6994544355\\
0.5	0.3797205493\\
1	0.160542889\\
1.5	0.0477885594\\
2	0.0115389267\\
2.5	0.0019885173\\
3	0.0002329554\\
3.5	2.06523e-05\\
};
\addlegendentry{Upper bound~(\ref{equ:approx-bound})}

\addplot [color=red, mark=o, mark options={solid, red}]
  table[row sep=crcr]{%
0	0.4739336493\\
0.5	0.2541296061\\
1	0.1121704992\\
1.5	0.0307834385\\
2	0.0066806961\\
2.5	0.0010303065\\
3	9.77761e-05\\
3.5	5.2666e-06\\
};
\addlegendentry{LC-OSD$(8, 2^{14})$}

\addplot [color=black!50!red, dashdotted]
  table[row sep=crcr]{%
0	0.5823785227\\
0.5	0.3068944059\\
1	0.1227660071\\
1.5	0.0337064648\\
2	0.0073661898\\
2.5	0.0010702053\\
3	0.0001005263\\
3.5	5.7596e-06\\
};
\addlegendentry{Upper bound~(\ref{equ:approx-bound})}

\addplot [color=gray, dashed]
  table[row sep=crcr]{%
0	0.4219409283\\
0.5	0.2361275089\\
1	0.0984251969\\
1.5	0.0271223217\\
2	0.0060134099\\
2.5	0.0008895729\\
3	8.17035e-05\\
3.5	4.4325e-06\\
};
\addlegendentry{MLD lower bound}

\addplot [color=black, line width=0.1pt, forget plot]
  table[row sep=crcr]{%
2.45	0.0009\\
2.55	0.0009\\
2.55	0.002\\
2.45	0.002\\
2.45	0.0009\\
1	0.01\\
};
\end{axis}

\begin{axis}[%
width=2.7877433cm,
height=3.2233282cm,
at={(0.8854588cm,2.5563532cm)},
scale only axis,
xmin=2.45,
xmax=2.55,
xtick={2.46,  2.5, 2.54},
ymode=log,
ymin=0.0009,
ymax=0.002,
ytick={    0.001, 0.0012589, 0.0015849, 0.0019953},
yminorticks=true,
axis background/.style={fill=white},
xmajorgrids,
ymajorgrids,
yminorgrids,
legend columns=2
]
\addplot [color=green, mark=square, mark options={solid, green}, forget plot]
  table[row sep=crcr]{%
0	0.5308056872\\
0.5	0.3062261753\\
1	0.1615255188\\
1.5	0.0624903802\\
2	0.0206099476\\
2.5	0.00517729\\
3	0.0010765152\\
3.5	0.0001610522\\
};
\addplot [color=black!50!green, dashdotted, forget plot]
  table[row sep=crcr]{%
0	0.8425878507\\
0.5	0.4937274808\\
1	0.232696298\\
1.5	0.0834441913\\
2	0.0252137344\\
2.5	0.005992905\\
3	0.001104644\\
3.5	0.0001628241\\
};
\addplot [color=blue, mark=asterisk, mark options={solid, blue}, forget plot]
  table[row sep=crcr]{%
0	0.4786729858\\
0.5	0.2579415502\\
1	0.1211441391\\
1.5	0.0361705402\\
2	0.0090189398\\
2.5	0.0016278842\\
3	0.0002180408\\
3.5	1.91177e-05\\
};
\addplot [color=black!50!blue, dashdotted, forget plot]
  table[row sep=crcr]{%
0	0.6994544355\\
0.5	0.3797205493\\
1	0.160542889\\
1.5	0.0477885594\\
2	0.0115389267\\
2.5	0.0019885173\\
3	0.0002329554\\
3.5	2.06523e-05\\
};
\addplot [color=red, mark=o, mark options={solid, red}, forget plot]
  table[row sep=crcr]{%
0	0.4739336493\\
0.5	0.2541296061\\
1	0.1121704992\\
1.5	0.0307834385\\
2	0.0066806961\\
2.5	0.0010303065\\
3	9.77761e-05\\
3.5	5.2666e-06\\
};
\addplot [color=black!50!red, dashdotted, forget plot]
  table[row sep=crcr]{%
0	0.5823785227\\
0.5	0.3068944059\\
1	0.1227660071\\
1.5	0.0337064648\\
2	0.0073661898\\
2.5	0.0010702053\\
3	0.0001005263\\
3.5	5.7596e-06\\
};
\addplot [color=gray, dashed, forget plot]
  table[row sep=crcr]{%
0	0.4219409283\\
0.5	0.2361275089\\
1	0.0984251969\\
1.5	0.0271223217\\
2	0.0060134099\\
2.5	0.0008895729\\
3	8.17035e-05\\
3.5	4.4325e-06\\
};
\end{axis}
\end{tikzpicture}%
} 
  \hfill
  \subfloat[$\ell_{\textrm{max}} = 2^{11}$ and $\delta = 5, 8, 11$.\label{fig:bch-delta-fer}]{
%
%
\begin{tikzpicture}[%
thick,scale=0.7, every node/.style={scale=0.7}
]

\begin{axis}[%
width=10.2666667cm,
height=7.547142cm,
at={(0cm,0cm)},
scale only axis,
xmin=0,
xmax=3.5,
xlabel style={font=\color{white!15!black}},
xlabel={$E_{\textrm{b}}/N_0~(\textrm{dB})$},
ymode=log,
ymin=1e-06,
ymax=1,
yminorticks=true,
ylabel style={font=\color{white!15!black}},
ylabel={FER},
axis background/.style={fill=white},
xmajorgrids,
ymajorgrids,
yminorgrids,
legend style={at={(0.03,0.03)}, anchor=south west, legend cell align=left, align=left, draw=white!15!black},
legend columns=2
]
\addplot [color=green, mark=square, mark options={solid, green}]
  table[row sep=crcr]{%
0	0.503649635\\
0.5	0.2881578947\\
1	0.1391549296\\
1.5	0.0467380721\\
2	0.0132085856\\
2.5	0.0031433452\\
3	0.0005812256\\
3.5	7.55845e-05\\
};
\addlegendentry{LC-OSD$(5, 2^{11})$}

\addplot [color=black!50!green, dashdotted]
  table[row sep=crcr]{%
0	0.7563140271\\
0.5	0.4291398558\\
1	0.1888281541\\
1.5	0.0618604064\\
2	0.0170120196\\
2.5	0.0036194211\\
3	0.0005921316\\
3.5	7.66593e-05\\
};
\addlegendentry{Upper bound~(\ref{equ:approx-bound})}

\addplot [color=blue, mark=asterisk, mark options={solid, blue}]
  table[row sep=crcr]{%
0	0.4786729858\\
0.5	0.2579415502\\
1	0.1211441391\\
1.5	0.0361705402\\
2	0.0090189398\\
2.5	0.0016278842\\
3	0.0002180408\\
3.5	1.91177e-05\\
};
\addlegendentry{LC-OSD$(8, 2^{11})$}

\addplot [color=black!50!blue, dashdotted]
  table[row sep=crcr]{%
0	0.6994544355\\
0.5	0.3797205493\\
1	0.160542889\\
1.5	0.0477885594\\
2	0.0115389267\\
2.5	0.0019885173\\
3	0.0002329554\\
3.5	2.06523e-05\\
};
\addlegendentry{Upper bound~(\ref{equ:approx-bound})}

\addplot [color=red, mark=o, mark options={solid, red}]
  table[row sep=crcr]{%
0	0.4739336493\\
0.5	0.2577319588\\
1	0.1144139091\\
1.5	0.0324088501\\
2	0.0072151518\\
2.5	0.0011610787\\
3	0.0001259448\\
3.5	7.94e-06\\
};
\addlegendentry{LC-OSD$(11, 2^{11})$}

\addplot [color=black!50!red, dashdotted]
  table[row sep=crcr]{%
0	0.6551801483\\
0.5	0.3449539968\\
1	0.1400509439\\
1.5	0.0397003747\\
2	0.0088503581\\
2.5	0.0013446125\\
3	0.000132877\\
3.5	8.4921e-06\\
};
\addlegendentry{Upper bound~(\ref{equ:approx-bound})}

\addplot [color=gray, dashed]
  table[row sep=crcr]{%
0	0.4219409283\\
0.5	0.2361275089\\
1	0.0984251969\\
1.5	0.0271223217\\
2	0.0060134099\\
2.5	0.0008895729\\
3	8.17035e-05\\
3.5	4.4325e-06\\
};
\addlegendentry{MLD lower bound}

\addplot [color=black, line width=0.1pt, forget plot]
  table[row sep=crcr]{%
2.45	0.0009\\
2.55	0.0009\\
2.55	0.002\\
2.45	0.002\\
2.45	0.0009\\
1	0.01\\
};
\end{axis}

\begin{axis}[%
width=2.7877433cm,
height=3.2233282cm,
at={(0.8854588cm,2.5563532cm)},
scale only axis,
xmin=2.45,
xmax=2.55,
xtick={2.46,  2.5, 2.54},
ymode=log,
ymin=0.0009,
ymax=0.002,
ytick={    0.001, 0.0012589, 0.0015849, 0.0019953},
yminorticks=true,
axis background/.style={fill=white},
xmajorgrids,
ymajorgrids,
yminorgrids,
legend columns=2
]
\addplot [color=green, mark=square, mark options={solid, green}, forget plot]
  table[row sep=crcr]{%
0	0.503649635\\
0.5	0.2881578947\\
1	0.1391549296\\
1.5	0.0467380721\\
2	0.0132085856\\
2.5	0.0031433452\\
3	0.0005812256\\
3.5	7.55845e-05\\
};
\addplot [color=black!50!green, dashdotted, forget plot]
  table[row sep=crcr]{%
0	0.7563140271\\
0.5	0.4291398558\\
1	0.1888281541\\
1.5	0.0618604064\\
2	0.0170120196\\
2.5	0.0036194211\\
3	0.0005921316\\
3.5	7.66593e-05\\
};
\addplot [color=blue, mark=asterisk, mark options={solid, blue}, forget plot]
  table[row sep=crcr]{%
0	0.4786729858\\
0.5	0.2579415502\\
1	0.1211441391\\
1.5	0.0361705402\\
2	0.0090189398\\
2.5	0.0016278842\\
3	0.0002180408\\
3.5	1.91177e-05\\
};
\addplot [color=black!50!blue, dashdotted, forget plot]
  table[row sep=crcr]{%
0	0.6994544355\\
0.5	0.3797205493\\
1	0.160542889\\
1.5	0.0477885594\\
2	0.0115389267\\
2.5	0.0019885173\\
3	0.0002329554\\
3.5	2.06523e-05\\
};
\addplot [color=red, mark=o, mark options={solid, red}, forget plot]
  table[row sep=crcr]{%
0	0.4739336493\\
0.5	0.2577319588\\
1	0.1144139091\\
1.5	0.0324088501\\
2	0.0072151518\\
2.5	0.0011610787\\
3	0.0001259448\\
3.5	7.94e-06\\
};
\addplot [color=black!50!red, dashdotted, forget plot]
  table[row sep=crcr]{%
0	0.6551801483\\
0.5	0.3449539968\\
1	0.1400509439\\
1.5	0.0397003747\\
2	0.0088503581\\
2.5	0.0013446125\\
3	0.000132877\\
3.5	8.4921e-06\\
};
\addplot [color=gray, dashed, forget plot]
  table[row sep=crcr]{%
0	0.4219409283\\
0.5	0.2361275089\\
1	0.0984251969\\
1.5	0.0271223217\\
2	0.0060134099\\
2.5	0.0008895729\\
3	8.17035e-05\\
3.5	4.4325e-06\\
};
\end{axis}
\end{tikzpicture}%
} 
  \caption{Performance and approximate upper bound of the eBCH code {$\mathscr{C}_{\textrm{eBCH}}[128,64]$}.}
  \label{fig:bch-fer}
\end{figure}
Fig.~\ref{fig:bch-fer} shows the performance of the RBL code {$\mathscr{C}_{\textrm{eBCH}}[128,64]$}. 
We also plotted the approximate upper bound~(\ref{equ:approx-bound}) of {$\mathscr{C}_{\textrm{eBCH}}[128,64]$}. 
\rchange{The values of $\widehat{\varepsilon}_{\textrm{MRB}}$ in this example are calculated by the saddlepoint approximation discussed in Example~\ref{ex:two-techs-Dr}.}
As shown in Fig.~\ref{fig:bch-fer} we can observe that the performance of {$\mathscr{C}_{\textrm{eBCH}}[128,64]$} improves as $\ell_{\textrm{max}}$ or $\delta$ increases. 
Additionally, we note that the approximate upper bounds are highly accurate due to the accuracy of~(\ref{equ:def-epsilon-mrb-hat}), which has already been verified in Example~\ref{ex:l-hist} and Example~\ref{ex:cc-rank64}.
\end{example}

\subsection{Decoding Parameters Tuning}
The tight performance upper bound deduced in Sec.~\ref{sec:performance-bound} for the LC-OSD becomes a helpful tool in many situations.
In this subsection, we will give examples that explain how we could tune the decoding parameters $(\delta, \ell_{\textrm{max}})$ to satisfy the diversified requirements depending on application scenarios.

To evaluate the time consumption of the LC-OSD, we need to write out the time factor~(in the case when the SLVA~(see Sec.~\ref{sec:lva} for details) is taken as the LGA)~(\ref{equ:Tavg}) and~(\ref{equ:Tmax}), namely, 
\begin{align}
  T_{\textrm{avg}}^{\textrm{tot}} &\approx \rho_1\cdot n(n-k)(n-k-\delta) + \rho_2\cdot 2^\delta(k+\delta) + \rho_3\cdot {\ell}_{\textrm{avg}}(n-k-\delta)(k+\delta), \\
  T_{\textrm{max}}^{\textrm{tot}} &\approx \rho_1\cdot n(n-k)(n-k-\delta) + \rho_2\cdot 2^\delta(k+\delta) + \rho_3\cdot {\ell}_{\textrm{max}}(n-k-\delta)(k+\delta),
\end{align}
where $\rho_i~(i=1,2,3)$ are called the time factor constants.
We can measure these time factor constants by implementing the LC-OSD on a specific computer~(or device). 
\begin{example}
Table~\ref{tab:time-factor} shows the values of these time factor constants for reference%
\footnote{We have measured these values on a desktop computer with a CPU of 11th Gen Intel(R) Core(TM) i7-11700F @ 2.50GHz.}.
\begin{table}[H]
  \centering
  \caption{Reference values of time factors.}
  \label{tab:time-factor}
  \begin{tabular}{c|ccc}
    \hline
    Time factor constants & $\rho_1$ & $\rho_2$ & $\rho_3$ \\
    \hline
    Measured values (ns) & $0.0816$ & $26.4$ & $0.728$ \\
    \hline
  \end{tabular}
\end{table}
\end{example}
Then, we may tune the decoding parameters as required by applications. 

\subsubsection{Minimum Average Time Complexity}
In the scenarios where the power consumption is concerned, we need to optimize $(\delta, \ell_{\textrm{max}})$ to minimize the average time complexity $T_{\textrm{avg}}^{\textrm{tot}}$.

\begin{example}
\begin{figure}[!t]
  \centering
  \subfloat[The average number of searches $\ell_{\textrm{avg}}$.\label{fig:delta-lavg}]{
%
%
\definecolor{mycolor1}{rgb}{0.00000,0.14286,0.85714}%
\definecolor{mycolor2}{rgb}{0.71429,0.00000,0.28571}%
\definecolor{mycolor3}{rgb}{0.00000,0.71429,0.28571}%
\begin{tikzpicture}[%
thick,scale=0.7, every node/.style={scale=0.7}
]

\begin{axis}[%
width=10.267cm,
height=7.695cm,
at={(0cm,0cm)},
scale only axis,
xmin=4,
xmax=11,
xlabel style={font=\color{white!15!black}},
xlabel={$\delta$},
ymode=log,
ymin=1,
ymax=8192,
yminorticks=true,
ylabel style={font=\color{white!15!black}},
ylabel={$\ell_{\textrm{avg}}$},
axis background/.style={fill=white},
xmajorgrids,
ymajorgrids,
yminorgrids,
legend style={legend cell align=left, align=left, draw=white!15!black},
legend columns=2
]
\addplot [color=red, mark=o, mark options={solid, red}]
  table[row sep=crcr]{%
4	22.63\\
5	15.06\\
6	10.23\\
7	7.08\\
8	5.07\\
9	3.73\\
10	2.84\\
11	2.25\\
};
\addlegendentry{$E_{\textrm{b}}/N_0 = 3.5~\textrm{dB}$}

\addplot [color=mycolor1, mark=o, mark options={solid, mycolor1}]
  table[row sep=crcr]{%
4	887.26\\
5	705.11\\
6	552.5\\
7	424.05\\
8	321.18\\
9	239.27\\
10	174.42\\
11	127.66\\
};
\addlegendentry{$E_{\textrm{b}}/N_0 = 1.5~\textrm{dB}$}

\addplot [color=mycolor2, mark=square, mark options={solid, mycolor2}]
  table[row sep=crcr]{%
4	69.6\\
5	49.01\\
6	34.43\\
7	24.31\\
8	17.26\\
9	12.34\\
10	8.88\\
11	6.48\\
};
\addlegendentry{$E_{\textrm{b}}/N_0 = 3.0~\textrm{dB}$}

\addplot [color=teal!80!mycolor1, mark=square, mark options={solid, teal!80!mycolor1}]
  table[row sep=crcr]{%
4	1426.96\\
5	1128.49\\
6	890.35\\
7	692.94\\
8	527.47\\
9	384.22\\
10	275.62\\
11	201.26\\
};
\addlegendentry{$E_{\textrm{b}}/N_0 = 1.0~\textrm{dB}$}

\addplot [color=blue!40!mycolor2, mark=+, mark options={solid, blue!40!mycolor2}]
  table[row sep=crcr]{%
4	185.37\\
5	137.54\\
6	101.27\\
7	75.6\\
8	55.81\\
9	40.52\\
10	29.82\\
11	21.7\\
};
\addlegendentry{$E_{\textrm{b}}/N_0 = 2.5~\textrm{dB}$}

\addplot [color=mycolor3, mark=+, mark options={solid, mycolor3}]
  table[row sep=crcr]{%
4	1535.88\\
5	1129.99\\
6	831.4\\
7	604.38\\
8	446.45\\
9	319.69\\
10	229.35\\
11	173.11\\
};
\addlegendentry{$E_{\textrm{b}}/N_0 = 0.5~\textrm{dB}$}

\addplot [color=blue!80!mycolor2, mark=asterisk, mark options={solid, blue!80!mycolor2}]
  table[row sep=crcr]{%
4	436.93\\
5	336.28\\
6	260.41\\
7	199.83\\
8	147.79\\
9	109.78\\
10	80.97\\
11	59.73\\
};
\addlegendentry{$E_{\textrm{b}}/N_0 = 2.0~\textrm{dB}$}

\addplot [color=green, mark=asterisk, mark options={solid, green}]
  table[row sep=crcr]{%
4	1737.3\\
5	1327.73\\
6	951.39\\
7	710.88\\
8	511.25\\
9	383.49\\
10	285.94\\
11	216.12\\
};
\addlegendentry{$E_{\textrm{b}}/N_0 = 0.0~\textrm{dB}$}

\end{axis}
\end{tikzpicture}%
}
  \hfill
  \subfloat[The average decoding time $T_{\textrm{avg}}^{\textrm{tot}}$.\label{fig:delta-tavg}]{
%
%
\definecolor{mycolor1}{rgb}{0.71429,0.00000,0.28571}%
\definecolor{mycolor2}{rgb}{0.00000,0.14286,0.85714}%
\definecolor{mycolor3}{rgb}{0.00000,0.71429,0.28571}%
\begin{tikzpicture}[%
thick,scale=0.7, every node/.style={scale=0.7}
]

\begin{axis}[%
width=10.267cm,
height=7.521cm,
at={(0cm,0cm)},
scale only axis,
xmin=4,
xmax=11,
xlabel style={font=\color{white!15!black}},
xlabel={$\delta$},
ymode=log,
ymin=0.1,
ymax=10,
yminorticks=true,
ylabel style={font=\color{white!15!black}},
ylabel={$T_{\textrm{avg}}~(\textrm{ms})$},
axis background/.style={fill=white},
xmajorgrids,
ymajorgrids,
yminorgrids,
legend style={at={(0.97,0.03)}, anchor=south east, legend cell align=left, align=left, draw=white!15!black}
]
\addplot [color=red, mark=o, mark options={solid, red}]
  table[row sep=crcr]{%
4	0.1360827270152\\
5	0.14241006259638\\
6	0.1873593671746\\
7	0.29903943220108\\
8	0.53923125582912\\
9	1.03502381318735\\
10	2.04612281964752\\
11	4.09957192273995\\
};
\addlegendentry{$E_{\textrm{b}}/N_0 = 3.5~\textrm{dB}$}

\addplot [color=mycolor1, mark=square, mark options={solid, mycolor1}]
  table[row sep=crcr]{%
4	0.275643384128\\
5	0.24306223744023\\
6	0.2589116809306\\
7	0.34982037552301\\
8	0.57502494903936\\
9	1.0601989003823\\
10	2.06369981753504\\
11	4.1118169507452\\
};
\addlegendentry{$E_{\textrm{b}}/N_0 = 3.0~\textrm{dB}$}

\addplot [color=blue!40!mycolor1, mark=+, mark options={solid, blue!40!mycolor1}]
  table[row sep=crcr]{%
4	0.6196275713528\\
5	0.50552871870462\\
6	0.4565379888418\\
7	0.5009843095324\\
8	0.68821993454016\\
9	1.1425954110854\\
10	2.12463729034376\\
11	4.1558758931187\\
};
\addlegendentry{$E_{\textrm{b}}/N_0 = 2.5~\textrm{dB}$}

\addplot [color=blue!80!mycolor1, mark=asterisk, mark options={solid, blue!80!mycolor1}]
  table[row sep=crcr]{%
4	1.3670808160472\\
5	1.09473676667124\\
6	0.927068369467\\
7	0.86711992119133\\
8	0.95830228905024\\
9	1.3451072042471\\
10	2.27348885357996\\
11	4.26596535293895\\
};
\addlegendentry{$E_{\textrm{b}}/N_0 = 2.0~\textrm{dB}$}

\addplot [color=mycolor2, mark=o, mark options={solid, mycolor2}]
  table[row sep=crcr]{%
4	2.7051338438864\\
5	2.18821369296453\\
6	1.7906929696732\\
7	1.52795004955135\\
8	1.46743013693568\\
9	1.72372765034165\\
10	2.54543760400856\\
11	4.4626095024417\\
};
\addlegendentry{$E_{\textrm{b}}/N_0 = 1.5~\textrm{dB}$}

\addplot [color=teal!80!mycolor2, mark=square, mark options={solid, teal!80!mycolor2}]
  table[row sep=crcr]{%
4	4.3087294718144\\
5	3.44341598413947\\
6	2.7896164904362\\
7	2.32043338384234\\
8	2.07316278437952\\
9	2.1475521438269\\
10	2.83993962159416\\
11	4.6756672001217\\
};
\addlegendentry{$E_{\textrm{b}}/N_0 = 1.0~\textrm{dB}$}

\addplot [color=mycolor3, mark=+, mark options={solid, mycolor3}]
  table[row sep=crcr]{%
4	4.6323604441952\\
5	3.44786306108397\\
6	2.6153186021752\\
7	2.05942581909538\\
8	1.8352624542816\\
9	1.95887056697555\\
10	2.7052893414194\\
11	4.59417842037045\\
};
\addlegendentry{$E_{\textrm{b}}/N_0 = 0.5~\textrm{dB}$}

\addplot [color=green, mark=asterisk, mark options={solid, green}]
  table[row sep=crcr]{%
4	5.230834098776\\
5	4.03410639108759\\
6	2.9700938967034\\
7	2.37330686748688\\
8	2.0255357373024\\
9	2.14541767069655\\
10	2.86997184315032\\
11	4.7186840124522\\
};
\addlegendentry{$E_{\textrm{b}}/N_0 = 0.0~\textrm{dB}$}

\end{axis}
\end{tikzpicture}%
}
  \caption{Average time complexity of the LC-OSD with $\ell_{\textrm{max}} = 2^{14}$ for the eBCH code $\mathscr{C}_{\textrm{eBCH}}[128, 64]$.}
  \label{fig:delta-complexity}
\end{figure}
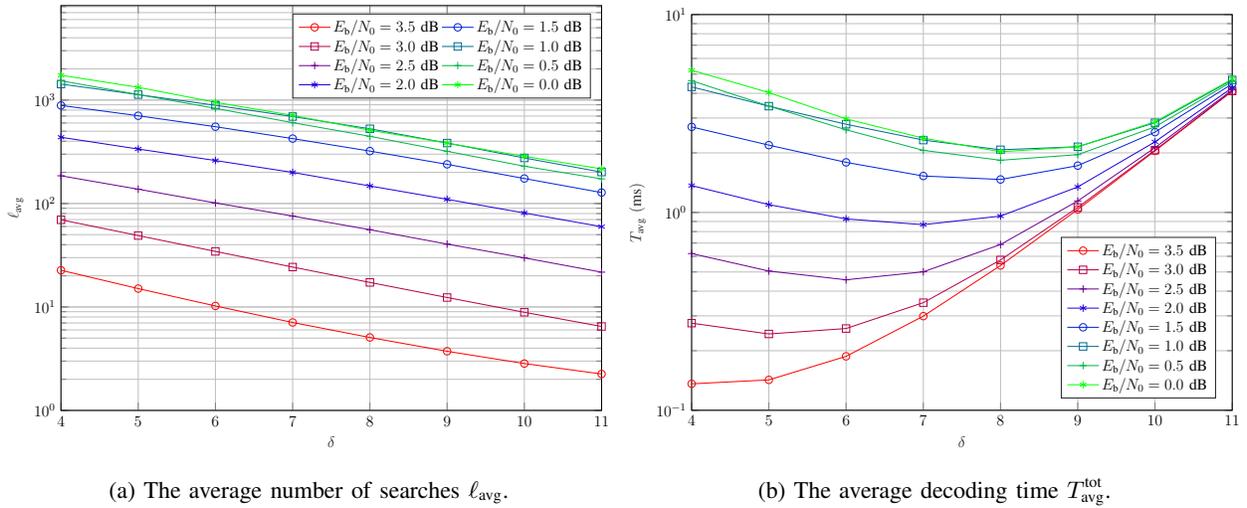
We have simulated the average number of searches of the LC-OSD for the eBCH code $\mathscr{C}_{\textrm{eBCH}}[128, 64]$ using different decoding parameters $(\delta, \ell_{\textrm{max}})$.
From Fig.~\ref{fig:delta-lavg}, we see that the worse the channel is, the higher complexity of the LC-OSD is, as in line with our intuition.
We have implemented the LC-OSD algorithm with SLVA on the desktop computer and tried to optimize the speed as much as possible. 
\end{example}

\subsubsection{Minimum Worst-case Time Complexity}
In the scenarios where the maximum delay is concerned, 
we need to limit the maximum list size $\ell_{\textrm{max}}$ to control the worst-case decoding time.

Recalling the definition in~(\ref{equ:def-epsilon-mrb-hat}), $\widehat{\varepsilon}_{\textrm{MRB}}$ is a function of $(\delta, \ell_{\textrm{max}})$, denoted as $\widehat{\varepsilon}_{\textrm{MRB}}(\delta, \ell_{\textrm{max}})$.
Intuitively, $\widehat{\varepsilon}_{\textrm{MRB}}(\delta, \ell_{\textrm{max}})$ is a non-increasing function of $\ell_{\textrm{max}}$ for a fixed $\delta$.
Therefore, given a target FER $\widehat{\varepsilon}_{\textrm{MRB}}^*$, let $\ell_{\textrm{max}}^*(\delta)$ be the minimum $\ell_{\textrm{max}}$ for a fixed $\delta$ under the constraint of the target FER, i.e, 
\begin{equation}
  \label{equ:lmax-star}
  \ell_{\textrm{max}}^*(\delta) \triangleq \underset{\widehat{\varepsilon}_{\textrm{MRB}}(\delta, \ell_{\textrm{max}}) \leq \widehat{\varepsilon}_{\textrm{MRB}}^*}{\min}\,\ell_{\textrm{max}}.
\end{equation}
To meet the requirement of minimum worst-case time complexity, we can adjust the decoding parameter to $(\ell_{\textrm{max}}^*, \delta)$.

\begin{figure}[!t]
  \centering
  \subfloat[The predicted lower bound of maximum list size $\ell_{\textrm{max}}^*$.\label{fig:delta-lmax}]{
%
%
\definecolor{mycolor1}{rgb}{0.00000,0.14286,0.85714}%
\definecolor{mycolor2}{rgb}{0.71429,0.00000,0.28571}%
\definecolor{mycolor3}{rgb}{0.00000,0.71429,0.28571}%
\begin{tikzpicture}[%
thick,scale=0.7, every node/.style={scale=0.7}
]

\begin{axis}[%
width=10.267cm,
height=7.695cm,
at={(0cm,0cm)},
scale only axis,
xmin=4,
xmax=11,
xlabel style={font=\color{white!15!black}},
xlabel={$\delta$},
ymode=log,
ymin=64,
ymax=16384,
yminorticks=true,
ylabel style={font=\color{white!15!black}},
ylabel={$\ell_{\textrm{max}}^*$},
axis background/.style={fill=white},
xmajorgrids,
ymajorgrids,
yminorgrids,
legend style={at={(0.03,0.03)}, anchor=south west, legend cell align=left, align=left, draw=white!15!black},
legend columns=2
]
\addplot [color=red, mark=o, mark options={solid, red}]
  table[row sep=crcr]{%
4	0\\
5	0\\
6	0\\
7	11583.2271856858\\
8	7695.94346352338\\
9	5088.57626292378\\
10	3412.13244815689\\
11	2313.76366665377\\
};
\addlegendentry{$E_{\textrm{b}}/N_0 = 3.5~\textrm{dB}$}

\addplot [color=mycolor1, mark=o, mark options={solid, mycolor1}]
  table[row sep=crcr]{%
4	4848.07311243677\\
5	3278.62202397684\\
6	2332.5986558164\\
7	1640.67966209574\\
8	1191.81898302694\\
9	875.988140827135\\
10	620.677648326822\\
11	456.199411402662\\
};
\addlegendentry{$E_{\textrm{b}}/N_0 = 1.5~\textrm{dB}$}

\addplot [color=mycolor2, mark=square, mark options={solid, mycolor2}]
  table[row sep=crcr]{%
4	0\\
5	14508.5468838012\\
6	9303.54024935695\\
7	6249.11554137633\\
8	4265.92591408185\\
9	2906.98626575052\\
10	2020.18807744131\\
11	1393.54647247399\\
};
\addlegendentry{$E_{\textrm{b}}/N_0 = 3.0~\textrm{dB}$}

\addplot [color=teal!80!mycolor1, mark=square, mark options={solid, teal!80!mycolor1}]
  table[row sep=crcr]{%
4	2283.81946994309\\
5	1658.75476533816\\
6	1221.5153631328\\
7	879.355154669866\\
8	628.744844189826\\
9	452.238903519397\\
10	328.522235168599\\
11	237.298773078267\\
};
\addlegendentry{$E_{\textrm{b}}/N_0 = 1.0~\textrm{dB}$}

\addplot [color=blue!40!mycolor2, mark=+, mark options={solid, blue!40!mycolor2}]
  table[row sep=crcr]{%
4	12478.7115658361\\
5	8478.41319101763\\
6	5742.22834351327\\
7	3876.42142362355\\
8	2699.79685199762\\
9	1857.81309168145\\
10	1284.90869237205\\
11	900.219964940889\\
};
\addlegendentry{$E_{\textrm{b}}/N_0 = 2.5~\textrm{dB}$}

\addplot [color=mycolor3, mark=+, mark options={solid, mycolor3}]
  table[row sep=crcr]{%
4	1377.56026716854\\
5	951.100901822585\\
6	687.093177580137\\
7	491.397229222841\\
8	355.445920619472\\
9	273.397270034213\\
10	199.664900504915\\
11	150.202032914155\\
};
\addlegendentry{$E_{\textrm{b}}/N_0 = 0.5~\textrm{dB}$}

\addplot [color=blue!80!mycolor2, mark=asterisk, mark options={solid, blue!80!mycolor2}]
  table[row sep=crcr]{%
4	7859.02325680749\\
5	5270.35660709557\\
6	3751.02592774651\\
7	2595.15096625933\\
8	1795.69152646287\\
9	1266.39821906912\\
10	893.617439896468\\
11	635.955588366034\\
};
\addlegendentry{$E_{\textrm{b}}/N_0 = 2.0~\textrm{dB}$}

\addplot [color=green, mark=asterisk, mark options={solid, green}]
  table[row sep=crcr]{%
4	799.486897223703\\
5	586.054580006705\\
6	425.792322876016\\
7	336.814255010455\\
8	254.54553104096\\
9	193.76579813978\\
10	148.974047035542\\
11	118.890760693634\\
};
\addlegendentry{$E_{\textrm{b}}/N_0 = 0.0~\textrm{dB}$}

\end{axis}
\end{tikzpicture}%
}
  \hfill
  \subfloat[The predicted maximum decoding time $T_{\textrm{max}}^{\textrm{tot}}$.\label{fig:delta-tmax}]{
%
%
\definecolor{mycolor1}{rgb}{0.00000,0.14286,0.85714}%
\definecolor{mycolor2}{rgb}{0.71429,0.00000,0.28571}%
\definecolor{mycolor3}{rgb}{0.00000,0.71429,0.28571}%
\begin{tikzpicture}[%
thick,scale=0.7, every node/.style={scale=0.7}
]

\begin{axis}[%
width=10.267cm,
height=7.462cm,
at={(0cm,0cm)},
scale only axis,
xmin=4,
xmax=11,
xlabel style={font=\color{white!15!black}},
xlabel={$\delta$},
ymode=log,
ymin=1,
ymax=100,
yminorticks=true,
ylabel style={font=\color{white!15!black}},
ylabel={$T_{\textrm{max}}~(\textrm{ms})$},
axis background/.style={fill=white},
xmajorgrids,
ymajorgrids,
yminorgrids,
legend style={legend cell align=left, align=left, draw=white!15!black},
legend columns=2
]
\addplot [color=red, mark=o, mark options={solid, red}]
  table[row sep=crcr]{%
4	0\\
5	0\\
6	0\\
7	34.4167222019418\\
8	23.1220667990721\\
9	15.9027878134222\\
10	11.9675013228249\\
11	10.7909548028094\\
};
\addlegendentry{$E_{\textrm{b}}/N_0 = 3.5~\textrm{dB}$}

\addplot [color=mycolor1, mark=o, mark options={solid, mycolor1}]
  table[row sep=crcr]{%
4	14.4737878926978\\
5	9.81795101844517\\
6	7.0539234464339\\
7	5.11364952225377\\
8	4.02390139660277\\
9	3.58545061363196\\
10	3.8440915347917\\
11	5.41366725503128\\
};
\addlegendentry{$E_{\textrm{b}}/N_0 = 1.5~\textrm{dB}$}

\addplot [color=mycolor2, mark=square, mark options={solid, mycolor2}]
  table[row sep=crcr]{%
4	0\\
5	43.1115109735067\\
6	27.6649565074159\\
7	18.6958155807852\\
8	13.0504510590845\\
9	9.52395873067332\\
10	7.91680541825323\\
11	8.12710484686515\\
};
\addlegendentry{$E_{\textrm{b}}/N_0 = 3.0~\textrm{dB}$}

\addplot [color=teal!80!mycolor1, mark=square, mark options={solid, teal!80!mycolor1}]
  table[row sep=crcr]{%
4	6.85469222843743\\
5	5.01550145908351\\
6	3.76877549737826\\
7	2.8698435639723\\
8	2.370537739494\\
9	2.34643505056392\\
10	2.99389036262124\\
11	4.7799924449487\\
};
\addlegendentry{$E_{\textrm{b}}/N_0 = 1.0~\textrm{dB}$}

\addplot [color=blue!40!mycolor2, mark=+, mark options={solid, blue!40!mycolor2}]
  table[row sep=crcr]{%
4	37.1464921027598\\
5	25.2338652952197\\
6	17.1352000251884\\
7	11.7029168278031\\
8	8.45180108567315\\
9	6.45624373185902\\
10	5.7770696238659\\
11	6.69902047323039\\
};
\addlegendentry{$E_{\textrm{b}}/N_0 = 2.5~\textrm{dB}$}

\addplot [color=mycolor3, mark=+, mark options={solid, mycolor3}]
  table[row sep=crcr]{%
4	4.16194941698887\\
5	2.91750733833252\\
6	2.1886455842034\\
7	1.72643849006523\\
8	1.56804571119489\\
9	1.82351359678032\\
10	2.61890276342007\\
11	4.52786430724763\\
};
\addlegendentry{$E_{\textrm{b}}/N_0 = 0.5~\textrm{dB}$}

\addplot [color=blue!80!mycolor2, mark=asterisk, mark options={solid, blue!80!mycolor2}]
  table[row sep=crcr]{%
4	23.4201404726905\\
5	15.7228823145456\\
6	11.2477975456575\\
7	7.92670542469948\\
8	5.79706206582613\\
9	4.72698461613246\\
10	4.63837334602346\\
11	5.93402645622116\\
};
\addlegendentry{$E_{\textrm{b}}/N_0 = 2.0~\textrm{dB}$}

\addplot [color=green, mark=asterisk, mark options={solid, green}]
  table[row sep=crcr]{%
4	2.44433606018852\\
5	1.8352479507179\\
6	1.4160554709596\\
7	1.27084538199706\\
8	1.27177027318284\\
9	1.59067628569965\\
10	2.471387362038\\
11	4.4372242587322\\
};
\addlegendentry{$E_{\textrm{b}}/N_0 = 0.0~\textrm{dB}$}

\end{axis}
\end{tikzpicture}%
}
  \caption{Tuning parameters on~(\ref{equ:lmax-star}) for the LC-OSD of the eBCH code $\mathscr{C}_{\textrm{eBCH}}[128, 64]$ to have near MLD performance~($\widehat{\varepsilon}_{\textrm{MRB}}^*={\textrm{FER}}_{\textrm{ML}}$).}
\end{figure}
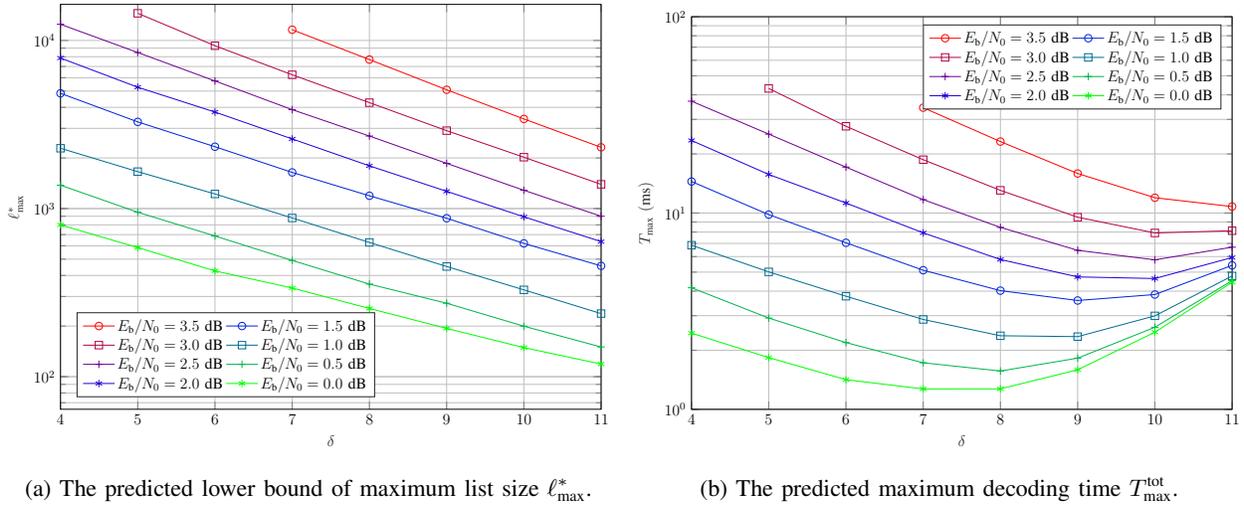

\begin{example}
Intuitively, as $\delta$ increases, the constraint specified by $\mathbf{P}_2$ becomes more ``global'', and hence, the number of searches required for the LC-OSD to have near MLD performance decreases. 
In the extreme case when $\delta$ reaches its maximum value $(n - k)$, $\ell_{\textrm{max}} = 1$ is sufficient. 
This intuition can be verified by the simulation. 
Fig.~\ref{fig:delta-lmax} shows the required $\ell_{\textrm{max}}$ versus $\delta$ for the LC-OSD to have near MLD performance.
As expected, the larger $\delta$ is, the smaller $\ell_{\textrm{max}}$ is required.
\end{example}

\begin{example}
At first glance, it seems that a larger $\delta$ has higher decoding efficiency. However, this is not always the case.
On the one hand, smaller $\delta$ values lead to larger $\ell_{\textrm{max}}$ values, resulting in slower searches. 
On the other hand, large $\delta$ values (which result in small $\ell_{\textrm{max}}$) can
have significant delays caused by initialization (as discussed in Sec.~\ref{sec:time-complexity}). 
As a result, the maximum time complexity is dominated by search when $\delta$ is small and by initialization when $\delta$ is large. 
Therefore, it is meaningful to carefully tune the values of $\delta$ and $\ell_{\textrm{max}}$ for achieving high decoding efficiency. 
This has been verified by the simulation shown in Fig.~\ref{fig:delta-tmax}.
\end{example}

\subsubsection{Limited Maximum Space Complexity}
In the scenarios where the maximum memory usage is concerned, we need to consider the problem of parameter tuning in the case of limited space.
The examples in these cases are similar to the previous scenarios and will not be repeated here.

\section{Two LGAs for LC-OSD}
\label{sec:2-list-alg}
The above analysis for LC-OSD is independent of the LGA.
In practice, we need an algorithm that can generate MRB sequences one by one in the order of increasing soft weight~(see Sec.~\ref{sec:osd-with-lc}).
In this section, we provide two LGAs for LC-OSD, which are applicable to general linear block codes and hence to the MRB code~(see Definition~\ref{def:mrb-code}) with length $N$ and dimension $K$.

\subsection{List Viterbi Decoding Algorithm}
\label{sec:lva}
The list Viterbi decoding algorithms~(LVAs)~(including the parallel LVA, see, for example,~\cite{xiao2003path}, and the serial LVA) produce an ordered list of the $\ell$ best candidates.

\subsubsection{Trellis and Viterbi Algorithm}
A code~(block or convolutional) can be represented graphically by a trellis~\cite{forney1973viterbi, bahl1974optimal, lin1998trellises}, which makes it possible~(also convenient) to implement the MLD with reduced complexity.
The most well-known trellis-based MLD algorithm is the Viterbi algorithm.

A binary linear code $\mathscr{C}[N, K]$ with a parity-check matrix $\mathbf{P}$ can be represented by a trellis with $N$ sections and at most $2^{N-K}$ states at each section.
The state space at level-$i$, written as $\Sigma_i(\mathscr{C})$, is the set of allowable states at the $i$-th level.
In the trellis, every vertex at level-$i$ has a path from the starting state.
Every vertex in the trellis has at least one incoming edge except for the starting state and at most two outgoing edges except for the states at the last level.
For $i > 0$, the two outgoing edges from a state at the $(i-1)$-th level represent the bit $0$ and bit $1$, leading the state $\sigma\in\Sigma_{i-1}(\mathscr{C})$ to $(\sigma + \mathbf{0})\in\Sigma_{i}(\mathscr{C})$ and $(\sigma + \mathbf{P}_i)\in\Sigma_{i}(\mathscr{C})$, respectively, where $\mathbf{P}_i$ is the $i$-th column of $\mathbf{P}$.

Once a trellis of a code is constructed, the ML codeword could be obtained by running the Viterbi algorithm over the trellis.
Equivalently, the Viterbi algorithm can be implemented to find the shortest path~(lightest error pattern) $\bm{e}$ from the starting state $\mathbf{0}$ to the ending state $\mathbf{P}\bm{z}^{\mathrm{T}}$ over the trellis, where the soft weights of the dashed edges are $0$ and that of the solid edges are the bit reliability $|r_i|$. See the discussion in Sec.~\ref{sec:lc-osd}.

\subsubsection{Parallel LVA}
The parallel LVA~(PLVA) searches for the $\ell_{\textrm{max}}$ best paths at the same time by updating the $\ell_{\textrm{max}}$ best paths into each state level by level.
\rchange{
  Therefore, the time complexity of the PLVA is
\begin{equation}
  T_{\text{PLVA}}
  = \mathcal{O}(\ell_{\textrm{max}} 2^{N-K} N),
\end{equation}
which could be significantly reduced by the ``calculate-when-requested'' technique presented in \rchange{below}.
}

\subsubsection{Serial LVA}
The serial LVA~(SLVA) searches for the $\ell_{\textrm{max}}$ most probable candidates, one by one, starting with the most probable path. 
The main advantage of this algorithm is that the $\ell$-th best path is only computed when the previously found $\ell - 1$ paths are not good enough, such as not fulfilling the \rchange{global} constraints. 
More importantly, after the first path is calculated, the SLVA can produce each following path in linear time and space complexity, which is done by using the intermediate results from previous calculations. 
This avoids many unnecessary computations of the PLVA.

Here, we present a recursive implementation of SLVA~(Algorithm~\ref{algo:SLVA})%
\footnote{
  The basic idea of the SLVA was initially presented by Seshadri and Sundberg in~\cite{seshadri1994list}. 
  The description in Sec.~\ref{sec:lva} is based on Wenchao Lin's thesis~\cite{lin2020thesis}, which is believed to be more general than that given in~\cite{seshadri1994list}.
}.
The algorithm first allocates an array to store paths and costs for all trellis vertexes.
Then, the algorithm visits the ending state vertex and queries the $\ell$-th best path.
When visiting a vertex $\sigma\in\Sigma_{i}(\mathscr{C})$, the algorithm is going to visit recursively the two incoming vertexes~(predecessors $\sigma\in\Sigma_{i - 1}(\mathscr{C})$ and $\sigma - \mathbf{P}_i\in\Sigma_{i - 1}(\mathscr{C})$) and select the better one.
If an unallowable state vertex is reached, it returns a path with the cost of $+\infty$.
Finally, the $\ell$-th best path at the ending state vertex is presented after a recursive search.

\begin{figure}[!t]
  \centering
  \begin{tikzpicture}[every node/.style={line join=round}]
  \tikzstyle{pred} = [shape=rectangle, draw=black, text width=4.5em,align=center]
  \tikzstyle{succ} = [shape=rectangle, draw=black, text width=2.2em,align=center]
  \tikzstyle{vsn}  = [diamond, draw=black, minimum height=1.5em, minimum width=2em]
  \tikzstyle{vsp}  = [thick, radius=1.5pt, fill=black,line cap=round]
  \tikzstyle{outp} = [double, double distance between line centers=2pt, rounded corners]

  \node (1A) [pred,                ] {$\overset{\text{1st}}{\leadsto}A\rightarrow C$};
  \node (2A) [pred, right=3.5em of 1A] {$\overset{\text{2nd}}{\leadsto}A\rightarrow C$};
  \node (3A) [pred, right=3.5em of 2A] {$\overset{\text{3rd}}{\leadsto}A\rightarrow C$};

  \node (1B) [pred, below=2em of 1A] {$\overset{\text{1st}}{\leadsto}B\rightarrow C$};
  \node (2B) [pred, right=3.5em of $(1B.east -| 3A.east)$] {$\overset{\text{2nd}}{\leadsto}B\rightarrow C$};
  \node (3B) [pred, right=3.5em of 2B, draw=none] {};

  \node (AC) [above=1.68em of 1A.north] {};
  \node (BC) [below=1.68em of 1B.south] {};

  \node (1C) [succ, right=.8em of $(AC-|1A.east)$] {$\overset{\text{1st}}{\leadsto}C$};
  \node (2C) [succ, right=.8em of $(AC-|2A.east)$] {$\overset{\text{2nd}}{\leadsto}C$};
  \node (3C) [succ, right=.8em of $(BC-|3A.east)$] {$\overset{\text{3rd}}{\leadsto}C$};
  \node (4C) [succ, right=.8em of $(BC-|2B.east)$] {$\overset{\text{4th}}{\leadsto}C$};

  \foreach \i in {1,...,3} {\coordinate (\i Ae) at (\i A.east);}
  \foreach \i in {1,...,3} {\coordinate (\i Be) at (\i B.east);}
  \foreach \i in {1,...,4} {\coordinate (\i Cn) at (\i C.north);}

  \node (1W) at ($(1Ae)!.5!(1Cn)$) {};
  \node (2W) at ($(2Ae)!.5!(2Cn)$) {};
  \node (3W) at ($(3Ae)!.5!(3Cn)$) {};
  \node (4W) at ($(2Be)!.5!(4Cn)$) {};

  \foreach \i in {1,...,4} {\coordinate (\i WA) at (\i W |- 1A);}
  \foreach \i in {1,...,4} {\coordinate (\i WB) at (\i W |- 1B);}

  \foreach \i/\a/\b in {1/win/lose, 2/win/lose,3/lose/win,4/lose/win}{
    \node (\i VS) [vsn] at ($(\i WA)!.5!(\i WB)$) {};
    \node at (\i VS) {\footnotesize{vs}};
    \draw [vsp] (\i VS.north) -- (\i WA) circle;
    \draw [vsp] (\i VS.south) -- (\i WB) circle;
    \node [above=0em of \i WA] {\a} ; 
    \node [below=0em of \i WB] {\b};
  }

  \node [right=of 4VS.center] {$\cdots$};

  \begin{scope}[on background layer]
    \draw [outp] (1A) -| (1C);
    \draw [outp] (2A) -| (2C);
    \draw [outp] (3A) -- (3A -| 3B);
    \draw [outp] (1B) -| (3C);
    \draw [outp] (2B) -| (4C);
  \end{scope}
\end{tikzpicture}
  \caption{A process of selecting $\ell$ best paths for vertex $C$, which is assumed to have two predecessors $A$ and $B$. The notation $\big[\overset{\text{$j$-th}}{\leadsto}X\rightarrow C\big]$ stands for the $j$-th path to $X$ and then one step to $C$.}
  \label{fig:pk}
\end{figure}
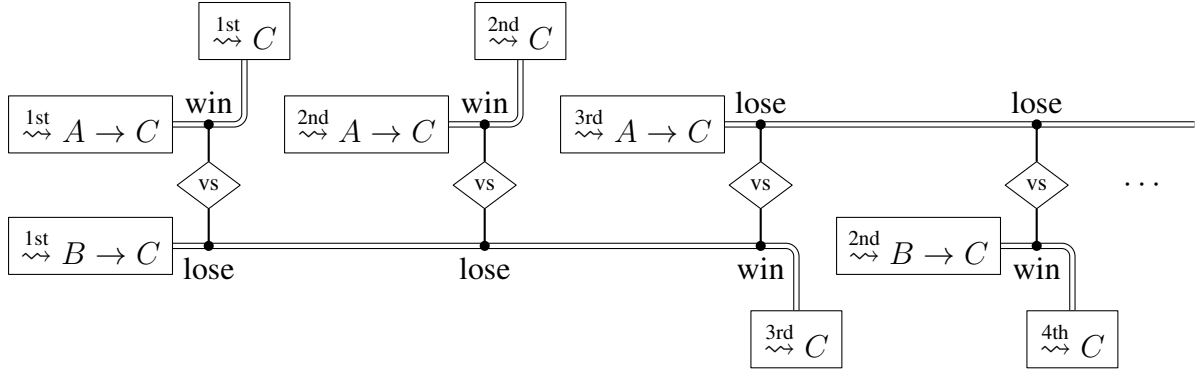

\rchange{
When $\ell \geq 2$, the vertex only requests the predecessor on the $(\ell -1)$-th path to generate a new candidate for comparison with the unused but available candidate at the other predecessor.
For example, let vertexes {$A$} and {$B$} be the predecessors of {$C$}.
The first path to {$C$}~(denoted {$\big[\overset{\textrm{1st}}{\leadsto}C\big]$}) is obtained by comparing {$\big[\overset{\textrm{1st}}{\leadsto}A\rightarrow C\big]$} and {$\big[\overset{\textrm{1st}}{\leadsto}B\rightarrow C\big]$}.
Say {$\big[\overset{\textrm{1st}}{\leadsto}A\rightarrow C\big]$} is better, then {$\big[\overset{\textrm{1st}}{\leadsto}C\big] = \big[\overset{\textrm{1st}}{\leadsto}A\rightarrow C\big]$}.
If required, {$\big[\overset{\textrm{2nd}}{\leadsto}C\big]$} is obtained by comparing {$\big[\overset{\textrm{2nd}}{\leadsto}A\rightarrow C\big]$}~(newly generated candidate) and {$\big[\overset{\textrm{1st}}{\leadsto}B\rightarrow C\big]$}~(unused but available candidate).
This is illustrated in Fig.~\ref{fig:pk}.
Therefore, the time complexity of the SLVA is
\begin{equation}
  T_{\text{avg}}^{\text{SLVA}}
  = \underbrace{\mathcal{O}(2^{N-K} N)}_\textrm{the first candidate}
  + \underbrace{\mathcal{O}((\ell_{\textrm{avg}} - 1)N)}_\textrm{other candidates}
  = \mathcal{O}((2^{N-K} + \ell_{\textrm{avg}}) N).
\end{equation}
}

\begin{algorithm}[H]
  \renewcommand{\algorithmicrequire}{\textbf{Input:}}
  \renewcommand{\algorithmicensure}{\textbf{Output:}}
  \caption{Serial list Viterbi Algorithm~(SLVA)}
  \label{algo:SLVA}
  \begin{algorithmic}[1]
    \Require The parity-check matrix $\mathbf{P}\in\mathbb{F}_2^{(N-K)\times N}$, the reliability vector $|\bm{r}|\in\mathbb{R}^N$, the ending state $\bm{s}_{\text{end}}\in\mathbb{F}_2^{N-K}$, the rank of TEP to search $\ell$. 
    \Ensure The $\ell$-th lightest TEP $\bm{e}^{(\ell)}$ that satisfies the parity-check constraints.

    \State ${nodes}\leftarrow$ a sequence array of $N\times\left|\mathbb{F}_2^{N-K}\right|$ of $(path, cost)$ pairs
    \Comment{Allocate globally.}

    \Function{SLVA}{$\mathbf{P},|\bm{r}|,\bm{s}_{\text{end}},\ell$}
    \State $\bm{s}_{\text{start}} \leftarrow \mathbf{0}$
    \Comment{Set the starting state.}
    
    \Function{SLVA-Impl}{$i, \bm{s},\ell$}
    \Comment{A recursive implementation.}
      \If {$i = 0$}
      \Comment{Boundary of recursion.}
        \If{$\bm{s}=\bm{s}_{\text{start}}$ \textbf{and} $\ell = 1$}
        \Comment{The starting state $\bm{s}_{\text{start}}$ has only $1$ legal path.}
          \State \Return $(\varepsilon, 0)$ 
          \Comment{Empty string and initial cost.}
          \EndIf
        \State \Return $(\varepsilon, +\infty)$ 
        \Comment{Infinity cost for unreachable states.}
      \EndIf

      \While{$\mathrm{length}({nodes}[i][\bm{s}]) < \ell$} 
      \Comment{Check whether it is calculated or not.}
        \For{$b=0,1$} 
        \Comment{enumerate the incoming edges.}
          \State $\bm{s}_b \leftarrow \bm{s} - b\cdot\mathbf{P}_{i}^{{\mathrm{T}}}$
          \Comment{Minus operator is defined in vector space $\mathbb{F}_2^{N-K}$.}
          \State $t_b \leftarrow$ the number of paths end with $b$ in ${nodes}[i][\bm{s}]$
          \State $({path}_b, {cost}_b)\leftarrow$ \Call{SLVA-Impl}{$i - 1, \bm{s}_b, t_b + 1$}
          \Comment{Recursive search.}
          \State ${cost}_b \leftarrow {cost}_b + b \cdot |r_i|$ 
          \Comment{Cost accumulation.}
          \State ${path}_b \leftarrow {path}_b \mathbin{\Vert} b$ 
          \Comment{Append a single bit.}
        \EndFor
        \State $b^* \leftarrow \arg\min_{b\in\{0, 1\}}\, {cost}_b$
        \Comment{The optimal transition.}
        \State ${nodes}[i][\bm{s}]\leftarrow {nodes}[i][\bm{s}] \mathbin{\Vert} ({path}_{b^*}, {cost}_{b^*})$
        \Comment{Append to information sequence.}
      \EndWhile 
      \State \Return ${nodes}[i][\bm{s}][\ell]$ 
      \Comment{The $\ell$-th information of ${nodes}[i][\bm{s}]$.}
    \EndFunction
    \State $(\bm{e}^{(\ell)}, cost)\leftarrow$ \Call{SLVA-Impl}{$N, \bm{s}_{\text{end}},\ell$}
    \State \Return $\bm{e}^{(\ell)}$
  \EndFunction
  \end{algorithmic}
\end{algorithm}

\begin{figure}[!t]
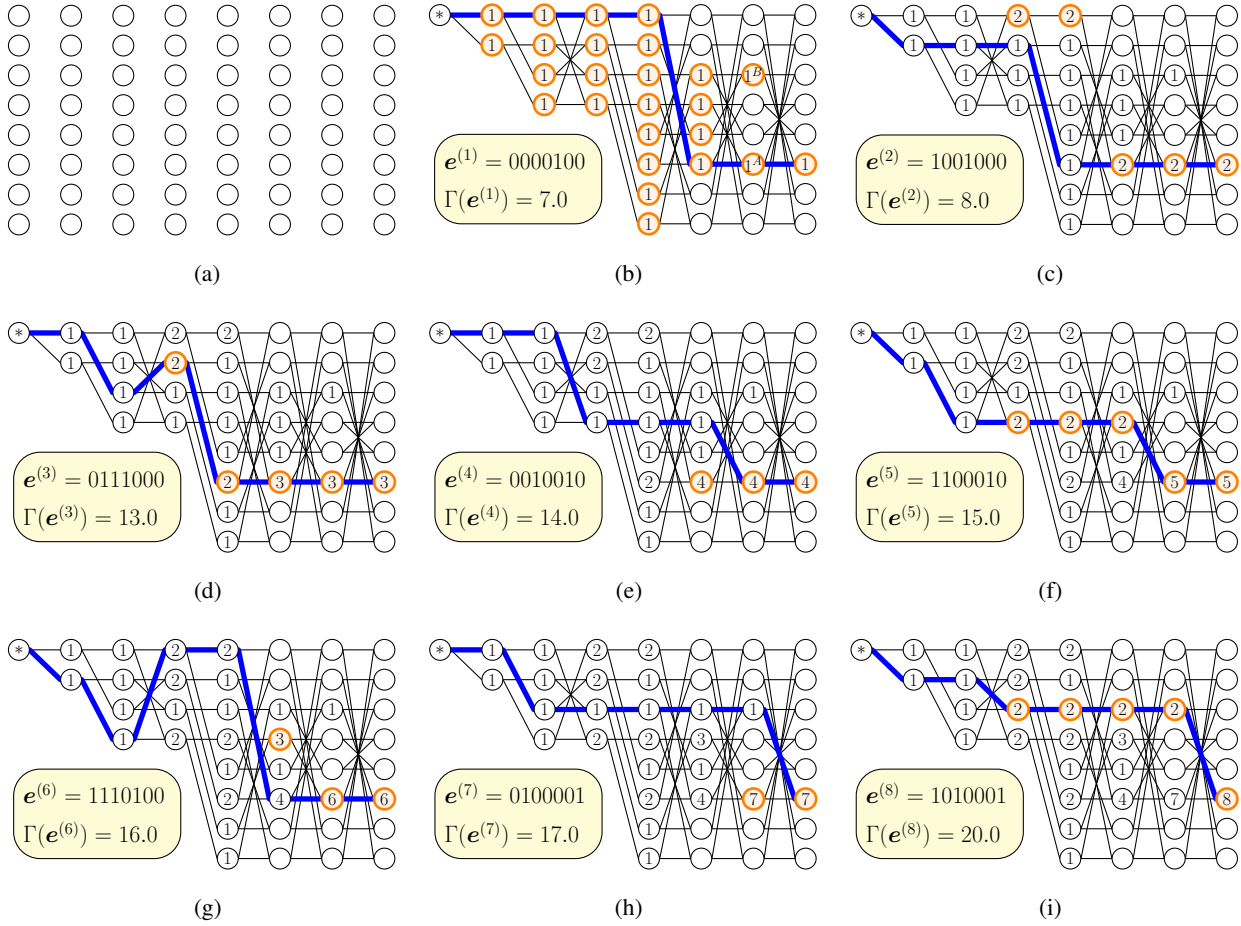

  \centering
  \subfloat[\label{fig:SLVA-path0}]{\begin{tikzpicture}[style={font=\Large}, scale=0.3967, every node/.style={scale=0.3967}]
  \draw (     0,     -0) node [fill=white, minimum size=0.7cm, draw, circle] (l00s00) {};
  \draw (     0,     -1) node [fill=white, minimum size=0.7cm, draw, circle] (l00s01) {};
  \draw (     0,     -2) node [fill=white, minimum size=0.7cm, draw, circle] (l00s02) {};
  \draw (     0,     -3) node [fill=white, minimum size=0.7cm, draw, circle] (l00s03) {};
  \draw (     0,     -4) node [fill=white, minimum size=0.7cm, draw, circle] (l00s04) {};
  \draw (     0,     -5) node [fill=white, minimum size=0.7cm, draw, circle] (l00s05) {};
  \draw (     0,     -6) node [fill=white, minimum size=0.7cm, draw, circle] (l00s06) {};
  \draw (     0,     -7) node [fill=white, minimum size=0.7cm, draw, circle] (l00s07) {};

  \draw (  1.75,     -0) node [fill=white, minimum size=0.7cm, draw, circle] (l01s00) {};
  \draw (  1.75,     -1) node [fill=white, minimum size=0.7cm, draw, circle] (l01s01) {};
  \draw (  1.75,     -2) node [fill=white, minimum size=0.7cm, draw, circle] (l01s02) {};
  \draw (  1.75,     -3) node [fill=white, minimum size=0.7cm, draw, circle] (l01s03) {};
  \draw (  1.75,     -4) node [fill=white, minimum size=0.7cm, draw, circle] (l01s04) {};
  \draw (  1.75,     -5) node [fill=white, minimum size=0.7cm, draw, circle] (l01s05) {};
  \draw (  1.75,     -6) node [fill=white, minimum size=0.7cm, draw, circle] (l01s06) {};
  \draw (  1.75,     -7) node [fill=white, minimum size=0.7cm, draw, circle] (l01s07) {};

  \draw (   3.5,     -0) node [fill=white, minimum size=0.7cm, draw, circle] (l02s00) {};
  \draw (   3.5,     -1) node [fill=white, minimum size=0.7cm, draw, circle] (l02s01) {};
  \draw (   3.5,     -2) node [fill=white, minimum size=0.7cm, draw, circle] (l02s02) {};
  \draw (   3.5,     -3) node [fill=white, minimum size=0.7cm, draw, circle] (l02s03) {};
  \draw (   3.5,     -4) node [fill=white, minimum size=0.7cm, draw, circle] (l02s04) {};
  \draw (   3.5,     -5) node [fill=white, minimum size=0.7cm, draw, circle] (l02s05) {};
  \draw (   3.5,     -6) node [fill=white, minimum size=0.7cm, draw, circle] (l02s06) {};
  \draw (   3.5,     -7) node [fill=white, minimum size=0.7cm, draw, circle] (l02s07) {};

  \draw (  5.25,     -0) node [fill=white, minimum size=0.7cm, draw, circle] (l03s00) {};
  \draw (  5.25,     -1) node [fill=white, minimum size=0.7cm, draw, circle] (l03s01) {};
  \draw (  5.25,     -2) node [fill=white, minimum size=0.7cm, draw, circle] (l03s02) {};
  \draw (  5.25,     -3) node [fill=white, minimum size=0.7cm, draw, circle] (l03s03) {};
  \draw (  5.25,     -4) node [fill=white, minimum size=0.7cm, draw, circle] (l03s04) {};
  \draw (  5.25,     -5) node [fill=white, minimum size=0.7cm, draw, circle] (l03s05) {};
  \draw (  5.25,     -6) node [fill=white, minimum size=0.7cm, draw, circle] (l03s06) {};
  \draw (  5.25,     -7) node [fill=white, minimum size=0.7cm, draw, circle] (l03s07) {};

  \draw (     7,     -0) node [fill=white, minimum size=0.7cm, draw, circle] (l04s00) {};
  \draw (     7,     -1) node [fill=white, minimum size=0.7cm, draw, circle] (l04s01) {};
  \draw (     7,     -2) node [fill=white, minimum size=0.7cm, draw, circle] (l04s02) {};
  \draw (     7,     -3) node [fill=white, minimum size=0.7cm, draw, circle] (l04s03) {};
  \draw (     7,     -4) node [fill=white, minimum size=0.7cm, draw, circle] (l04s04) {};
  \draw (     7,     -5) node [fill=white, minimum size=0.7cm, draw, circle] (l04s05) {};
  \draw (     7,     -6) node [fill=white, minimum size=0.7cm, draw, circle] (l04s06) {};
  \draw (     7,     -7) node [fill=white, minimum size=0.7cm, draw, circle] (l04s07) {};

  \draw (  8.75,     -0) node [fill=white, minimum size=0.7cm, draw, circle] (l05s00) {};
  \draw (  8.75,     -1) node [fill=white, minimum size=0.7cm, draw, circle] (l05s01) {};
  \draw (  8.75,     -2) node [fill=white, minimum size=0.7cm, draw, circle] (l05s02) {};
  \draw (  8.75,     -3) node [fill=white, minimum size=0.7cm, draw, circle] (l05s03) {};
  \draw (  8.75,     -4) node [fill=white, minimum size=0.7cm, draw, circle] (l05s04) {};
  \draw (  8.75,     -5) node [fill=white, minimum size=0.7cm, draw, circle] (l05s05) {};
  \draw (  8.75,     -6) node [fill=white, minimum size=0.7cm, draw, circle] (l05s06) {};
  \draw (  8.75,     -7) node [fill=white, minimum size=0.7cm, draw, circle] (l05s07) {};

  \draw (  10.5,     -0) node [fill=white, minimum size=0.7cm, draw, circle] (l06s00) {};
  \draw (  10.5,     -1) node [fill=white, minimum size=0.7cm, draw, circle] (l06s01) {};
  \draw (  10.5,     -2) node [fill=white, minimum size=0.7cm, draw, circle] (l06s02) {};
  \draw (  10.5,     -3) node [fill=white, minimum size=0.7cm, draw, circle] (l06s03) {};
  \draw (  10.5,     -4) node [fill=white, minimum size=0.7cm, draw, circle] (l06s04) {};
  \draw (  10.5,     -5) node [fill=white, minimum size=0.7cm, draw, circle] (l06s05) {};
  \draw (  10.5,     -6) node [fill=white, minimum size=0.7cm, draw, circle] (l06s06) {};
  \draw (  10.5,     -7) node [fill=white, minimum size=0.7cm, draw, circle] (l06s07) {};

  \draw ( 12.25,     -0) node [fill=white, minimum size=0.7cm, draw, circle] (l07s00) {};
  \draw ( 12.25,     -1) node [fill=white, minimum size=0.7cm, draw, circle] (l07s01) {};
  \draw ( 12.25,     -2) node [fill=white, minimum size=0.7cm, draw, circle] (l07s02) {};
  \draw ( 12.25,     -3) node [fill=white, minimum size=0.7cm, draw, circle] (l07s03) {};
  \draw ( 12.25,     -4) node [fill=white, minimum size=0.7cm, draw, circle] (l07s04) {};
  \draw ( 12.25,     -5) node [fill=white, minimum size=0.7cm, draw, circle] (l07s05) {};
  \draw ( 12.25,     -6) node [fill=white, minimum size=0.7cm, draw, circle] (l07s06) {};
  \draw ( 12.25,     -7) node [fill=white, minimum size=0.7cm, draw, circle] (l07s07) {};
\end{tikzpicture}} \hfill
  \subfloat[\label{fig:SLVA-path1}]{\input{Figures/trellis-cyz/path1.tex}} \hfill
  \subfloat[\label{fig:SLVA-path2}]{\input{Figures/trellis-cyz/path2.tex}} \\
  \subfloat[\label{fig:SLVA-path3}]{\input{Figures/trellis-cyz/path3.tex}} \hfill
  \subfloat[\label{fig:SLVA-path4}]{\input{Figures/trellis-cyz/path4.tex}} \hfill
  \subfloat[\label{fig:SLVA-path5}]{\input{Figures/trellis-cyz/path5.tex}} \\
  \subfloat[\label{fig:SLVA-path6}]{\input{Figures/trellis-cyz/path6.tex}} \hfill
  \subfloat[\label{fig:SLVA-path7}]{\input{Figures/trellis-cyz/path7.tex}} \hfill
  \subfloat[\label{fig:SLVA-path8}]{\input{Figures/trellis-cyz/path8.tex}} \\
  
  \caption{
    An illustration example of SLVA over a trellis for the $[7,4]$ Hamming code.
    The digit in each circle counts the number of candidates, while empty circles imply no candidate.
    The updated states are highlighted in orange.
    }
  \label{fig:SLVA-process}
\end{figure}
\rchange{
\begin{example}
  \label{ex:SLVA-process}
  To better understand the SLVA, we present
  as an example to show how to generate candidates through the trellis defined by the parity check matrix of the $(7, 4)$ Hamming code
  \begin{equation}
    \label{equ:hamming}
    \mathbf{P}=
    \begin{bNiceMatrix}
      1 & 0 & 1 & 0 & 1 & 0 & 1 \\
      0 & 1 & 1 & 0 & 0 & 1 & 1 \\
      0 & 0 & 0 & 1 & 1 & 1 & 1 \\
    \end{bNiceMatrix}
    .
  \end{equation}
  The trellis representation%
  \footnote{The edges are directed from left to right, while the arrows are omitted for the sake of clarity.}
  of the Hamming code is shown in Fig.~\ref{fig:SLVA-path1}.
  It is worth noting that the trellis does not correspond to the Hamming code but its cosets.
  
  Suppose that the LLR vector is 
  \begin{equation}
    \bm{r} = [-2.0, 3.0, 4.0, -6.0, 7.0, 10.0, 14.0],
  \end{equation}
  resulting in the bit-wise hard-decision vector 
  \begin{equation}
    \bm{z} = [1, 0, 0, 1, 0, 0, 0].
  \end{equation}
  Initially, the space of size $2^{N-K} \times N$ for the trellis is ``blanked'', meaning that no branches exist between the states, as shown in Fig.~\ref{fig:SLVA-path0}. 
  Then the syndrome, i.e., the ending state of the trellis, is given by  
  \begin{equation}
    \bm{s}_{\text{end}} \triangleq \bm{z}\mathbf{P}^{{\mathrm{T}}} = [1, 0, 1].
  \end{equation}
  By the SLVA, the best eight candidates $e^{(i)}, 1 \leq i \leq 8$, are generated serially, shown in the yellow boxes along with their soft weights, and highlighted in blue paths through the trellis in Figs.~\ref{fig:SLVA-path1}-\ref{fig:SLVA-path8}.
  The first candidate $e^{(1)}$ is generated by the VA with the complexity of order $\mathcal{O}(2^{N-K} N)$, which starts from the initial state $\bm{s}_{\text{start}} = [0, 0, 0]$ and implicitly generates all~(prefix) best paths, labeled by \circled{$1$} in the orange circles.
  Now, to generate the second (best) path, the ending state compares the two paths, the second path entering the state \circled{$1^A$} and the first path entering the state \circled{$1^B$}~(since \circled{$1^B$} has not been used). 
  That is, the state \circled{$1^A$} needs to generate the second path, while the state \circled{$1^B$} does not have to.
  Then, the state \circled{$1^A$} takes on the same role as the ending state. This process continues recursively and requires the complexity of order $\mathcal{O}(N)$.
  To avoid updating unnecessary states, the SLVA uses the same processes to generate the subsequent candidates, reducing the time complexity compared to the PLVA.
\end{example}
}

\rchange{
We conclude~(without proof) that when searching for the $\ell$-th candidate~($\ell > 1$), the states to be updated must lay on the $(\ell -1)$-th path.
Therefore, searching for the $\ell$-th candidate consumes only $\mathcal{O}(N)$ time whenever $\ell > 1$.
}

\begin{example}
  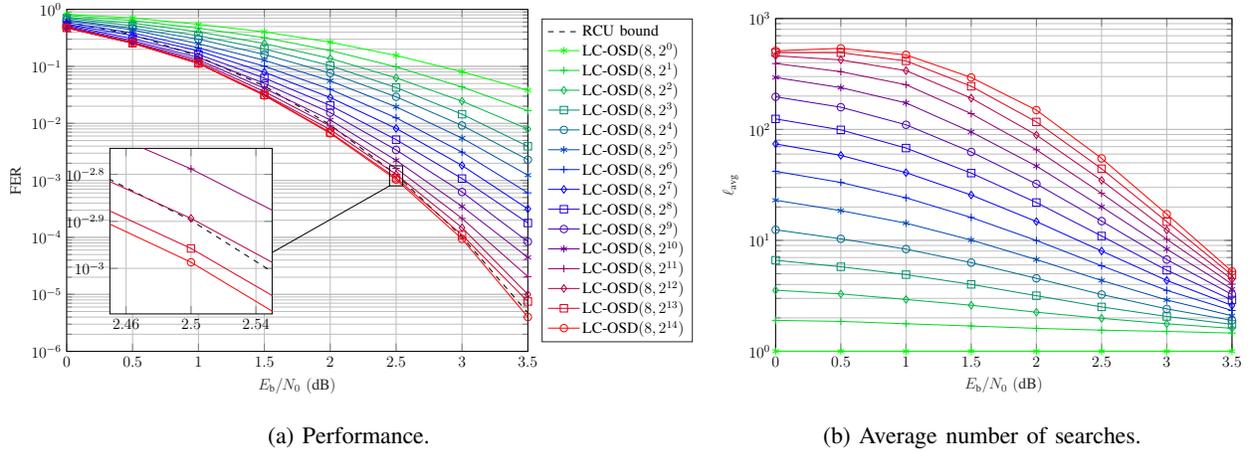
\begin{figure}[!t]
    \centering
    \subfloat[Performance.\label{fig:bch64-vl-fer}]{
%
%
\definecolor{mycolor1}{rgb}{0.00000,0.85714,0.14286}%
\definecolor{mycolor2}{rgb}{0.00000,0.28571,0.71429}%
\definecolor{mycolor3}{rgb}{0.14286,0.00000,0.85714}%
\definecolor{mycolor4}{rgb}{0.71429,0.00000,0.28571}%
\begin{tikzpicture}[%
thick,scale=0.7, every node/.style={scale=0.7}
]

\begin{axis}[%
width=8.759cm,
height=6.5cm,
at={(0cm,0cm)},
scale only axis,
xmin=0,
xmax=3.5,
xtick={  0, 0.5,   1, 1.5,   2, 2.5,   3, 3.5},
xlabel style={font=\color{white!15!black}},
xlabel={$E_{\textrm{b}}/N_0~(\textrm{dB})$},
ymode=log,
ymin=1e-06,
ymax=1,
yminorticks=true,
ylabel style={font=\color{white!15!black}},
ylabel={FER},
axis background/.style={fill=white},
xmajorgrids,
ymajorgrids,
yminorgrids,
legend style={at={(1.03,0.5)}, anchor=west, legend cell align=left, align=left, draw=white!15!black}
]
\addplot [color=black, dashed]
  table[row sep=crcr]{%
0	0.5981292\\
0.5	0.3642932\\
1	0.1496848\\
1.5	0.04529887\\
2	0.009435829\\
2.5	0.001264726\\
3	0.0001024066\\
3.5	4.751623e-06\\
};
\addlegendentry{RCU bound}

\addplot [color=green, mark=asterisk, mark options={solid, green}]
  table[row sep=crcr]{%
0	0.8199052133\\
0.5	0.7072243346\\
1	0.5462703309\\
1.5	0.4028931979\\
2	0.2659251094\\
2.5	0.1553547603\\
3	0.0805105\\
3.5	0.0381255\\
};
\addlegendentry{LC-OSD$(8,2^{0})$}

\addplot [color=mycolor1, mark=+, mark options={solid, mycolor1}]
  table[row sep=crcr]{%
0	0.7843601896\\
0.5	0.6362484157\\
1	0.4677509815\\
1.5	0.3191751308\\
2	0.1899655944\\
2.5	0.0978997203\\
3	0.043555\\
3.5	0.016689\\
};
\addlegendentry{LC-OSD$(8,2^{1})$}

\addplot [color=teal!40!mycolor1, mark=diamond, mark options={solid, teal!40!mycolor1}]
  table[row sep=crcr]{%
0	0.7322274882\\
0.5	0.5741444867\\
1	0.403813797\\
1.5	0.2514619883\\
2	0.1375221298\\
2.5	0.0628383913\\
3	0.0245235\\
3.5	0.007954\\
};
\addlegendentry{LC-OSD$(8,2^{2})$}

\addplot [color=teal!80!mycolor1, mark=square, mark options={solid, teal!80!mycolor1}]
  table[row sep=crcr]{%
0	0.7014218009\\
0.5	0.5171102662\\
1	0.3449242849\\
1.5	0.2022160665\\
2	0.1028159134\\
2.5	0.0425207478\\
3	0.0144905\\
3.5	0.0039605\\
};
\addlegendentry{LC-OSD$(8,2^{3})$}

\addplot [color=blue!50!mycolor1, mark=o, mark options={solid, blue!50!mycolor1}]
  table[row sep=crcr]{%
0	0.6587677725\\
0.5	0.4689480355\\
1	0.3000560852\\
1.5	0.1617420745\\
2	0.0762601463\\
2.5	0.0292658551\\
3	0.009191\\
3.5	0.002304\\
};
\addlegendentry{LC-OSD$(8,2^{4})$}

\addplot [color=mycolor2, mark=asterisk, mark options={solid, mycolor2}]
  table[row sep=crcr]{%
0	0.6327014218\\
0.5	0.4372623574\\
1	0.2495793606\\
1.5	0.1278855032\\
2	0.0557838127\\
2.5	0.0195243075\\
3	0.0055025\\
3.5	0.0012335\\
};
\addlegendentry{LC-OSD$(8,2^{5})$}

\addplot [color=blue!50!mycolor2, mark=+, mark options={solid, blue!50!mycolor2}]
  table[row sep=crcr]{%
0	0.5900473934\\
0.5	0.3865652725\\
1	0.2114413909\\
1.5	0.1008002462\\
2	0.039616528\\
2.5	0.0125182235\\
3	0.0031285\\
3.5	0.0006045\\
};
\addlegendentry{LC-OSD$(8,2^{6})$}

\addplot [color=blue, mark=diamond, mark options={solid, blue}]
  table[row sep=crcr]{%
0	0.5521327014\\
0.5	0.3472750317\\
1	0.1873247336\\
1.5	0.0791012619\\
2	0.028426362\\
2.5	0.0081600272\\
3	0.001827\\
3.5	0.0003155\\
};
\addlegendentry{LC-OSD$(8,2^{7})$}

\addplot [color=mycolor3, mark=square, mark options={solid, mycolor3}]
  table[row sep=crcr]{%
0	0.5308056872\\
0.5	0.3067173638\\
1	0.1615255188\\
1.5	0.0624807633\\
2	0.0206099476\\
2.5	0.00517729\\
3	0.001076\\
3.5	0.0001785\\
};
\addlegendentry{LC-OSD$(8,2^{8})$}

\addplot [color=violet!40!mycolor3, mark=o, mark options={solid, violet!40!mycolor3}]
  table[row sep=crcr]{%
0	0.5071090047\\
0.5	0.2826362484\\
1	0.1430173864\\
1.5	0.0503231764\\
2	0.0153990046\\
2.5	0.0034154659\\
3	0.000626\\
3.5	8.4e-05\\
};
\addlegendentry{LC-OSD$(8,2^{9})$}

\addplot [color=violet!80!mycolor3, mark=asterisk, mark options={solid, violet!80!mycolor3}]
  table[row sep=crcr]{%
0	0.490521327\\
0.5	0.2661596958\\
1	0.1306786315\\
1.5	0.0412434595\\
2	0.0114907973\\
2.5	0.0022460681\\
3	0.000349\\
3.5	4.45e-05\\
};
\addlegendentry{LC-OSD$(8,2^{10})$}

\addplot [color=red!50!mycolor3, mark=+, mark options={solid, red!50!mycolor3}]
  table[row sep=crcr]{%
0	0.4786729858\\
0.5	0.2585551331\\
1	0.1211441391\\
1.5	0.0361649738\\
2	0.0090189398\\
2.5	0.0016278842\\
3	0.000215\\
3.5	2.05e-05\\
};
\addlegendentry{LC-OSD$(8,2^{11})$}

\addplot [color=mycolor4, mark=diamond, mark options={solid, mycolor4}]
  table[row sep=crcr]{%
0	0.4739336493\\
0.5	0.2560202788\\
1	0.1138530566\\
1.5	0.0324715297\\
2	0.0077496075\\
2.5	0.00127758\\
3	0.000148\\
3.5	1e-05\\
};
\addlegendentry{LC-OSD$(8,2^{12})$}

\addplot [color=red!50!mycolor4, mark=square, mark options={solid, red!50!mycolor4}]
  table[row sep=crcr]{%
0	0.4739336493\\
0.5	0.2547528517\\
1	0.1127313517\\
1.5	0.0315481687\\
2	0.0069145205\\
2.5	0.0011024279\\
3	0.00011\\
3.5	7.5e-06\\
};
\addlegendentry{LC-OSD$(8,2^{13})$}

\addplot [color=red, mark=o, mark options={solid, red}]
  table[row sep=crcr]{%
0	0.4739336493\\
0.5	0.2547528517\\
1	0.1121704992\\
1.5	0.0307787011\\
2	0.0066806961\\
2.5	0.0010303065\\
3	9.4e-05\\
3.5	4e-06\\
};
\addlegendentry{LC-OSD$(8,2^{14})$}

\addplot [color=black, line width=0.1pt, forget plot]
  table[row sep=crcr]{%
2.45	0.0008\\
2.55	0.0008\\
2.55	0.0018\\
2.45	0.0018\\
2.45	0.0008\\
1	1e-05\\
};
\end{axis}

\begin{axis}[%
width=3.082cm,
height=3.147cm,
at={(0.821cm,0.71cm)},
scale only axis,
xmin=2.45,
xmax=2.55,
xtick={2.46,  2.5, 2.54},
ymode=log,
ymin=0.0008,
ymax=0.0018,
ytick={0.00079433,      0.001,  0.0012589,  0.0015849,  0.0019953},
yminorticks=true,
axis background/.style={fill=white},
xmajorgrids,
ymajorgrids,
yminorgrids
]
\addplot [color=black, dashed, forget plot]
  table[row sep=crcr]{%
0	0.5981292\\
0.5	0.3642932\\
1	0.1496848\\
1.5	0.04529887\\
2	0.009435829\\
2.5	0.001264726\\
3	0.0001024066\\
3.5	4.751623e-06\\
};
\addplot [color=green, mark=asterisk, mark options={solid, green}, forget plot]
  table[row sep=crcr]{%
0	0.8199052133\\
0.5	0.7072243346\\
1	0.5462703309\\
1.5	0.4028931979\\
2	0.2659251094\\
2.5	0.1553547603\\
3	0.0805105\\
3.5	0.0381255\\
};
\addplot [color=mycolor1, mark=+, mark options={solid, mycolor1}, forget plot]
  table[row sep=crcr]{%
0	0.7843601896\\
0.5	0.6362484157\\
1	0.4677509815\\
1.5	0.3191751308\\
2	0.1899655944\\
2.5	0.0978997203\\
3	0.043555\\
3.5	0.016689\\
};
\addplot [color=green!60!mycolor2, mark=diamond, mark options={solid, green!60!mycolor2}, forget plot]
  table[row sep=crcr]{%
0	0.7322274882\\
0.5	0.5741444867\\
1	0.403813797\\
1.5	0.2514619883\\
2	0.1375221298\\
2.5	0.0628383913\\
3	0.0245235\\
3.5	0.007954\\
};
\addplot [color=green!40!mycolor2, mark=square, mark options={solid, green!40!mycolor2}, forget plot]
  table[row sep=crcr]{%
0	0.7014218009\\
0.5	0.5171102662\\
1	0.3449242849\\
1.5	0.2022160665\\
2	0.1028159134\\
2.5	0.0425207478\\
3	0.0144905\\
3.5	0.0039605\\
};
\addplot [color=green!20!mycolor2, mark=o, mark options={solid, green!20!mycolor2}, forget plot]
  table[row sep=crcr]{%
0	0.6587677725\\
0.5	0.4689480355\\
1	0.3000560852\\
1.5	0.1617420745\\
2	0.0762601463\\
2.5	0.0292658551\\
3	0.009191\\
3.5	0.002304\\
};
\addplot [color=mycolor2, mark=asterisk, mark options={solid, mycolor2}, forget plot]
  table[row sep=crcr]{%
0	0.6327014218\\
0.5	0.4372623574\\
1	0.2495793606\\
1.5	0.1278855032\\
2	0.0557838127\\
2.5	0.0195243075\\
3	0.0055025\\
3.5	0.0012335\\
};
\addplot [color=blue!50!mycolor2, mark=+, mark options={solid, blue!50!mycolor2}, forget plot]
  table[row sep=crcr]{%
0	0.5900473934\\
0.5	0.3865652725\\
1	0.2114413909\\
1.5	0.1008002462\\
2	0.039616528\\
2.5	0.0125182235\\
3	0.0031285\\
3.5	0.0006045\\
};
\addplot [color=blue, mark=diamond, mark options={solid, blue}, forget plot]
  table[row sep=crcr]{%
0	0.5521327014\\
0.5	0.3472750317\\
1	0.1873247336\\
1.5	0.0791012619\\
2	0.028426362\\
2.5	0.0081600272\\
3	0.001827\\
3.5	0.0003155\\
};
\addplot [color=mycolor3, mark=square, mark options={solid, mycolor3}, forget plot]
  table[row sep=crcr]{%
0	0.5308056872\\
0.5	0.3067173638\\
1	0.1615255188\\
1.5	0.0624807633\\
2	0.0206099476\\
2.5	0.00517729\\
3	0.001076\\
3.5	0.0001785\\
};
\addplot [color=blue!60!mycolor4, mark=o, mark options={solid, blue!60!mycolor4}, forget plot]
  table[row sep=crcr]{%
0	0.5071090047\\
0.5	0.2826362484\\
1	0.1430173864\\
1.5	0.0503231764\\
2	0.0153990046\\
2.5	0.0034154659\\
3	0.000626\\
3.5	8.4e-05\\
};
\addplot [color=blue!40!mycolor4, mark=asterisk, mark options={solid, blue!40!mycolor4}, forget plot]
  table[row sep=crcr]{%
0	0.490521327\\
0.5	0.2661596958\\
1	0.1306786315\\
1.5	0.0412434595\\
2	0.0114907973\\
2.5	0.0022460681\\
3	0.000349\\
3.5	4.45e-05\\
};
\addplot [color=red!50!mycolor3, mark=+, mark options={solid, red!50!mycolor3}, forget plot]
  table[row sep=crcr]{%
0	0.4786729858\\
0.5	0.2585551331\\
1	0.1211441391\\
1.5	0.0361649738\\
2	0.0090189398\\
2.5	0.0016278842\\
3	0.000215\\
3.5	2.05e-05\\
};
\addplot [color=mycolor4, mark=diamond, mark options={solid, mycolor4}, forget plot]
  table[row sep=crcr]{%
0	0.4739336493\\
0.5	0.2560202788\\
1	0.1138530566\\
1.5	0.0324715297\\
2	0.0077496075\\
2.5	0.00127758\\
3	0.000148\\
3.5	1e-05\\
};
\addplot [color=red!50!mycolor4, mark=square, mark options={solid, red!50!mycolor4}, forget plot]
  table[row sep=crcr]{%
0	0.4739336493\\
0.5	0.2547528517\\
1	0.1127313517\\
1.5	0.0315481687\\
2	0.0069145205\\
2.5	0.0011024279\\
3	0.00011\\
3.5	7.5e-06\\
};
\addplot [color=red, mark=o, mark options={solid, red}, forget plot]
  table[row sep=crcr]{%
0	0.4739336493\\
0.5	0.2547528517\\
1	0.1121704992\\
1.5	0.0307787011\\
2	0.0066806961\\
2.5	0.0010303065\\
3	9.4e-05\\
3.5	4e-06\\
};
\end{axis}
\end{tikzpicture}%
} 
    \hfill
    \subfloat[Average number of searches.\label{fig:bch64-vl-timing}]{
%
%
\definecolor{mycolor1}{rgb}{0.00000,0.85714,0.14286}%
\definecolor{mycolor2}{rgb}{0.00000,0.28571,0.71429}%
\definecolor{mycolor3}{rgb}{0.14286,0.00000,0.85714}%
\definecolor{mycolor4}{rgb}{0.71429,0.00000,0.28571}%
\begin{tikzpicture}[%
thick,scale=0.7, every node/.style={scale=0.7}
]

\begin{axis}[%
width=8.667cm,
height=6.324cm,
at={(0cm,0cm)},
scale only axis,
xmin=0,
xmax=3.5,
xtick={  0, 0.5,   1, 1.5,   2, 2.5,   3, 3.5},
xlabel style={font=\color{white!15!black}},
xlabel={$E_{\textrm{b}}/N_0~(\textrm{dB})$},
ymode=log,
ymin=1,
ymax=1000,
yminorticks=true,
ylabel style={font=\color{white!15!black}},
ylabel={$\ell_{\textrm{avg}}$},
axis background/.style={fill=white},
xmajorgrids,
ymajorgrids,
yminorgrids
]
\addplot [color=black, forget plot]
  table[row sep=crcr]{%
0	0\\
0.5	0\\
1	0\\
1.5	0\\
2	0\\
2.5	0\\
3	0\\
3.5	0\\
};
\addplot [color=green, mark=asterisk, mark options={solid, green}, forget plot]
  table[row sep=crcr]{%
0	1\\
0.5	1\\
1	1\\
1.5	1\\
2	1\\
2.5	1\\
3	1\\
3.5	1\\
};
\addplot [color=mycolor1, mark=+, mark options={solid, mycolor1}, forget plot]
  table[row sep=crcr]{%
0	1.9\\
0.5	1.86\\
1	1.77\\
1.5	1.69\\
2	1.61\\
2.5	1.55\\
3	1.51\\
3.5	1.46\\
};
\addplot [color=teal!40!mycolor1, mark=diamond, mark options={solid, teal!40!mycolor1}, forget plot]
  table[row sep=crcr]{%
0	3.55\\
0.5	3.29\\
1	2.93\\
1.5	2.6\\
2	2.25\\
2.5	1.98\\
3	1.77\\
3.5	1.61\\
};
\addplot [color=teal!80!mycolor1, mark=square, mark options={solid, teal!80!mycolor1}, forget plot]
  table[row sep=crcr]{%
0	6.62\\
0.5	5.78\\
1	4.91\\
1.5	4.01\\
2	3.17\\
2.5	2.51\\
3	2.06\\
3.5	1.75\\
};
\addplot [color=blue!50!mycolor1, mark=o, mark options={solid, blue!50!mycolor1}, forget plot]
  table[row sep=crcr]{%
0	12.45\\
0.5	10.32\\
1	8.34\\
1.5	6.3\\
2	4.55\\
2.5	3.25\\
3	2.41\\
3.5	1.9\\
};
\addplot [color=mycolor2, mark=asterisk, mark options={solid, mycolor2}, forget plot]
  table[row sep=crcr]{%
0	23\\
0.5	18.55\\
1	14.29\\
1.5	10.07\\
2	6.71\\
2.5	4.35\\
3	2.9\\
3.5	2.09\\
};
\addplot [color=blue!50!mycolor2, mark=+, mark options={solid, blue!50!mycolor2}, forget plot]
  table[row sep=crcr]{%
0	41.95\\
0.5	33.28\\
1	24.2\\
1.5	16.08\\
2	9.98\\
2.5	5.9\\
3	3.54\\
3.5	2.33\\
};
\addplot [color=blue, mark=diamond, mark options={solid, blue}, forget plot]
  table[row sep=crcr]{%
0	74.08\\
0.5	58.37\\
1	40.78\\
1.5	25.63\\
2	14.78\\
2.5	8.02\\
3	4.35\\
3.5	2.59\\
};
\addplot [color=mycolor3, mark=square, mark options={solid, mycolor3}, forget plot]
  table[row sep=crcr]{%
0	124.42\\
0.5	99.14\\
1	67.96\\
1.5	40.4\\
2	21.94\\
2.5	10.95\\
3	5.38\\
3.5	2.9\\
};
\addplot [color=violet!40!mycolor3, mark=o, mark options={solid, violet!40!mycolor3}, forget plot]
  table[row sep=crcr]{%
0	196.98\\
0.5	158.73\\
1	110.19\\
1.5	62.68\\
2	32.3\\
2.5	14.91\\
3	6.71\\
3.5	3.25\\
};
\addplot [color=violet!80!mycolor3, mark=asterisk, mark options={solid, violet!80!mycolor3}, forget plot]
  table[row sep=crcr]{%
0	294.06\\
0.5	238.43\\
1	173.58\\
1.5	95.05\\
2	46.75\\
2.5	20.1\\
3	8.33\\
3.5	3.63\\
};
\addplot [color=red!50!mycolor3, mark=+, mark options={solid, red!50!mycolor3}, forget plot]
  table[row sep=crcr]{%
0	392.88\\
0.5	331.95\\
1	253.08\\
1.5	138.79\\
2	65.62\\
2.5	26.56\\
3	10.22\\
3.5	4.03\\
};
\addplot [color=mycolor4, mark=diamond, mark options={solid, mycolor4}, forget plot]
  table[row sep=crcr]{%
0	461.71\\
0.5	423.77\\
1	339.07\\
1.5	191.5\\
2	89.09\\
2.5	34.74\\
3	12.36\\
3.5	4.45\\
};
\addplot [color=red!50!mycolor4, mark=square, mark options={solid, red!50!mycolor4}, forget plot]
  table[row sep=crcr]{%
0	492.8\\
0.5	492\\
1	414.96\\
1.5	245.19\\
2	116.91\\
2.5	44.23\\
3	14.73\\
3.5	4.86\\
};
\addplot [color=red, mark=o, mark options={solid, red}, forget plot]
  table[row sep=crcr]{%
0	511.25\\
0.5	538.81\\
1	470.48\\
1.5	294.21\\
2	149.48\\
2.5	54.84\\
3	17.23\\
3.5	5.25\\
};
\end{axis}
\end{tikzpicture}%
} 
    \caption{Simulation results of the eBCH code $\mathscr{C}_{\textrm{eBCH}}[128,64]$.}
    \label{fig:bch64-vl}
  \end{figure}
  Fig.~\ref{fig:bch64-vl} shows that the performance of the eBCH code $\mathscr{C}_{\textrm{eBCH}}[128,64]$ improves as $\ell_{\textrm{max}}$ increases but saturates at a certain level of $\ell_{\textrm{max}}$. 
  This is reasonable because the performance of the LC-OSD is lower bound by that of the MLD, regardless of how large $\ell_{\textrm{max}}$ is.
  \end{example}

\rchange{
\subsection{Flipping Pattern Tree and Its Variants}
\label{sec:fpt}
This subsection presents the flipping pattern tree~(FPT) algorithm and its variants.

\subsubsection{Introduction of FPT}
FPT is a tree taking vectors in $\mathbb{F}_2^N$ as vertexes; the FPT algorithm is an algorithm that runs on the FPT to generate an ordered list of the $\ell$ best candidates. 
Without causing ambiguity, we also refer to the FPT algorithm simply as the FPT.
Unlike the LVA, the candidates of the FPT are not constrained.
Formally, let $N$ be the candidate's length, and $|\bm{r}|\in \mathbb{R}^N$ is the reliability vector.
Without loss of generality, we assume that the reliability vector is non-decreasing, i.e., 
\begin{equation}
  |r_1| \leq |r_2| \leq \cdots \leq |r_N|.
\end{equation}
The FPT can output one candidate at a time~(and at most $2^N$ candidates in total), with non-decreasing soft weight, i.e.,
\begin{equation}
  \Gamma({{\bm{e}}}^{(1)}) 
  \leq \Gamma({{\bm{e}}}^{(2)})
  \leq \cdots
  \leq \Gamma({{\bm{e}}}^{(\ell)})
  \leq \cdots
  \leq \Gamma({{\bm{e}}}^{(2^N)}).
\end{equation}

A brute-force implementation of FPT is to generate all $2^N$ candidates, sort them by soft weights in ascending order, and output them one by one. This brute-force method is suitable for small $N$ but does not work for large $N$ due to its exponential complexity.
To circumvent such a brute-force search, we give the following definition.

\begin{definition}({Precedence})%
\footnote{This is investigated in constructions of polar codes, such as~\cite{He2017Beta} and~\cite{Mondelli2019Construction}.}
\label{def:prec}
For two TEP $\bm{e}, \bm{e}' \in \mathbb{F}_2^N$, we say $\bm{e}$ \emph{precedes} $\bm{e}'$~(denoted $\bm{e} \preceq \bm{e}'$) if and only if there exists an injection $\pi: \supp(\bm{e}) \rightarrow \supp(\bm{e}')$ such that $i \leq \pi(i)$, where the support $\supp(\cdot)$ is defined as
\begin{equation}
  \label{equ:def-supp}
  \supp(\bm{e}) \triangleq \{i\in\{1,\ldots,N\}: e_i \neq 0\}.
\end{equation}
Here, the injection $\pi$ is called a \emph{witness} of $\bm{e} \preceq \bm{e}'$.
\end{definition}

\begin{proposition}
  \label{prop:def2-prec}
  If for all positive integer $j$, the inequality
  \begin{equation}
    \label{equ:def2-prec}
    |\supp_{\geq j}(\bm{e})| \leq |\supp_{\geq j}(\bm{e}')|.
  \end{equation}
  holds, then $\bm{e} \preceq \bm{e}'$ with a witness $\pi$ that maps the $t$-th greatest element of $\supp(\bm{e})$ to the $t$-th greatest element of $\supp(\bm{e}')$.
  To be precise, let 
  \begin{align}
    \supp(\bm{e})  &= \{i_1, i_2, \ldots, i_{w}\},     &w &=|\supp(\bm{e})|,  &i_{t}  &> i_{t+1}~(1\leq t < w),\\
    \supp(\bm{e}') &= \{i'_1, i'_2, \ldots, i'_{w'}\}, &w'&=|\supp(\bm{e}')|, &i'_{t} &> i'_{t+1}~(1\leq t < w').
  \end{align}
  Then $\pi(i_{t}) = i'_{t}~(1\leq t \leq w)$.

  Here $\supp_{\geq j}(\cdot)$ is defined as
  \begin{equation}
    \supp_{\geq j}(\bm{e}) \triangleq \{i \in \supp(\bm{e}): i \geq j\}.
  \end{equation}
\end{proposition}
\begin{IEEEproof}
  By substituting $j=1$ in~(\ref{equ:def2-prec}), we know that $w \leq w'$ and hence $\pi$ is a well-defined injection.
  For $1\leq t \leq w$, we have $t = |\supp_{\geq i_{t}}(\bm{e})| \leq |\supp_{\geq i_{t}}(\bm{e}')|$, implying that $i'_{t} \in \supp_{\geq i_{t}}(\bm{e}')$ and hence $i_{t} \leq i'_{t} = \pi(i_{t})$.
  Therefore, $\pi$ is a witness of $\bm{e} \preceq \bm{e}'$.
\end{IEEEproof}

Taking into account the fact that
\begin{equation}
  \pi(\supp_{\geq j}(\bm{e})) \subseteq \supp_{\geq j}(\bm{e}'),
\end{equation}
we see that Proposition~\ref{prop:def2-prec} is actually an equivalent definition for ``$\preceq$''~(Definition~\ref{def:prec}).
That is,
\begin{equation}
  \bm{e} \preceq \bm{e}' \Longleftrightarrow \forall j \in \mathbb{Z}^+\,\big(|\supp_{\geq j}(\bm{e})| \leq |\supp_{\geq j}(\bm{e}')|\big).
\end{equation}

\begin{proposition}
  $(\mathbb{F}_2^N, \preceq)$ is a partial order set.
\end{proposition}

\begin{IEEEproof}
  Let $\bm{e}^{(a)}, \bm{e}^{(b)}, \bm{e}^{(c)} \in \mathbb{F}_2^N$.
  \begin{enumerate}
    \item Reflexivity. The identity map is a witness of $\bm{e}^{(a)} \preceq \bm{e}^{(a)}$. 
    
    \item Transitivity. 
    Assume $\pi^{(ab)}$ and $\pi^{(bc)}$ are witnesses of $\bm{e}^{(a)} \preceq \bm{e}^{(b)}$ and $\bm{e}^{(b)} \preceq \bm{e}^{(c)}$, respectively.
    Let $\pi^{(ac)} \triangleq \pi^{(bc)}\pi^{(ab)}$.
    Then $\pi^{(ac)}: \supp(\bm{e}^{(a)}) \rightarrow \supp(\bm{e}^{(c)})$ is an injection.
    By Definition~\ref{def:prec}, we have
    \begin{equation}
      i \leq \pi^{(ab)}(i) \leq \pi^{(bc)}(\pi^{(ab)}(i)) = \pi^{(ac)}(i).
    \end{equation}
    Therefore $\pi^{(ac)}$ is a witness of $\bm{e}^{(a)} \preceq \bm{e}^{(c)}$.

    \item Antisymmetry.
    Assume $\pi^{(ab)}$ and $\pi^{(ba)}$ are witnesses of $\bm{e}^{(a)} \preceq \bm{e}^{(b)}$ and $\bm{e}^{(b)} \preceq \bm{e}^{(a)}$, respectively.
    Let $\pi^{(aa)} \triangleq \pi^{(ba)}\pi^{(ab)}$.
    Then $\pi^{(aa)}: \supp(\bm{e}^{(a)}) \rightarrow \supp(\bm{e}^{(a)})$ is a bijection.
    Moreover, because $\pi^{(aa)}$ is a witness of $\bm{e}^{(a)} \preceq \bm{e}^{(a)}$, $\pi^{(aa)}$ must be an identity map~(which can be proven by induction).
    By Definition~\ref{def:prec}, we have
    \begin{equation}
      i \leq \pi^{(ab)}(i) \leq \pi^{(ba)}(\pi^{(ab)}(i)) = \pi^{(aa)}(i) = i.
    \end{equation}
    Hence, $\pi^{(ab)}(i) = i$ and thus $\supp(\bm{e}^{(a)}) \subseteq \supp(\bm{e}^{(b)})$.
    Analogously, $\supp(\bm{e}^{(b)}) \subseteq \supp(\bm{e}^{(a)})$.
    Therefore, $\supp(\bm{e}^{(a)}) = \supp(\bm{e}^{(b)})$ and thus $\bm{e}^{(a)}=\bm{e}^{(b)}$.
  \end{enumerate}
\end{IEEEproof}

\begin{example}
  Let $N=4$. We have
  \begin{align}
    0000 &\preceq 1000 \preceq 0100 \preceq 0010 \preceq 0001, \\
    1000 &\preceq 1100 \preceq 1010 \preceq 1001 \preceq 0111, \\
    1001 &\not\preceq 0110, 0110 \not\preceq 1001,\\
    0001 &\not\preceq 1010, 1010 \not\preceq 0001,\\
    0000 &\preceq \bm{e} \preceq 1111, ~\forall \bm{e} \in \mathbb{F}_2^N.
  \end{align}
\end{example}

\begin{proposition}
  \label{prop:preserve-order}
  If $\bm{e} \preceq \bm{e}'$ and $|\bm{r}|$ is a non-decreasing vector, then $\Gamma(\bm{e}) \leq \Gamma(\bm{e}')$.
\end{proposition}
\begin{IEEEproof}
  Let $\pi$ be a witness of $\bm{e} \preceq \bm{e}'$.
  By Definition~\ref{def:soft-weight}, we have
  \begin{equation}
    \Gamma(\bm{e}) 
    = \sum_{i\in\supp(\bm{e})} |r_i| 
    \leq \sum_{i\in\supp(\bm{e})} |r_{\pi(i)}| 
    = \sum_{j\in\pi(\supp(\bm{e}))} |r_{j}| 
    \leq \sum_{j\in\supp(\bm{e}')} |r_j| 
    = \Gamma(\bm{e}').
  \end{equation}
\end{IEEEproof}
Proposition~\ref{prop:preserve-order} says that the precedence preserves the order of soft weights.
Conversely, we have the following proposition.

\begin{proposition}
  \label{prop:max-relation}
  If $\bm{e} \not\preceq \bm{e}'$, then there exits a non-decreasing vector $|\bm{r}|$ such that $\Gamma(\bm{e}) > \Gamma(\bm{e}')$.
\end{proposition}
\begin{IEEEproof}
  From Proposition~\ref{prop:def2-prec}, we know that there exists a positive integer $j$ such that
  \begin{equation}
    |\supp_{\geq j}(\bm{e})| > |\supp_{\geq j}(\bm{e}')|.
  \end{equation}
  Assigning
  \begin{equation}
    r_i =
    \begin{cases} 
      0, & \mbox{if } i < j\\
      1, & \mbox{if } i \geq j\\
    \end{cases},
  \end{equation}
  we have $\Gamma(\bm{e}) = |\supp_{\geq j}(\bm{e})| > |\supp_{\geq j}(\bm{e}')| = \Gamma(\bm{e}')$.
\end{IEEEproof}
}

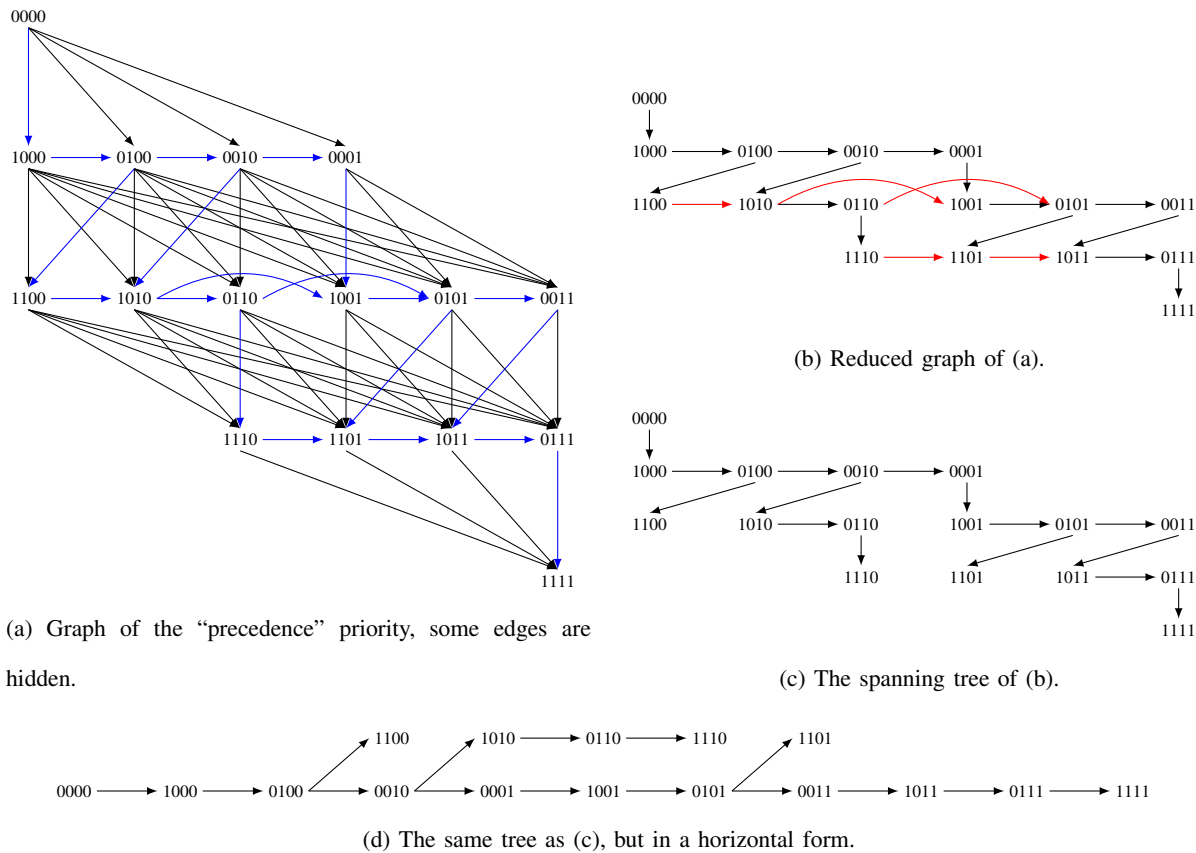
\begin{figure}[!t]
  \centering

  \begin{minipage}[t][.58\textwidth][t]{.5\textwidth}
    \vspace*{\fill}
    \centering
    \subfloat[\label{fig:fpt-full}Graph of the ``precedence'' priority, some edges are hidden.]{\begin{tikzpicture}[style={font=\large}, scale=0.46902, every node/.style={scale=0.46902}]
  \def\xfac{3pt}
  \def\yfac{4pt}
  \node at($\xfac*(0, 0) - \yfac*(0, 0)$) (0000) {0000};
  \node at($\xfac*(0, 0) - \yfac*(0, 1)$) (1000) {1000};
  \node at($\xfac*(1, 0) - \yfac*(0, 1)$) (0100) {0100};
  \node at($\xfac*(2, 0) - \yfac*(0, 1)$) (0010) {0010};
  \node at($\xfac*(3, 0) - \yfac*(0, 1)$) (0001) {0001};
  \node at($\xfac*(0, 0) - \yfac*(0, 2)$) (1100) {1100};
  \node at($\xfac*(1, 0) - \yfac*(0, 2)$) (1010) {1010};
  \node at($\xfac*(2, 0) - \yfac*(0, 2)$) (0110) {0110};
  \node at($\xfac*(3, 0) - \yfac*(0, 2)$) (1001) {1001};
  \node at($\xfac*(4, 0) - \yfac*(0, 2)$) (0101) {0101};
  \node at($\xfac*(5, 0) - \yfac*(0, 2)$) (0011) {0011};
  \node at($\xfac*(2, 0) - \yfac*(0, 3)$) (1110) {1110};
  \node at($\xfac*(3, 0) - \yfac*(0, 3)$) (1101) {1101};
  \node at($\xfac*(4, 0) - \yfac*(0, 3)$) (1011) {1011};
  \node at($\xfac*(5, 0) - \yfac*(0, 3)$) (0111) {0111};
  \node at($\xfac*(5, 0) - \yfac*(0, 4)$) (1111) {1111};

  \draw[-latex](0000.south) -- (0100.north);
  \draw[-latex](0000.south) -- (0010.north);
  \draw[-latex](0000.south) -- (0001.north);

  
  \draw[-latex](1000.south) -- (1100.north);
  \draw[-latex](1000.south) -- (1010.north);
  \draw[-latex](1000.south) -- (0110.north);
  \draw[-latex](1000.south) -- (1001.north);
  \draw[-latex](1000.south) -- (0101.north);
  \draw[-latex](1000.south) -- (0011.north);
  
  \draw[-latex](0100.south) -- (1010.north);
  \draw[-latex](0100.south) -- (0110.north);
  \draw[-latex](0100.south) -- (1001.north);
  \draw[-latex](0100.south) -- (0101.north);
  \draw[-latex](0100.south) -- (0011.north);
  
  \draw[-latex](0010.south) -- (0110.north);
  \draw[-latex](0010.south) -- (1001.north);
  \draw[-latex](0010.south) -- (0101.north);
  \draw[-latex](0010.south) -- (0011.north);
  
  \draw[-latex](0001.south) -- (0101.north);
  \draw[-latex](0001.south) -- (0011.north);


  \draw[-latex](1100.south) -- (1110.north);
  \draw[-latex](1100.south) -- (1101.north);
  \draw[-latex](1100.south) -- (1011.north);
  \draw[-latex](1100.south) -- (0111.north);

  \draw[-latex](1010.south) -- (1110.north);
  \draw[-latex](1010.south) -- (1101.north);
  \draw[-latex](1010.south) -- (1011.north);
  \draw[-latex](1010.south) -- (0111.north);

  \draw[-latex](0110.south) -- (1101.north);
  \draw[-latex](0110.south) -- (1011.north);
  \draw[-latex](0110.south) -- (0111.north);

  \draw[-latex](1001.south) -- (1101.north);
  \draw[-latex](1001.south) -- (1011.north);
  \draw[-latex](1001.south) -- (0111.north);

  \draw[-latex](0101.south) -- (1011.north);
  \draw[-latex](0101.south) -- (0111.north);

  \draw[-latex](0011.south) -- (0111.north);

  
  \draw[-latex](1110.south) -- (1111.north);
  \draw[-latex](1101.south) -- (1111.north);
  \draw[-latex](1011.south) -- (1111.north);
  

  \draw[-latex, blue](0000.south) -- (1000.north);
  \draw[-latex, blue](1000.east) -- (0100.west);
  \draw[-latex, blue](0100.east) -- (0010.west);
  \draw[-latex, blue](0010.east) -- (0001.west);
  \draw[-latex, blue](0100.south) -- (1100.north);
  \draw[-latex, blue](0010.south) -- (1010.north);
  \draw[-latex, blue](0001.south) -- (1001.north);
  \draw[-latex, blue](1100.east) -- (1010.west);
  \draw[-latex, blue](1010.east) -- (0110.west);
  \draw[-latex, blue](1001.east) -- (0101.west);
  \draw[-latex, blue](0101.east) -- (0011.west);
  \draw[-latex, blue](0110.south) -- (1110.north);
  \draw[-latex, blue](0101.south) -- (1101.north);
  \draw[-latex, blue](0011.south) -- (1011.north);
  \draw[-latex, blue](1110.east) -- (1101.west);
  \draw[-latex, blue](1101.east) -- (1011.west);
  \draw[-latex, blue](1011.east) -- (0111.west);
  \draw[-latex, blue](0111.south) -- (1111.north);
  \draw[-latex, blue](1010.east) to[bend left] (1001.west);
  \draw[-latex, blue](0110.east) to[bend left] (0101.west);


\end{tikzpicture}}
  \end{minipage}%
  \begin{minipage}[t][.58\textwidth][t]{.5\textwidth}
    \vspace*{\fill}
    \centering
    \subfloat[\label{fig:fpt-reduced}Reduced graph of~(a).]{\begin{tikzpicture}[style={font=\large}, scale=0.46902, every node/.style={scale=0.46902}]
  \def\xfac{3pt}
  \def\yfac{1.5pt}
  \node at($\xfac*(0, 0) - \yfac*(0, 0)$) (0000) {0000};
  \node at($\xfac*(0, 0) - \yfac*(0, 1)$) (1000) {1000};
  \node at($\xfac*(1, 0) - \yfac*(0, 1)$) (0100) {0100};
  \node at($\xfac*(2, 0) - \yfac*(0, 1)$) (0010) {0010};
  \node at($\xfac*(3, 0) - \yfac*(0, 1)$) (0001) {0001};
  \node at($\xfac*(0, 0) - \yfac*(0, 2)$) (1100) {1100};
  \node at($\xfac*(1, 0) - \yfac*(0, 2)$) (1010) {1010};
  \node at($\xfac*(2, 0) - \yfac*(0, 2)$) (0110) {0110};
  \node at($\xfac*(3, 0) - \yfac*(0, 2)$) (1001) {1001};
  \node at($\xfac*(4, 0) - \yfac*(0, 2)$) (0101) {0101};
  \node at($\xfac*(5, 0) - \yfac*(0, 2)$) (0011) {0011};
  \node at($\xfac*(2, 0) - \yfac*(0, 3)$) (1110) {1110};
  \node at($\xfac*(3, 0) - \yfac*(0, 3)$) (1101) {1101};
  \node at($\xfac*(4, 0) - \yfac*(0, 3)$) (1011) {1011};
  \node at($\xfac*(5, 0) - \yfac*(0, 3)$) (0111) {0111};
  \node at($\xfac*(5, 0) - \yfac*(0, 4)$) (1111) {1111};

  \draw[-latex](0000.south) -- (1000.north);

  \draw[-latex](1000.east) -- (0100.west);
  \draw[-latex](0100.east) -- (0010.west);
  \draw[-latex](0010.east) -- (0001.west);
  
  \draw[-latex](0100.south) -- (1100.north);
  \draw[-latex](0010.south) -- (1010.north);
  \draw[-latex](0001.south) -- (1001.north);

  \draw[-latex, red](1100.east) -- (1010.west);
  \draw[-latex](1010.east) -- (0110.west);
  \draw[-latex](1001.east) -- (0101.west);
  \draw[-latex](0101.east) -- (0011.west);

  \draw[-latex](0110.south) -- (1110.north);
  \draw[-latex](0101.south) -- (1101.north);
  \draw[-latex](0011.south) -- (1011.north);

  \draw[-latex, red](1110.east) -- (1101.west);
  \draw[-latex, red](1101.east) -- (1011.west);
  \draw[-latex](1011.east) -- (0111.west);
  
  \draw[-latex](0111.south) -- (1111.north);
  
  \draw[-latex, red](1010.east) to[bend left] (1001.west);
  \draw[-latex, red](0110.east) to[bend left] (0101.west);


\end{tikzpicture}}
    \par
    \subfloat[\label{fig:fpt-tree}The spanning tree of~(b).]{\begin{tikzpicture}[style={font=\large}, scale=0.46902, every node/.style={scale=0.46902}]
  \def\xfac{3pt}
  \def\yfac{1.5pt}
  \node at($\xfac*(0, 0) - \yfac*(0, 0)$) (0000) {0000};
  \node at($\xfac*(0, 0) - \yfac*(0, 1)$) (1000) {1000};
  \node at($\xfac*(1, 0) - \yfac*(0, 1)$) (0100) {0100};
  \node at($\xfac*(2, 0) - \yfac*(0, 1)$) (0010) {0010};
  \node at($\xfac*(3, 0) - \yfac*(0, 1)$) (0001) {0001};
  \node at($\xfac*(0, 0) - \yfac*(0, 2)$) (1100) {1100};
  \node at($\xfac*(1, 0) - \yfac*(0, 2)$) (1010) {1010};
  \node at($\xfac*(2, 0) - \yfac*(0, 2)$) (0110) {0110};
  \node at($\xfac*(3, 0) - \yfac*(0, 2)$) (1001) {1001};
  \node at($\xfac*(4, 0) - \yfac*(0, 2)$) (0101) {0101};
  \node at($\xfac*(5, 0) - \yfac*(0, 2)$) (0011) {0011};
  \node at($\xfac*(2, 0) - \yfac*(0, 3)$) (1110) {1110};
  \node at($\xfac*(3, 0) - \yfac*(0, 3)$) (1101) {1101};
  \node at($\xfac*(4, 0) - \yfac*(0, 3)$) (1011) {1011};
  \node at($\xfac*(5, 0) - \yfac*(0, 3)$) (0111) {0111};
  \node at($\xfac*(5, 0) - \yfac*(0, 4)$) (1111) {1111};

  \draw[-latex](0000.south) -- (1000.north);
  \draw[-latex](1000.east) --  (0100.west);
  \draw[-latex](0100.east) --  (0010.west);
  \draw[-latex](0010.east) --  (0001.west);
  \draw[-latex](0100.south) -- (1100.north);
  \draw[-latex](0010.south) -- (1010.north);
  \draw[-latex](1010.east) --  (0110.west);
  \draw[-latex](0001.south) -- (1001.north);
  \draw[-latex](1001.east) --  (0101.west);
  \draw[-latex](0101.east) --  (0011.west);
  \draw[-latex](0110.south) -- (1110.north);
  \draw[-latex](0101.south) -- (1101.north);
  \draw[-latex](0011.south) -- (1011.north);
  \draw[-latex](1011.east) --  (0111.west);
  \draw[-latex](0111.south) -- (1111.north);
\end{tikzpicture}}
  \end{minipage}
  \vspace{.5em}
  \subfloat[\label{fig:fpt-tree-laid}The same tree as~(c), but in a horizontal form.]{\begin{tikzpicture}[style={font=\large}, scale=0.46902, every node/.style={scale=0.46902}]
  \def\xfac{3pt}
  \def\yfac{1.5pt}
  \node at($\xfac*(0, 0) - \yfac*(0, 0)$) (0000) {0000};
  \node at($\xfac*(1, 0) - \yfac*(0, 0)$) (1000) {1000};
  \node at($\xfac*(2, 0) - \yfac*(0, 0)$) (0100) {0100};
  \node at($\xfac*(3, 0) - \yfac*(0, 0)$) (0010) {0010};
  \node at($\xfac*(4, 0) - \yfac*(0, 0)$) (0001) {0001};
  \node at($\xfac*(3, 0) - \yfac*(0, -1)$) (1100) {1100};
  \node at($\xfac*(4, 0) - \yfac*(0, -1)$) (1010) {1010};
  \node at($\xfac*(5, 0) - \yfac*(0, -1)$) (0110) {0110};
  \node at($\xfac*(5, 0) - \yfac*(0, 0)$) (1001) {1001};
  \node at($\xfac*(6, 0) - \yfac*(0, 0)$) (0101) {0101};
  \node at($\xfac*(7, 0) - \yfac*(0, 0)$) (0011) {0011};
  \node at($\xfac*(6, 0) - \yfac*(0, -1)$) (1110) {1110};
  \node at($\xfac*(7, 0) - \yfac*(0, -1)$) (1101) {1101};
  \node at($\xfac*(8, 0) - \yfac*(0, 0)$) (1011) {1011};
  \node at($\xfac*(9, 0) - \yfac*(0, 0)$) (0111) {0111};
  \node at($\xfac*(10, 0) - \yfac*(0, 0)$) (1111) {1111};

  \draw[-latex](0000.east) -- (1000.west);
  \draw[-latex](1000.east) -- (0100.west);
  \draw[-latex](0100.east) -- (0010.west);
  \draw[-latex](0010.east) -- (0001.west);
  \draw[-latex](0100.east) -- (1100.west);
  \draw[-latex](0010.east) -- (1010.west);
  \draw[-latex](1010.east) -- (0110.west);
  \draw[-latex](0001.east) -- (1001.west);
  \draw[-latex](1001.east) -- (0101.west);
  \draw[-latex](0101.east) -- (0011.west);
  \draw[-latex](0110.east) -- (1110.west);
  \draw[-latex](0101.east) -- (1101.west);
  \draw[-latex](0011.east) -- (1011.west);
  \draw[-latex](1011.east) -- (0111.west);
  \draw[-latex](0111.east) -- (1111.west);
\end{tikzpicture}}
  
  \caption{Graph representations of the FPT.}
  \label{fig:fpt}
\end{figure}

\rchange{
If one can build a graph~(see Fig.~\ref{fig:fpt-full} for example) on the ``precedence'' partial order, i.e., take TEPs as vertexes and ``$\preceq$'' as directed edges, one can generate candidates through the topological sort algorithm~\cite{Knuth1978art}. The time complexity is generally proportional to the average degree of the graph, which is $(2^N-1)/2$ in our case~(the proof is omitted here). For reducing the complexity~(or the average degree), one can equivalently run the topological sort on the reduced graph~(see Fig.~\ref{fig:fpt-reduced} for example).

However, the reduced graph has an average degree of $(N+1)/4$~(the proof is omitted here).
If we build a spanning tree~(see Fig.~\ref{fig:fpt-tree}) from the reduced graph, the average degree can be reduced to less than one.
Here, we give the method of systematically building a spanning tree from the reduced graph on the precedence partial order.

\begin{definition}{(Flipping pattern tree)%
  \footnote{
    This idea of building a tree for TEPs was proposed in~\cite{TANG2022Chase}~(with an earlier version~\cite{TANG2013Chase}) for decoding of RS codes and was applied to semi-LDPC-CCs~\cite{Chen2020SemiLDPC}. 
    Similar trees for binary codes were also constructed in~\cite[Algorithm~1]{yue2021probability} and~\cite[Algorithm~2]{solomon2020soft}. 
    Because the FPT is a spanning tree~(Fig.~\ref{fig:fpt-tree}) of the reduced graph~(Fig.~\ref{fig:fpt-reduced}) instead of the full graph~(Fig.~\ref{fig:fpt-full}), a careful inspection~(the details are omitted here) can find that the tree in this paper is the most effective in memory load.}}
  \label{def:fpt}
  The flipping pattern tree~(FPT) is a directed graph with TEPs as vertexes and edges defined in the following.
  For a TEP $\bm{e}$, denote ${\mathcal{S}} \triangleq \supp(\bm{e})$.
  If $1\notin {\mathcal{S}}$, the TEP $\bm{e}$ has an out-edge to the TEP with ${\mathcal{S}}_{\textrm{L}}(\bm{e})$ as the support set, where 
  \begin{equation}
    \label{equ:def-left-child}
    {\mathcal{S}}_{\textrm{L}}(\bm{e}) \triangleq {\mathcal{S}} \cup \{1\};
  \end{equation}
  if ${\mathcal{S}} \neq \varnothing$, $\min{\mathcal{S}} < N$ and $(\min{\mathcal{S}} + 1)\notin \mathcal{S}$, the TEP $\bm{e}$ has an out-edge to the TEP with ${\mathcal{S}}_{\textrm{R}}(\bm{e})$ as the support set, where 
  \begin{equation}
    \label{equ:def-right-child}
    {\mathcal{S}}_{\textrm{R}}(\bm{e}) \triangleq {\mathcal{S}} \setminus \{\min{\mathcal{S}}\} \cup \{\min{\mathcal{S}} + 1\}.
  \end{equation}
\end{definition}

We will see in the next example~(Example~\ref{ex:tep-cat}) that the flipping pattern tree is a rooted binary tree with the all-zero TEP as the root.
For a TEP $\bm{e}$ with ${\mathcal{S}} = \supp(\bm{e})$, we may refer to the TEP specified by ${\mathcal{S}}_{\textrm{L}}$ as the left child and ${\mathcal{S}}_{\textrm{R}}$ as the right child.

\newcommand\myX{\mathrm{X}}
\begin{example}
  \label{ex:tep-cat}
  From Definition~\ref{def:fpt}, we see that the left child~(if it exists) of a TEP is generated by setting its \emph{leftmost position} to $1$, and the right child~(if it exists) is generated by shifting its \emph{leftmost one} to the right.
  Therefore, we can classify the TEPs into several types by inspecting their parent and children.
  From the table below~(Table~\ref{tab:fpt-types}), we can conclude that the FPT takes $(00\cdots0)$ as root, vectors of the form $(11\myX \cdots \myX)$ as leaves, and the remaining vectors as inner vertexes.
  An inner vertex has two children if and only if it is in the form of $(00\cdots010\myX \cdots \myX)$, i.e., its leftmost one occurs in the context of the form $010$.
  \begin{table}[H]
    \centering
    \caption{Parent and children for different types$^*$ of TEPs.}
\begin{tabular}{c|c|c|c|c|c|c}
  \hline
  Type pattern
  & $(00\cdots0)$ 
  & \multicolumn{3}{c|}{$(00\cdots01 \myX \cdots \myX)$} 
  & \multicolumn{2}{c}{$(1 \myX \cdots \myX)$}
  \\
  \hline
  Parent
  & $-$ 
  & \multicolumn{3}{c|}{$(00\cdots\textcolor{red}{\underline{10}} \myX \cdots \myX)$}
  & \multicolumn{2}{c}{$(\textcolor{blue}{\underline{0}} \myX \cdots \myX)$}
  \\
  \hline
  \hline
  Type pattern
  & $(00\cdots0)$ 
  & $(00\cdots01)$ 
  & $(00\cdots010 \myX \cdots \myX)$ 
  & $(00\cdots011 \myX \cdots \myX)$ 
  & $(10 \myX \cdots \myX)$ 
  & $(11 \myX \cdots \myX)$ 
  \\
  \hline
  Left child 
  & $(\textcolor{blue}{\underline{1}}0\cdots0)$ 
  & $(\textcolor{blue}{\underline{1}}0\cdots01)$ 
  & $(\textcolor{blue}{\underline{1}}0\cdots010 \myX \cdots \myX)$
  & $(\textcolor{blue}{\underline{1}}0\cdots011 \myX \cdots \myX)$ 
  & $-$
  & $-$
  \\
  Right child 
  & $-$
  & $-$
  & $(00\cdots0\textcolor{red}{\underline{01}} \myX \cdots \myX)$
  & $-$
  & $(\textcolor{red}{\underline{01}} \myX \cdots \myX)$
  & $-$
  \\
  \hline
  \multicolumn{5}{l}{$^*$Here, `$\myX$' is a placeholder for either $0$ or $1$.} \\
\end{tabular}
    \label{tab:fpt-types}
  \end{table}
\end{example}

Based on the FPT, we proposed the FPT algorithm~(Algorithm~\ref{algo:FPT}).
In fact, the FPT algorithm is the specialization of topological sort on the FPT.
}

\begin{algorithm}[H]
  \renewcommand{\algorithmicrequire}{\textbf{Input:}}
  \renewcommand{\algorithmicensure}{\textbf{Output:}}
  \caption{Flipping pattern tree~(FPT) algorithm}
  \label{algo:FPT}
  \begin{algorithmic}[1]
    \Require the ascending sorted reliability vector $|\bm{r}|\in\mathbb{R}^N$, the rank of TEP to search $\ell$. 
    \Ensure The $\ell$-th lightest TEP $\bm{e}^{(\ell)}$.

    \State $\mathcal{Q}\leftarrow \{\bm{0}\}$
    \Comment{A priority queue allocated globally.}

    \Function{FPT}{$|\bm{r}|,\ell$}
    \Comment{This should be called in order of $\ell = 1, 2, \ldots, \ell_{\textrm{max}}$.}
      \State $\bm{e}^{(\ell)} \leftarrow \underset{\bm{e}\in\mathcal{Q}}{\arg\min}\, \Gamma(\bm{e})$
      \label{line:priority-queue}
      \State $\mathcal{Q}\leftarrow \mathcal{Q} \setminus \{\bm{e}^{(\ell)}\}$
      \If {${\mathcal{S}}_{\textrm{L}}(\bm{e}^{(\ell)})$ exists}
      \Comment{Left child, defined in~(\ref{equ:def-left-child}).}
        \State $\mathcal{Q}\leftarrow \mathcal{Q} \cup \{\mbox{the TEP specified by } {\mathcal{S}}_{\textrm{L}}(\bm{e}^{(\ell)})\}$
      \EndIf
      \If {${\mathcal{S}}_{\textrm{R}}(\bm{e}^{(\ell)})$ exists}
      \Comment{Right child, defined in~(\ref{equ:def-right-child}).}
        \State $\mathcal{Q}\leftarrow \mathcal{Q} \cup \{\mbox{the TEP specified by } {\mathcal{S}}_{\textrm{R}}(\bm{e}^{(\ell)})\}$
      \EndIf
      \State \Return $\bm{e}^{(\ell)}$
    \EndFunction
  \end{algorithmic}
\end{algorithm}

\rchange{
The time complexity of line~\ref{line:priority-queue} is proportional to $\log |\mathcal{Q}|$ if we use the priority queue as the data structure.
Since the FPT is a binary tree, the FPT algorithm removes $1$ element and adds at most $2$ to the set $\mathcal{Q}$ at each call.
Hence, the cardinality $|\mathcal{Q}| \leq \ell$.
Other lines of the FPT algorithm are some naive sequence manipulations with a computational time of $\mathcal{O}(N)$.
Therefore, the total time complexity%
\footnote{Because the priority queue can only store the soft weights and references to TEPs, its time cost is irrelevant to $N$.}
of the FPT algorithm is 
\begin{equation}
  T_{\text{avg}}^{\text{FPT}}
  = \underbrace{\mathcal{O}\bigg(\mathbb{E}_{\ell}\bigg[\sum_{j = 1}^{\ell} \log j \bigg]\bigg)}_\textrm{priority queue}
  + \underbrace{\mathcal{O}(\ell_{\textrm{avg}}N)}_\textrm{sequence manipulation}
  = \mathcal{O}(\ell_{\textrm{avg}} (\log \ell_{\textrm{max}} + N)),
\end{equation}
where the first term is simplified by
\begin{equation}
  \mathbb{E}_{\ell}\bigg[\sum_{j = 1}^{\ell} \log j \bigg]
  \leq \mathbb{E}_{\ell}\bigg[\sum_{j = 1}^{\ell} \log \ell_{\textrm{max}} \bigg]
  = \mathbb{E}_{\ell}[{\ell} \log \ell_{\textrm{max}}]
  = \ell_{\textrm{avg}} \log \ell_{\textrm{max}}.
\end{equation}

\subsubsection{Two-way FPT}
The FPT algorithm presented above is aimed at generating an un-constrained sequence over $\mathbb{F}_2^N$.
We can use the FPT under constraints using the BMA~\cite{Valembois2004Box} algorithm.

The two-way FPT~(tFPT), like the SLVA, can generate $\ell$ best codewords of a binary linear code $\mathscr{C}[N, K]$ with a parity-check matrix $\mathbf{P}$.
The basic idea of the tFPT is to separate the TEP into two halves, denoted by part~A and part~B.
Part~A~(part~B) is searched by the FPT algorithm, generating a list, called list~A~(list~B), of ordered but un-constrained candidates.
Then, two parts of TEPs, one from list~A and the other from list~B, are merged into a valid TEP if and only if the sum of their (partial) syndromes matches the total syndrome. 
Finally, all valid TEPs will be sorted before output.
In practice, the above procedure can be implemented in a ``calculate-when-request'' manner to reach a low
complexity.

In the following example~(Example~\ref{ex:tfpt-thr}), we use the tFPT as the LGA for the LC-OSD and measure the~(simulated) throughput of LC-OSD at different SNRs.
}
\begin{example}
  \label{ex:tfpt-thr}
  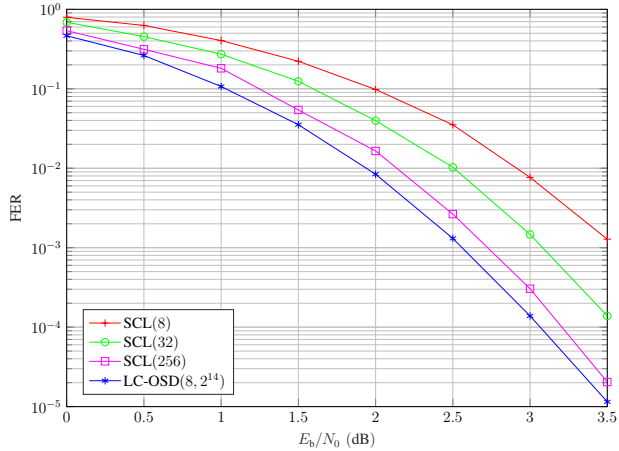
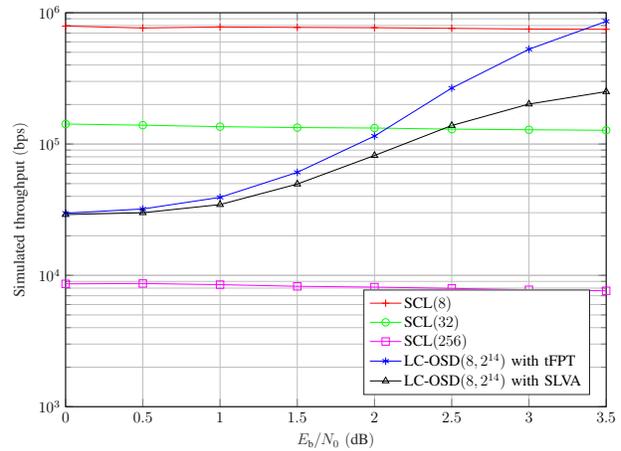
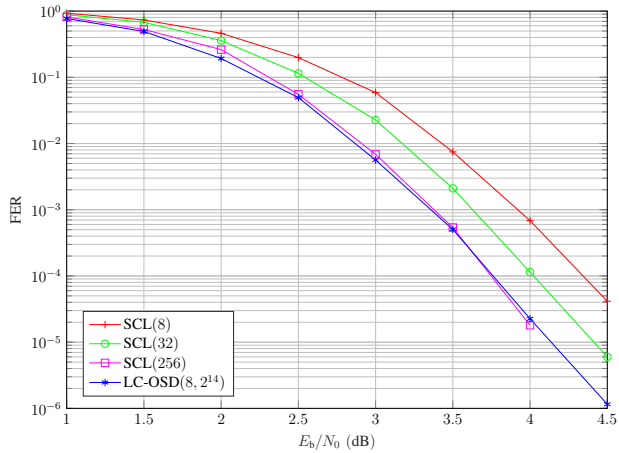
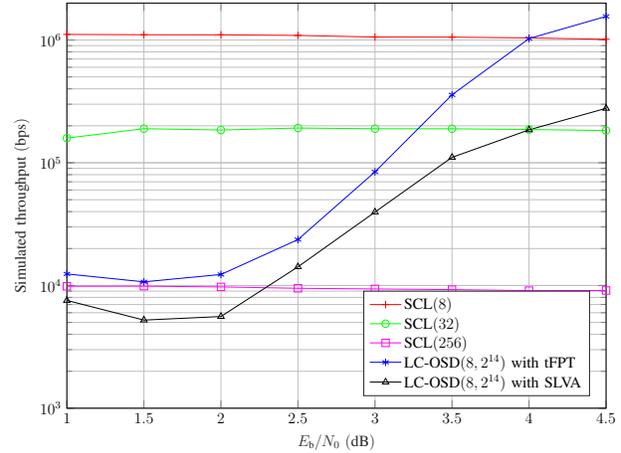
\begin{figure}[!t]
    \centering
    \subfloat[Performance of the CA-polar code {$\mathscr{C}_{\textrm{CA-polar}}[128,64]$}.\label{fig:capolar64-fer}]{
%
%
\definecolor{mycolor1}{rgb}{1.00000,0.00000,1.00000}%
\begin{tikzpicture}[%
thick,scale=0.7, every node/.style={scale=0.7}
]

\begin{axis}[%
width=10.267cm,
height=7.551cm,
at={(0cm,0cm)},
scale only axis,
xmin=0,
xmax=3.5,
xlabel style={font=\color{white!15!black}},
xlabel={$E_{\textrm{b}}/N_0~(\textrm{dB})$},
ymode=log,
ymin=1e-05,
ymax=1,
yminorticks=true,
ylabel style={font=\color{white!15!black}},
ylabel={FER},
axis background/.style={fill=white},
xmajorgrids,
ymajorgrids,
yminorgrids,
legend style={at={(0.03,0.03)}, anchor=south west, legend cell align=left, align=left, draw=white!15!black}
]
\addplot [color=red, mark=+, mark options={solid, red}]
  table[row sep=crcr]{%
0	0.7936507937\\
0.5	0.6269592476\\
1	0.4048582996\\
1.5	0.2219755827\\
2	0.0980872977\\
2.5	0.0352112676\\
3	0.0076731249\\
3.5	0.0012792141\\
};
\addlegendentry{SCL$(8)$}

\addplot [color=green, mark=o, mark options={solid, green}]
  table[row sep=crcr]{%
0	0.6872852234\\
0.5	0.4535147392\\
1	0.272851296\\
1.5	0.124300808\\
2	0.039698293\\
2.5	0.010292831\\
3	0.0014696265\\
3.5	0.0001385856\\
};
\addlegendentry{SCL$(32)$}

\addplot [color=mycolor1, mark=square, mark options={solid, mycolor1}]
  table[row sep=crcr]{%
0	0.539\\
0.5	0.314\\
1	0.181\\
1.5	0.0541\\
2	0.0165\\
2.5	0.00265\\
3	0.000305\\
3.5	2.04e-05\\
};
\addlegendentry{SCL$(256)$}

\addplot [color=blue, mark=asterisk, mark options={solid, blue}]
  table[row sep=crcr]{%
0	0.4651162791\\
0.5	0.2624671916\\
1	0.1068947087\\
1.5	0.0353169698\\
2	0.0084160916\\
2.5	0.0013096119\\
3	0.0001385302\\
3.5	1.15e-05\\
};
\addlegendentry{LC-OSD$(8, 2^{14})$}

\end{axis}
\end{tikzpicture}%
} 
    \hfill
    \subfloat[\rchange{Throughput of the CA-polar code {$\mathscr{C}_{\textrm{CA-polar}}[128,64]$}}.\label{fig:capolar64-thr}]{
%
%
\definecolor{mycolor1}{rgb}{1.00000,0.00000,1.00000}%
\begin{tikzpicture}[%
thick,scale=0.7, every node/.style={scale=0.7}
]

\begin{axis}[%
width=10.2666667cm,
height=7.4880713cm,
at={(0cm,0cm)},
scale only axis,
xmin=0,
xmax=3.5,
xlabel style={font=\color{white!15!black}},
xlabel={$E_{\textrm{b}}/N_0~(\textrm{dB})$},
ymode=log,
ymin=1000,
ymax=1000000,
yminorticks=true,
ylabel style={font=\color{white!15!black}},
ylabel={Simulated throughput $(\textrm{bps})$},
axis background/.style={fill=white},
xmajorgrids,
ymajorgrids,
yminorgrids,
legend style={at={(0.97,0.03)}, anchor=south east, legend cell align=left, align=left, draw=white!15!black}
]
\addplot [color=red, mark=+, mark options={solid, red}]
  table[row sep=crcr]{%
0	791603.2\\
0.5	765657.6\\
1	778969.6\\
1.5	773593.6\\
2	769273.6\\
2.5	760876.8\\
3	751500.8\\
3.5	748064\\
};
\addlegendentry{SCL$(8)$}

\addplot [color=green, mark=o, mark options={solid, green}]
  table[row sep=crcr]{%
0	142073.6\\
0.5	139244.8\\
1	135654.4\\
1.5	133465.6\\
2	132428.8\\
2.5	129888\\
3	128512\\
3.5	127308.8\\
};
\addlegendentry{SCL$(32)$}

\addplot [color=mycolor1, mark=square, mark options={solid, mycolor1}]
  table[row sep=crcr]{%
0	8620.8\\
0.5	8672\\
1	8499.2\\
1.5	8256\\
2	8140.8\\
2.5	7961.6\\
3	7769.6\\
3.5	7635.2\\
};
\addlegendentry{SCL$(256)$}

\addplot [color=blue, mark=asterisk, mark options={solid, blue}]
  table[row sep=crcr]{%
0	29750\\
0.5	32056\\
1	39287\\
1.5	60889\\
2	115001\\
2.5	266947\\
3	527280\\
3.5	859799\\
};
\addlegendentry{LC-OSD$(8, 2^{14})$ with tFPT}

\addplot [color=black, mark=triangle, mark options={solid, black}]
  table[row sep=crcr]{%
0	29063\\
0.5	30007\\
1	34612\\
1.5	49602\\
2	81836\\
2.5	138440\\
3	201663\\
3.5	250751\\
};
\addlegendentry{LC-OSD$(8, 2^{14})$ with SLVA}

\end{axis}
\end{tikzpicture}%
} 
    \\
    \subfloat[Performance of the CA-polar code {$\mathscr{C}_{\textrm{CA-polar}}[256,192]$}.\label{fig:capolar192-fer}]{
%
%
\definecolor{mycolor1}{rgb}{1.00000,0.00000,1.00000}%
\begin{tikzpicture}[%
thick,scale=0.7, every node/.style={scale=0.7}
]

\begin{axis}[%
width=10.267cm,
height=7.551cm,
at={(0cm,0cm)},
scale only axis,
xmin=1,
xmax=4.5,
xlabel style={font=\color{white!15!black}},
xlabel={$E_{\textrm{b}}/N_0~(\textrm{dB})$},
ymode=log,
ymin=1e-06,
ymax=1,
yminorticks=true,
ylabel style={font=\color{white!15!black}},
ylabel={FER},
axis background/.style={fill=white},
xmajorgrids,
ymajorgrids,
yminorgrids,
legend style={at={(0.03,0.03)}, anchor=south west, legend cell align=left, align=left, draw=white!15!black}
]
\addplot [color=red, mark=+, mark options={solid, red}]
  table[row sep=crcr]{%
1	0.9302325581\\
1.5	0.7352941176\\
2	0.4618937644\\
2.5	0.1986097319\\
3	0.0587889477\\
3.5	0.0074721662\\
4	0.0006836578\\
4.5	4.15e-05\\
};
\addlegendentry{SCL$(8)$}

\addplot [color=green, mark=o, mark options={solid, green}]
  table[row sep=crcr]{%
1	0.8810572687\\
1.5	0.6779661017\\
2	0.3590664273\\
2.5	0.1148765078\\
3	0.0227014756\\
3.5	0.0021065493\\
4	0.0001144078\\
4.5	6e-06\\
};
\addlegendentry{SCL$(32)$}

\addplot [color=mycolor1, mark=square, mark options={solid, mycolor1}]
  table[row sep=crcr]{%
1	0.816\\
1.5	0.528\\
2	0.262\\
2.5	0.0554\\
3	0.00686\\
3.5	0.000539\\
4	1.8e-05\\
4.5	0\\
};
\addlegendentry{SCL$(256)$}

\addplot [color=blue, mark=asterisk, mark options={solid, blue}]
  table[row sep=crcr]{%
1	0.7722007722\\
1.5	0.4926108374\\
2	0.1923076923\\
2.5	0.049382716\\
3	0.0056331681\\
3.5	0.0005038697\\
4	2.25764e-05\\
4.5	1.1528e-06\\
};
\addlegendentry{LC-OSD$(8, 2^{14})$}

\end{axis}
\end{tikzpicture}%
} 
    \hfill
    \subfloat[\rchange{Throughput of the CA-polar code {$\mathscr{C}_{\textrm{CA-polar}}[256,192]$}}.\label{fig:capolar192-thr}]{
%
%
\definecolor{mycolor1}{rgb}{1.00000,0.00000,1.00000}%
\begin{tikzpicture}[%
thick,scale=0.7, every node/.style={scale=0.7}
]

\begin{axis}[%
width=10.240255599999999cm,
height=7.699999999999999cm,
at={(0cm,0cm)},
scale only axis,
xmin=1,
xmax=4.5,
xlabel style={font=\color{white!15!black}},
xlabel={$E_{\textrm{b}}/N_0~(\textrm{dB})$},
ymode=log,
ymin=1000,
ymax=2000000,
yminorticks=true,
ylabel style={font=\color{white!15!black}},
ylabel={Simulated throughput $(\textrm{bps})$},
axis background/.style={fill=white},
xmajorgrids,
ymajorgrids,
yminorgrids,
legend style={at={(0.97,0.03)}, anchor=south east, legend cell align=left, align=left, draw=white!15!black}
]
\addplot [color=red, mark=+, mark options={solid, red}]
  table[row sep=crcr]{%
1	1111238.4\\
1.5	1104057.6\\
2	1102694.4\\
2.5	1090272\\
3	1057497.6\\
3.5	1056000\\
4	1040275.2\\
4.5	1013760\\
};
\addlegendentry{SCL$(8)$}

\addplot [color=green, mark=o, mark options={solid, green}]
  table[row sep=crcr]{%
1	158841.6\\
1.5	189465.6\\
2	184857.6\\
2.5	191884.8\\
3	189292.8\\
3.5	189100.8\\
4	186585.6\\
4.5	183072\\
};
\addlegendentry{SCL$(32)$}

\addplot [color=mycolor1, mark=square, mark options={solid, mycolor1}]
  table[row sep=crcr]{%
1	9849.6\\
1.5	9868.8\\
2	9734.4\\
2.5	9484.8\\
3	9369.6\\
3.5	9254.4\\
4	9081.6\\
4.5	9100.8\\
};
\addlegendentry{SCL$(256)$}

\addplot [color=blue, mark=asterisk, mark options={solid, blue}]
  table[row sep=crcr]{%
1	12417\\
1.5	10707\\
2	12286\\
2.5	23668\\
3	84302\\
3.5	357703\\
4	1027586\\
4.5	1557623\\
};
\addlegendentry{LC-OSD$(8, 2^{14})$ with tFPT}

\addplot [color=black, mark=triangle, mark options={solid, black}]
  table[row sep=crcr]{%
1	7554\\
1.5	5224\\
2	5587\\
2.5	14194\\
3	39655\\
3.5	110669\\
4	185417\\
4.5	277582\\
};
\addlegendentry{LC-OSD$(8, 2^{14})$ with SLVA}

\end{axis}
\end{tikzpicture}%
}
    \caption{Simulation results of the CA-polar codes, comparing the LC-OSD with the SCL decoding~\cite{tal2015list}.}
    \label{fig:capolar-perf}
  \end{figure}
  \rchange{In Fig.~\ref{fig:capolar-perf}}, we present the performance of two CRC-aided polar~(CA-polar) codes with different lengths, namely {$\mathscr{C}_{\textrm{CA-polar}}[128,64]$} and {$\mathscr{C}_{\textrm{CA-polar}}[256,192]$}, both of which are with $11$ bits CRCs.
  \rchange{
  The simulated throughput of the LC-OSD and the SCL decoders are also presented, where both the LC-OSD and the SCL decoders~\cite{Cassagne2019aff3ct} are implemented in C++. All of the above decoders are run on the same computer with the same environment%
  \footnote{We have measured the throughput in a single thread, respectively, on a desktop computer with a CPU of 11th Gen Intel(R) Core(TM) i7-11700F @ 2.50GHz.}.
  }
  \rchange{
  From Figs.~\ref{fig:capolar64-thr} and~\ref{fig:capolar192-thr}, we can see that, as the SNR increases, the throughputs of LC-OSDs~(including the tFPT and SLVA versions) in the presented examples keep increasing, which is consistent with our predictions.
  In contrast, the throughput of a~(particular) SCL decoder is relatively steady, no matter how high the SNR is.
  Comparing with the performance as well as the throughput, we conclude that
  } the LC-OSD outperforms the successive cancellation list~(SCL) decoding~\cite{tal2015list}, \rchange{especially in the high SNR region}.
  \rchange{From the figure, we also see that the LC-OSD with tFPT has a higher throughput than the SLVA in the high SNR region. This is due to the great portion of time consumed by the first candidate of the SLVA, which is also consistent with our predictions.}
\end{example}

\section{Conclusion}
In this paper, we have presented an extended version of OSD called the OSD with local constraints~(LC-OSD) algorithm. 
The LC-OSD uses more MRBs than the original OSD and solves the local constraint by list-generating algorithms~(LGAs). 
\rchange{The two presented LGAs, namely the SLVA and the tFPT, are suitable for different scenarios, where the SLVA is suitable for low SNR~(typically requiring a significant amount of searches), and tFPT is for high SNR~(typically requiring a small number of searches). }
We have also presented several early stopping criteria for the LC-OSD that significantly reduce computational complexity without noticeable performance loss. 
To analyze LC-OSD performance, we have proposed a method that uses random code to approximate the local constraint over MRBs. 
Several examples show that the approximation is quite accurate and does not depend on the structure of the codes. 
Based on the high accuracy of the approximation, we have proposed a tight approximate upper bound of the LC-OSD, which is applicable to different codes and decoding parameters. 
The random coding approach can lead to many research and applications related to the LC-OSD, one of which is tuning the decoding parameters according to applications. 

\section*{Acknowledgment}
The authors would like to express their gratitude to their colleague Dr. S Cai for his enlightening discussions and to Dr. M Fossorier for bringing~\cite{greenberger1979efficient} to their attention, of which they were previously unaware.

\bibliographystyle{IEEEtranTCOM}
\bibliography{ref}

\begin{thebibliography}{10}
\baselineskip 12pt
\providecommand{\url}[1]{#1}
\csname url@samestyle\endcsname
\providecommand{\newblock}{\relax}
\providecommand{\bibinfo}[2]{#2}
\providecommand{\BIBentrySTDinterwordspacing}{\spaceskip=0pt\relax}
\providecommand{\BIBentryALTinterwordstretchfactor}{4}
\providecommand{\BIBentryALTinterwordspacing}{\spaceskip=\fontdimen2\font plus
\BIBentryALTinterwordstretchfactor\fontdimen3\font minus
  \fontdimen4\font\relax}
\providecommand{\BIBforeignlanguage}[2]{{%
\expandafter\ifx\csname l@#1\endcsname\relax
\typeout{** WARNING: IEEEtran.bst: No hyphenation pattern has been}%
\typeout{** loaded for the language `#1'. Using the pattern for}%
\typeout{** the default language instead.}%
\else
\language=\csname l@#1\endcsname
\fi
#2}}
\providecommand{\BIBdecl}{\relax}
\BIBdecl

\bibitem{berlekamp1978inherent}
E.~Berlekamp, R.~McEliece, and H.~Van~Tilborg, ``On the inherent intractability
  of certain coding problems (corresp.),'' \emph{IEEE Transactions on
  Information Theory}, vol.~24, no.~3, pp. 384--386, 1978.

\bibitem{Fossorier1995OSD}
M.~Fossorier and S.~Lin, ``Soft-decision decoding of linear block codes based
  on ordered statistics,'' \emph{IEEE Transactions on Information Theory},
  vol.~41, no.~5, pp. 1379--1396, 1995.

\bibitem{Yue2019Segmentation}
C.~Yue, M.~Shirvanimoghaddam, Y.~Li, and B.~Vucetic, ``Segmentation-discarding
  ordered-statistic decoding for linear block codes,'' in \emph{2019 IEEE
  Global Communications Conference (GLOBECOM)}.\hskip 1em plus 0.5em minus
  0.4em\relax IEEE, 2019, pp. 1--6.

\bibitem{yue2022linear}
C.~Yue, M.~Shirvanimoghaddam, G.~Park, O.-S. Park, B.~Vucetic, and Y.~Li,
  ``Linear-equation ordered-statistics decoding,'' \emph{IEEE Transactions on
  Communications}, vol.~70, no.~11, pp. 7105--7123, 2022.

\bibitem{yue2021probability}
------, ``Probability-based ordered-statistics decoding for short block
  codes,'' \emph{IEEE Communications Letters}, vol.~25, no.~6, pp. 1791--1795,
  2021.

\bibitem{Wang2022ITW}
Y.~Wang, J.~Liang, and X.~Ma, ``Local constraint-based ordered statistics
  decoding for short block codes,'' in \emph{2022 IEEE Information Theory
  Workshop (ITW)}, 2022, pp. 107--112.

\bibitem{Liang2022CommL}
J.~Liang, Y.~Wang, S.~Cai, and X.~Ma, ``A low-complexity ordered statistic
  decoding of short block codes,'' \emph{IEEE Communications Letters}, vol.~27,
  no.~2, pp. 400--403, 2023.

\bibitem{greenberger1979efficient}
H.~Greenberger, ``An efficient soft decision decoding algorithm for block
  codes,'' Jet Propulsion Laboratory, Pasadena, Calif., Tech. Rep., January and
  February 1979.

\bibitem{may2011decoding}
A.~May, A.~Meurer, and E.~Thomae, ``Decoding random linear codes in
  {$O(2^{0.054n})$},'' in \emph{Advances in Cryptology--ASIACRYPT 2011: 17th
  International Conference on the Theory and Application of Cryptology and
  Information Security, Seoul, South Korea, December 4-8, 2011. Proceedings
  17}.\hskip 1em plus 0.5em minus 0.4em\relax Springer, 2011, pp. 107--124.

\bibitem{becker2012decoding}
A.~Becker, A.~Joux, A.~May, and A.~Meurer, ``Decoding random binary linear
  codes in {$2^{n/20}$}: How 1 + 1 = 0 improves information set decoding,'' in
  \emph{Advances in Cryptology--EUROCRYPT 2012: 31st Annual International
  Conference on the Theory and Applications of Cryptographic Techniques,
  Cambridge, UK, April 15-19, 2012. Proceedings 31}.\hskip 1em plus 0.5em minus
  0.4em\relax Springer, 2012, pp. 520--536.

\bibitem{may2015computing}
A.~May and I.~Ozerov, ``On computing nearest neighbors with applications to
  decoding of binary linear codes,'' in \emph{Advances in Cryptology--EUROCRYPT
  2015: 34th Annual International Conference on the Theory and Applications of
  Cryptographic Techniques, Sofia, Bulgaria, April 26-30, 2015, Proceedings,
  Part I}.\hskip 1em plus 0.5em minus 0.4em\relax Springer, 2015, pp. 203--228.

\bibitem{Martinez2011Saddlepoint}
A.~Martinez and A.~G. i~Fàbregas, ``Saddlepoint approximation of random-coding
  bounds,'' in \emph{2011 Information Theory and Applications Workshop}, 2011,
  pp. 1--6.

\bibitem{Font2018Saddlepoint}
J.~Font-Segura, G.~Vazquez-Vilar, A.~Martinez, A.~Guillén~i Fàbregas, and
  A.~Lancho, ``Saddlepoint approximations of lower and upper bounds to the
  error probability in channel coding,'' in \emph{2018 52nd Annual Conference
  on Information Sciences and Systems (CISS)}, 2018, pp. 1--6.

\bibitem{seshadri1994list}
N.~Seshadri and C.-E. Sundberg, ``List {V}iterbi decoding algorithms with
  applications,'' \emph{IEEE Transactions on Communications}, vol.~42, no. 234,
  pp. 313--323, 1994.

\bibitem{Nill1995List}
C.~Nill and C.-E. Sundberg, ``List and soft symbol output {V}iterbi algorithms:
  extensions and comparisons,'' \emph{IEEE Transactions on Communications},
  vol.~43, no. 2/3/4, pp. 277--287, 1995.

\bibitem{Valembois2004Box}
A.~Valembois and M.~Fossorier, ``Box and match techniques applied to
  soft-decision decoding,'' \emph{IEEE Transactions on Information Theory},
  vol.~50, no.~5, pp. 796--810, 2004.

\bibitem{yue2021revisit}
C.~Yue, M.~Shirvanimoghaddam, B.~Vucetic, and Y.~Li, ``A revisit to ordered
  statistics decoding: Distance distribution and decoding rules,'' \emph{IEEE
  Transactions on Information Theory}, vol.~67, no.~7, pp. 4288--4337, 2021.

\bibitem{valembois2002comparison}
A.~Valembois and M.~Fossorier, ``A comparison between ``most-reliable-basis
  reprocessing'' strategies,'' \emph{IEICE transactions on fundamentals of
  electronics, communications and computer sciences}, vol.~85, no.~7, pp.
  1727--1741, 2002.

\bibitem{Baldi2016OSD}
M.~Baldi, N.~Maturo, E.~Paolini, and F.~Chiaraluce, ``On the use of ordered
  statistics decoders for low-density parity-check codes in space telecommand
  links,'' \emph{EURASIP Journal on Wireless Communications and Networking},
  vol. 2016, no.~1, pp. 1--15, 2016.

\bibitem{TANG2022Chase}
S.~Tang, S.~Cai, and X.~Ma, ``A new {C}hase-type soft-decision decoding
  algorithm for {R}eed-{S}olomon codes,'' \emph{Alexandria Engineering
  Journal}, vol.~61, no.~12, pp. 13\,067--13\,077, 2022.

\bibitem{xiao2003path}
X.~Ma and A.~Kavcic, ``Path partitions and forward-only trellis algorithms,''
  \emph{IEEE Transactions on Information Theory}, vol.~49, no.~1, pp. 38--52,
  2003.

\bibitem{forney1973viterbi}
G.~Forney, ``The viterbi algorithm,'' \emph{Proceedings of the IEEE}, vol.~61,
  no.~3, pp. 268--278, 1973.

\bibitem{bahl1974optimal}
L.~Bahl, J.~Cocke, F.~Jelinek, and J.~Raviv, ``Optimal decoding of linear codes
  for minimizing symbol error rate (corresp.),'' \emph{IEEE Transactions on
  Information Theory}, vol.~20, no.~2, pp. 284--287, 1974.

\bibitem{lin1998trellises}
S.~Lin, T.~Kasami, T.~Fujiwara, and M.~Fossorier, \emph{Trellises and
  trellis-based decoding algorithms for linear block codes}.\hskip 1em plus
  0.5em minus 0.4em\relax Springer Science \& Business Media, 1998.

\bibitem{lin2020thesis}
W.~Lin, ``Research on block {M}arkov superposition transmission for
  ultra-reliable and low-latency communiccation,'' Master's thesis, Sun Yat-sen
  University, 2020.

\bibitem{He2017Beta}
G.~He, J.-C. Belfiore, I.~Land, G.~Yang, X.~Liu, Y.~Chen, R.~Li, J.~Wang,
  Y.~Ge, R.~Zhang, and W.~Tong, ``Beta-expansion: A theoretical framework for
  fast and recursive construction of polar codes,'' in \emph{GLOBECOM 2017 -
  2017 IEEE Global Communications Conference}, 2017, pp. 1--6.

\bibitem{Mondelli2019Construction}
M.~Mondelli, S.~H. Hassani, and R.~L. Urbanke, ``Construction of polar codes
  with sublinear complexity,'' \emph{IEEE Transactions on Information Theory},
  vol.~65, no.~5, pp. 2782--2791, 2019.

\bibitem{Knuth1978art}
D.~E. Knuth, \emph{The art of computer programming. {V}ol. 1: Fundamental
  algorithms}.\hskip 1em plus 0.5em minus 0.4em\relax Addison-Wesley, 1978.

\bibitem{TANG2013Chase}
S.~Tang, S.~Cai, and X.~M. Member, ``A new {C}hase-type soft-decision decoding
  algorithm for {R}eed-{S}olomon codes,'' \emph{arXiv preprint
  arXiv:1309.1555}, 2013.

\bibitem{Chen2020SemiLDPC}
Z.~Chen, S.~Cai, L.~Chen, and X.~Ma, ``Semi-{LDPC} convolutional codes with
  low-latency decoding algorithm,'' in \emph{2020 IEEE 6th International
  Conference on Computer and Communications (ICCC)}, 2020, pp. 105--109.

\bibitem{solomon2020soft}
A.~Solomon, K.~R. Duffy, and M.~Médard, ``Soft maximum likelihood decoding
  using {GRAND},'' in \emph{ICC 2020 - 2020 IEEE International Conference on
  Communications (ICC)}, 2020, pp. 1--6.

\bibitem{tal2015list}
I.~Tal and A.~Vardy, ``List decoding of polar codes,'' \emph{IEEE Transactions
  on Information Theory}, vol.~61, no.~5, pp. 2213--2226, 2015.

\bibitem{Cassagne2019aff3ct}
A.~Cassagne, O.~Hartmann, M.~Leonardon, K.~He, C.~Leroux, R.~Tajan, O.~Aumage,
  D.~Barthou, T.~Tonnellier, V.~Pignoly \emph{et~al.}, ``{AFF3CT}: A fast
  forward error correction toolbox!'' \emph{SoftwareX}, vol.~10, p. 100345,
  2019.

\end{thebibliography}
\vfill
\end{document}